\documentclass[twocolumn]{aastex701}
\usepackage{amsmath}
\usepackage{inconsolata} 
\usepackage[T1]{fontenc}  
\usepackage{xcolor}

\newcommand{\vlsr}     {v_\mathrm{lsr}}
\newcommand{\vsys}     {v_\mathrm{sys}}
\newcommand{\vout}     {v_\mathrm{out}}
\newcommand{\voutc}     {v_\mathrm{out,corr}}
\newcommand{\vrot}     {v_\mathrm{rot}}
\newcommand{\vinf}     {v_\mathrm{inf}}
\newcommand{\pa} {\mathrm{P.A.}}

\newcommand{\rcb}     {R_\mathrm{CB}}

\newcommand{\h} {^{\mathrm{h}}}
\newcommand{\m} {^{\mathrm{m}}}
\newcommand{\s} {^{\mathrm{s}}}

\newcommand{\mJybeam}  {\mbox{mJy}~\mbox{beam}^{-1}}

\newcommand{\mJybeamkms}  {\mbox{mJy}~\mbox{beam}^{-1}~\mbox{km s}^{-1}}

\newcommand{\kms}	{\mbox{km s}^{-1}}

\newcommand{\K}	{{\rm K}}

\newcommand{\au} {\mbox{au}}

\newcommand{\mJy} {\mbox{mJy}}

\newcommand*\chem[1]{\ensuremath{\mathrm{#1}}}

\begin{document}

\title{ALMA High-resolution Observation of the HH46/47 Outflow/disk/envelope System}

\correspondingauthor{Yichen Zhang}

\author[orcid=0009-0006-7369-1318]{Heyi Zhang}
\affiliation{State Key Laboratory of Dark Matter Physics, School of Physics and Astronomy, Shanghai Jiao Tong University, Shanghai 200240, People’s Republic of China}
\affiliation{Department of Astronomy, School of Physics and Astronomy, Shanghai Jiao Tong University, 800 Dongchuan Road, Shanghai 200240, People’s Republic of China}
\affiliation{Key Laboratory for Particle Astrophysics and Cosmology (MOE) / Shanghai Key Laboratory for Particle Physics and Cosmology, Shanghai 200240, People’s Republic of China}
\email{z1h2y3456@sjtu.edu.cn}  

\author[orcid=0000-0001-7511-0034]{Yichen Zhang}
\affiliation{State Key Laboratory of Dark Matter Physics, School of Physics and Astronomy, Shanghai Jiao Tong University, Shanghai 200240, People’s Republic of China}
\affiliation{Department of Astronomy, School of Physics and Astronomy, Shanghai Jiao Tong University, 800 Dongchuan Road, Shanghai 200240, People’s Republic of China}
\affiliation{Key Laboratory for Particle Astrophysics and Cosmology (MOE) / Shanghai Key Laboratory for Particle Physics and Cosmology, Shanghai 200240, People’s Republic of China}
\email[show]{yczhang.astro@gmail.com}  

\author[orcid=0000-0001-5653-7817]{H\'ector G. Arce}
\affiliation{Department of Astronomy, Yale University, New Haven, CT 06511, USA}
\email{hector.arce@yale.edu}  

\author[orcid=0000-0002-5065-9175]{Diego Mardones}
\affiliation{Departamento de Astronom\'ia, Universidad de Chile, Las Condes, 7591245 Santiago, Chile}
\email{d.mardones67@gmail.com}  

\author[orcid=0000-0002-1593-3693]{Sylvie Cabrit}
\affiliation{LERMA, Observatoire de Paris-PSL, Sorbonne Universit\'e, CNRS, F-75014 Paris, France}
\affiliation{IPAG, Observatoire de Grenoble, Universit\'e Grenoble-Alpes, France}
\email{sylvie.cabrit@obspm.fr}  

\author[orcid=0000-0003-0749-9505]{Michael M. Dunham}
\affiliation{Department of Physics, Middlebury College, Middlebury, VT 05753, USA}
\email{mdunham@middlebury.edu}  

\author[orcid=0000-0003-1252-9916]{Stella S. R. Offner}
\affiliation{Department of Astronomy, The University of Texas at Austin, Austin, TX 78712, USA}
\email{soffner@astro.as.utexas.edu}  

\author[orcid=0000-0001-8385-9838]{Hsien Shang}
\affiliation{Institute of Astronomy and Astrophysics, Academia Sinica, Taipei 106319, Taiwan}
\email{}  

\begin{abstract}
We present $0.1\arcsec$ ($\sim 50~\au$) resolution Atacama Large Millimeter/submillimeter Array (ALMA) observations of 
the HH 46/47 molecular outflow and its envelope-disk system. 
The 1.3 mm continuum emission reveals a compact central source surrounded by a circumbinary disk with substructures. 
The companion, identified in optical and infrared observations, is not detected in the millimeter continuum 
but coincides with a local intensity minimum. 
Two spur-like features extending from the primary source toward the companion are 
identified and are likely induced by gravitational perturbations from the companion.
The envelope-disk system is traced by $\chem{C^{18}O}$, SO, $\chem{H_2CO}$, and $\chem{CH_3OH}$. 
$\chem{C^{18}O}$ primarily traces the extended envelope, while SO probes the inner envelope, 
and $\chem{H_2CO}$ and $\chem{CH_3OH}$ trace compact, faster-rotating structures near the centrifugal barrier. 
The observations are well reproduced by a rotating-infalling envelope transitioning to an inner disk 
at a radius of $\sim 30$ au around a $0.3~M_\odot$ protostar.
The $\chem{^{12}CO}$ emission, together with JWST NIRCam images, reveals multiple shell structures in the outflow. 
Using $\chem{C^{18}O}$ and $\chem{^{13}CO}$ to correct for optical depth, 
we derive the spatial distributions of outflow mass, momentum, and kinetic energy, as well as their corresponding rates. 
A model-independent analysis of a well-defined redshifted shell yields its three-dimensional velocity field, 
showing that the shell expands radially rather than flowing along its surface. 
Although a transverse velocity gradient is detected, interpreting it as rotation implies an unphysically large magnetic lever arm, 
disfavoring a direct disk-wind origin. 
Instead, the shell kinematics support an entrainment scenario.
\end{abstract}

\section{Introduction}
\label{sec:intro}

Low-mass stars like the Sun form through the collapse of gravitationally bound molecular cloud cores. 
A typical embedded protostellar system consists of a protostellar disk, an infalling–rotating envelope, and bipolar outflows 
(e.g., \citealt{Shu1987,Frank2014}). 
The disk and envelope form as a consequence of angular momentum conservation during mass accretion (e.g., \citealt{Ulrich1976}), 
while outflows play a crucial role in regulating star formation by removing angular momentum and 
injecting energy and momentum into the surrounding environment (e.g., \citealt{Frank2014}).

Protostellar outflows are commonly observed to consist of two components: a high-velocity, collimated jet and a slower, wide-angle molecular outflow 
(e.g., \citealt{Frank2014,Arce2007,Ray2023,Hsieh2023}).
The jets are widely believed to be launched directly from the inner star–disk system through 
magneto-hydrodynamic (MHD) processes (e.g., \citealt{Lee2017}). 
In contrast, the origin of the slower molecular outflows remains debated. 
In the classical entrainment scenario, molecular outflows consist of ambient gas swept up by a fast jet or a wide-angle wind, 
forming shell-like structures and cavities (e.g., \citealt{Shu1991,Lee2001,Arce2013,Zhang2016,Shang2023}). 
Alternatively, disk-wind models propose that at least part of the molecular outflow, particularly near its base, 
directly traces material launched from the disk via magneto-centrifugal or thermal processes 
(e.g., \citealt{Blandford1982,Pudritz2007}). 
In this picture, observed transverse velocity gradients across the outflow axis are often interpreted as signatures of rotation 
(e.g., \citealt{Bjerkeli2016,Hirota2017,Zhang2018,deValon2022,Bacciotti2025,Kim2026}). 
Distinguishing between these scenarios is challenging, as molecular line emission (e.g., CO) 
often traces entrained material rather than the primary wind itself.

High-resolution observations have revealed that molecular outflows frequently exhibit complex structures, 
including nested shells and cavities (e.g., \citealt{Zhang2019,devalon2020,Vazquez2024}).
These features are often interpreted as the result of episodic mass ejection events, potentially linked to accretion variability. 
The kinematics of such shells provide key diagnostics of the underlying driving mechanism, 
as disk winds and entrainment models predict different velocity structures. 
Distinct shells enable a cleaner measurement of the velocity field, 
whereas in more continuous flows emission is blended along the line of sight (\citealt[]{Zhang2019,deValon2022}).
However, many previous studies rely on simplified analytic models, and direct, model-independent measurements of outflow kinematics remain limited.

In addition, multiplicity can significantly affect disk structure and outflow morphology. 
Gravitational interactions in binary systems may induce substructures, warping, and misalignment between inner and outer disk regions, 
potentially leading to differences in the orientations of jets and wide-angle outflows. 
Understanding how binarity influences both disk evolution and outflow launching is therefore essential 
for building a comprehensive picture of star formation.

In this study, we present high spatial resolution ($\sim 0.1\arcsec$, $\sim 45~\au$) observations 
of the HH 46/47 molecular outflow and its disk–envelope system, 
focusing on the circumstellar environment and the base of the outflow. 
The HH 46/47 outflow is driven by HH 46 IRS (2MASS J08254384$-$5100326), 
a low-mass early Class I protostar located in a Bok globule at the edge of the Gum Nebula 
at a distance of 450 pc (\citealt{Reipurth2000,Schwartz1977,Noriega-Crespo2004}). 
HH 46 IRS has been identified as a binary system with an apparent separation of $\sim 0.23\arcsec$ ($\sim 100$ au) 
based on infrared observations from HST and JWST (\citealt{Reipurth2000,Nisini2024}).

The HH 46/47 outflow has been extensively studied over the past decade. 
ALMA observations (\citealt{Arce2013,Zhang2016,Zhang2019}) revealed a highly asymmetric, 
wide-angle CO outflow, with the redshifted lobe extending significantly farther than the blueshifted lobe. 
Well-defined, coherent shell structures have been identified on both sides of the outflow (\citealt{Zhang2019}), 
and were interpreted as entrained shells produced by episodic wide-angle winds, 
while jet-driven entrainment may dominate on larger scales. 
More recent infrared observations with JWST and VLT (\citealt{Nisini2024,Birney2024,Navarro2025}) 
have revealed a highly collimated jet in the immediate vicinity of the central source, 
as well as complex structures including cavities, molecular shells, and jet-driven bow shocks.

Despite these advances, the disk and envelope system of HH 46 IRS has only been studied at relatively low resolution 
($\sim 1\arcsec$, $\sim 450$ au; \citealt{Zhang2016}), 
revealing a large-scale flattened infalling and rotating envelope and placing an upper limit of $\sim 380$ au on the disk radius. 
The detailed structure of the inner disk–envelope system and its connection to the binary companion remain unclear.

In this paper, we present the highest-resolution millimeter observations of HH 46/47 obtained to date. 
Using multiple molecular tracers together with a model-independent analysis of the outflow shells, 
we investigate the disk–envelope system, resolve the circumbinary structures, 
and characterize the morphology, kinematics, and origin of the outflow shells.
Section \ref{sec:obs} describes the observational setup and data reduction procedures. 
Section \ref{sec:result} presents the continuum and molecular line data, 
and Section \ref{sec:discussion} provides a detailed analysis and discussion of the disk, envelope, and outflow properties. 
Finally, Section \ref{sec:conclusion} summarizes our main findings.


\begin{table*}[ht!]
\scriptsize
\begin{center}
\caption{Parameters of the Observed Lines\footnote{Line information taken from the 
CDMS database (\citealt[]{Muller05})} \label{tab:lines}}
\begin{tabular}{lccccccc}
\hline
\hline
Molecule & Transition & Frequency & $E_u/k$ & $A_{ul}$ & Velocity Resolution & Synthesized Beam & Channel rms\\
 & & (GHz) & (K) & (s$^{-1}$) & ($\kms$) & & ($\mJybeam$)\\
\hline
$^{12}$CO & $2-1$ & 230.5380000 & 16.6 & $6.910 \times 10^{-7}$ & 0.3 & $0.12\arcsec\times 0.11\arcsec$  ($\pa=-53.5^\circ$) & 0.85 \\
$^{13}$CO & $2-1$ & 220.3986842 & 15.9 & $5.066 \times 10^{-7}$ & 0.3 & $0.14\arcsec\times 0.13\arcsec$  
($\pa=-59.9^\circ$) & 1.07\\
C$^{18}$O & $2-1$ & 219.5603577 & 15.8 & $6.011 \times 10^{-7}$  & 0.3 & $0.14\arcsec\times 0.13\arcsec$  ($\pa=-51.0^\circ$) & 0.80 \\

SO & $6_5 - 5_4$ & 219.9494420 & 35.0 & $1.335 \times 10^{-4}$  & 0.3 & $0.11\arcsec\times 0.10\arcsec$  ($\pa=-47.6^\circ$) & 1.20\\
CH$_{3}$OH & $4_{2,2}-3_{1,2}$; E & 218.4400630 & 45.5 & $4.686 \times 10^{-5}$ & 0.3 & $0.13\arcsec\times 0.12\arcsec$  ($\pa= -54.8^\circ$) & 0.65 \\
H$_{2}$CO & $3_{2,1} - 2_{2,0}$ & 218.7600660 & 68.1 & $1.577 \times 10^{-4}$  & 0.3 & $0.11\arcsec\times 0.10\arcsec$  ($\pa=-46.6^\circ$) & 0.75 \\
\hline
\end{tabular}
\end{center}
\end{table*}

\section{Observations}
\label{sec:obs}

The $\sim$0.1$\arcsec$ resolution ALMA Band 6 (1.3 mm) observations were obtained 
on 2019 August 20, 22, and 25 in the C43-7 configuration (hereafter C7; project ID: 2018.1.01625.S, PI: Mardones).
A total of 45 antennas were used, providing baseline lengths from 41.4 m to 3.6 km. 
J1107$-$4449 and J0538$-$4405 were used for bandpass and flux calibration, 
while J0845$-$5458 served as the phase calibrator. 
The total on-source integration time was 198 minutes. 
Data calibration was performed using the CASA pipeline (version 5.4.0).
Following pipeline calibration, self-calibration was carried out 
using the 1.8 GHz-wide continuum spectral window. 
We performed two iterations of phase-only self-calibration with solution intervals of 30 s and 6 s, 
followed by one iteration of amplitude self-calibration with solution intervals equal to the scan length. 
The derived continuum self-calibration solutions were subsequently applied to the spectral line data.

We further combine the $\sim$0.1$\arcsec$ resolution continuum and line data with previously obtained 
lower-resolution Band 6 observations presented by \citet[]{Zhang2019}. 
Here we briefly summarize the earlier observations and data reduction and 
refer the reader to that work for further details.
The lower-resolution Band 6 data were obtained with ALMA on 2016 January 6 in the C36-2 configuration, 
and on 2016 June 21, 30, and July 6 in the C36-4 configuration 
(hereafter C2+C4; project ID: 2015.1.01068.S, PI: Zhang). 
In the C36-2 observations, 36 antennas provided baselines ranging from 15 to 310 m, 
with a total on-source integration time of 75 minutes. 
J1107$-$4449 and J0538$-$4405 were used for bandpass and flux calibration, 
while J0811$-$4929 and J0904$-$5735 served as phase calibrators. 
In the C36-4 observations, 36 antennas provided baselines from 15 to 704 m, 
with a total integration time of 150 minutes. 
J1107$-$4449 and J0538$-$4405 were used for bandpass and flux calibration, 
and J0811$-$4929 was used as the phase calibrator. 
The data were calibrated using CASA version 4.5.3, 
followed by self-calibration on the continuum after standard calibration.

Additional self-calibration was performed to align the Band 6 data obtained in the three configurations 
prior to joint imaging. 
For the continuum and most spectral lines, data from all configurations are available, and the final images 
are produced from the combined data sets (hereafter C2+C4+C7). 
Imaging was performed using the CASA {\it tclean} task with Briggs weighting
(\citealt{Briggs1995}), which provides a compromise between natural weighting
(maximum sensitivity) and uniform weighting (maximum angular resolution). Unless
otherwise noted, we adopted a robust parameter of 0.5, which provides high
sensitivity while maintaining good angular resolution.
The resulting continuum image has a synthesized beam of
$0.12\arcsec \times 0.11\arcsec$ ($\pa=-50.8^\circ$)
and an rms sensitivity of $1\sigma = 0.009~\mJybeam$.
To better resolve the compact circumstellar structures, we also produced
images with a robust parameter of $-0.5$, which places greater weight on the
longer baselines and thus improves the angular resolution at the expense of
reduced sensitivity.
The corresponding synthesized beam is $0.09\arcsec \times 0.08\arcsec$ ($\pa=-46.2^\circ$), 
and the rms sensitivity is $1\sigma = 0.018~\mJybeam$. 
These higher-resolution images are used where noted
below.
The angular resolutions and other properties of the spectral line data are summarized in Table~\ref{tab:lines}.
Throughout this paper, 
we adopt a systemic velocity of $\vsys = 5.3~\kms$ (\citealt[]{vanKempen09}) 
and define the outflow velocity as $\vout = \vlsr - \vsys$.


\section{Results}
\label{sec:result}

\begin{figure*}[ht!]
\includegraphics[width=\textwidth]{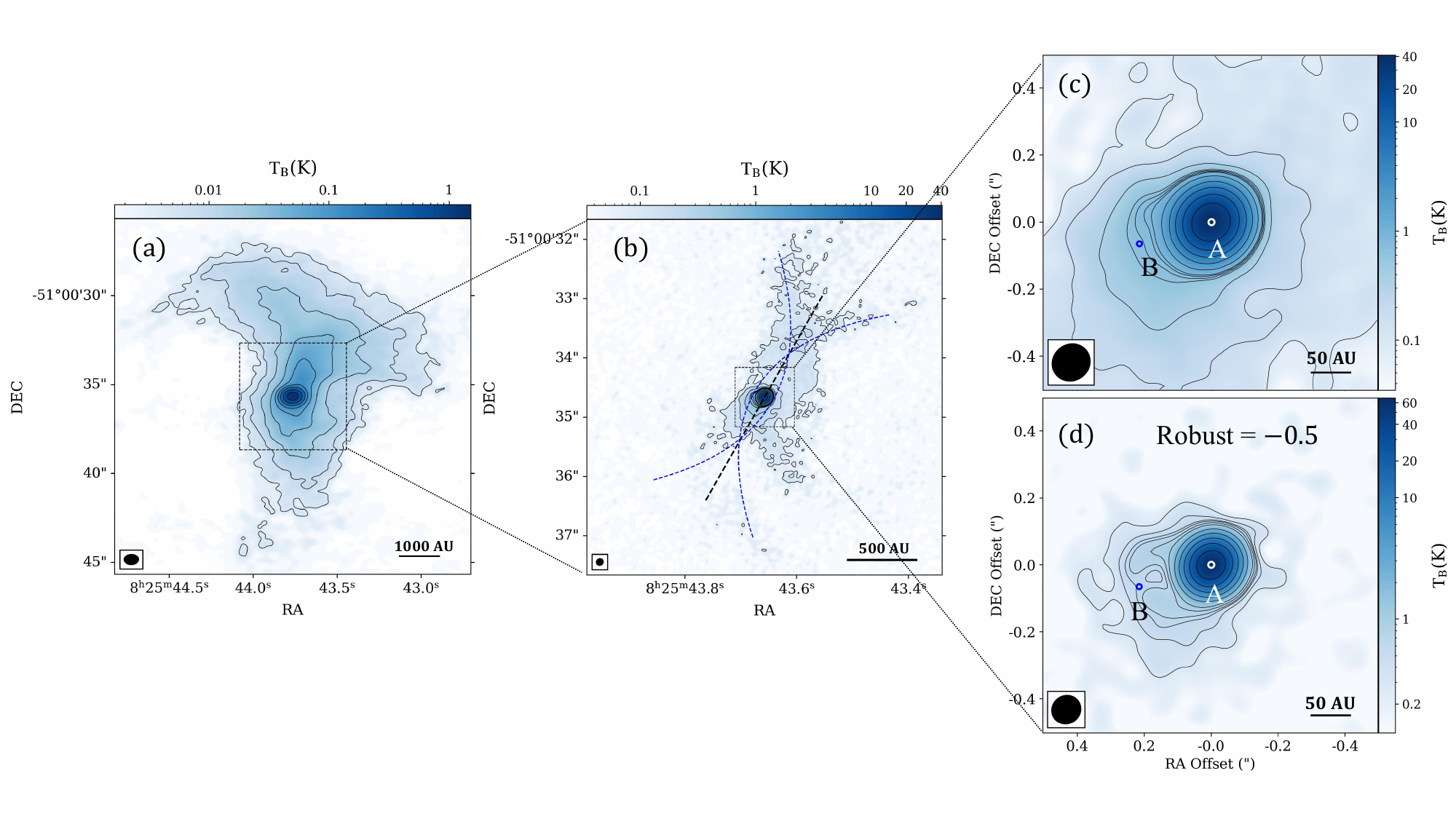}
\caption{
{\bf (a):} Low-resolution (C2+C4) 1.3 mm continuum image (\citealt[]{Zhang2019}). 
Contours start at $5\sigma$ and increase by factors of 2 up to $1280\sigma$. 
Here $1\sigma = 0.016~\mJybeam$ ( 8$\times10^{-4}$ K).
{\bf (b):} High-resolution (C2+C4+C7) 1.3 mm continuum image. 
Contours start at $5\sigma$ and increase by factors of 2 up to $2560\sigma$, 
with $1\sigma = 0.009~\mJybeam$ (0.017 K). 
The black dashed line indicates the cut used for the position–velocity diagrams shown in Figure~\ref{fig:PV}. 
The blue dashed curve traces the outflow cavity shape derived from H$_2$CO (Figure~\ref{fig:eye}).
{\bf (c):} Zoomed-in view of the central region. 
{\bf (d):} shows an image produced with a Briggs robust parameter of $-0.5$. 
The synthesized beam is $0.09\arcsec \times 0.08\arcsec$, 
and the rms sensitivity is $1\sigma = 0.018~\mJybeam$ (0.06 K). 
The positions of the binary system identified in near-infrared observations are marked.
}
\label{fig:Conti}
\end{figure*}

\begin{figure*}[ht!]
\includegraphics[width=\textwidth]{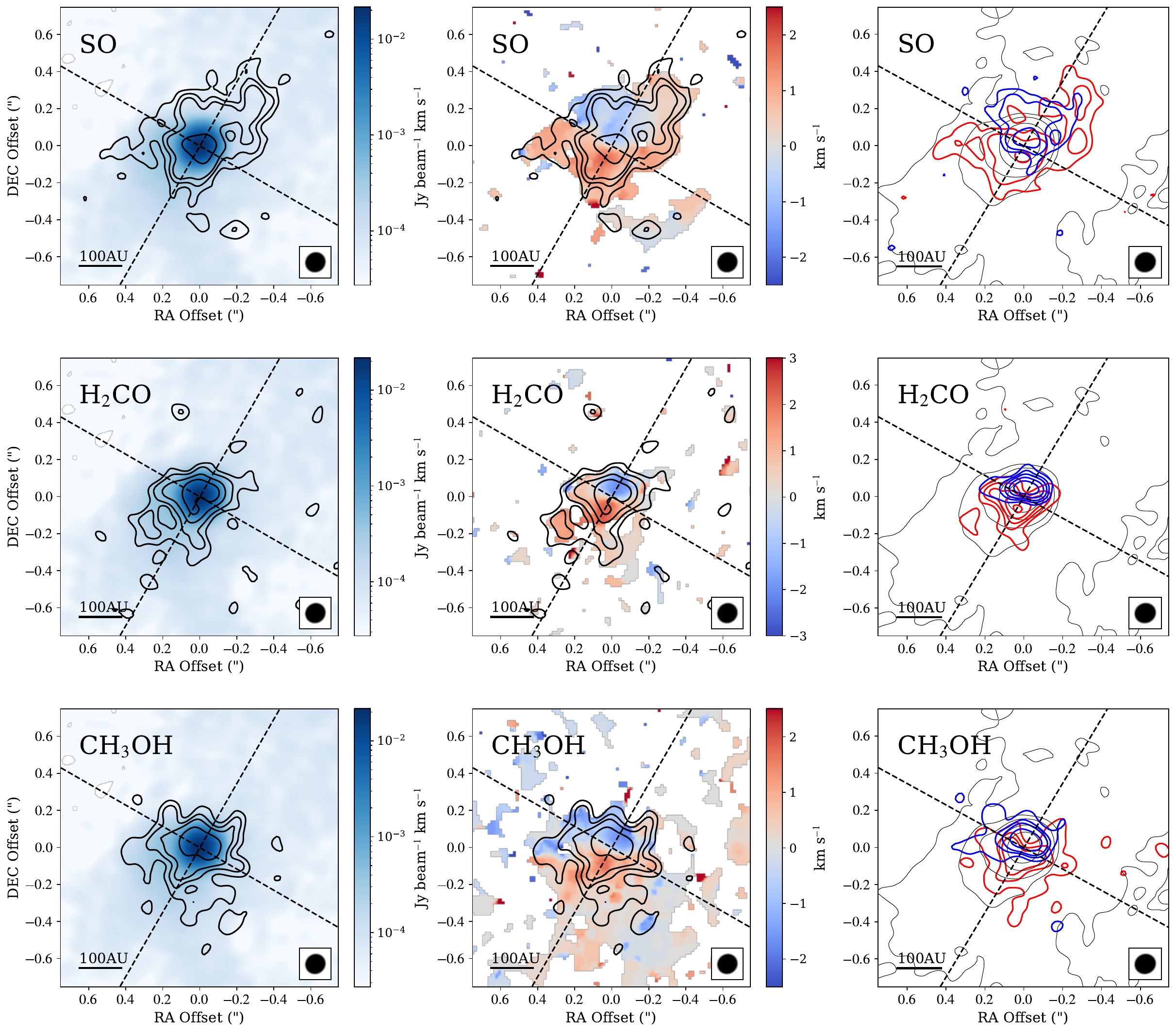}
\caption{
{\bf Left column:} Moment 0 maps of SO, H$_2$CO, and CH$_3$OH (contours) overlaid on the 1.3 mm continuum image (color scale). 
Black contour levels are [3$\sigma$, 5$\sigma$, 7$\sigma$, 14$\sigma$] for SO and H$_2$CO, 
and [5$\sigma$, 7$\sigma$, 9$\sigma$, 14$\sigma$] for CH$_3$OH, 
chosen to highlight both extended structures and emission peaks. 
The rms values are $1\sigma = 2.05$ (SO), 1.27 (H$_2$CO), and 1.05 (CH$_3$OH) $\mJybeamkms$.
{\bf Middle:} Moment 1 maps (color scale) of the three molecular lines, overlaid with the corresponding moment 0 contours. 
Velocities are given relative to the systemic velocity.
{\bf Right:} Blueshifted (blue) and redshifted (red) moment 0 maps overlaid on the 1.3 mm continuum (black contours). 
The velocity ranges with respect to the systemic velocity are $[-3.9, 0.3]~\kms$ (blueshifted) and $[0.3, 3.9]~\kms$ (redshifted). 
Contour levels are [5$\sigma$, 8$\sigma$, 11$\sigma$, 15$\sigma$] for all three lines, 
with $1\sigma = 1.3$ (SO), 0.8 (H$_2$CO), and 0.4 (CH$_3$OH) $\mJybeamkms$. 
The dashed lines indicate the disk major axis ($\pa = 147^\circ$) and the outflow axis ($\pa = 57^\circ$).
}
\label{fig:Envelope_M0}
\end{figure*}

\begin{figure*}[ht!]
\includegraphics[width=\textwidth]{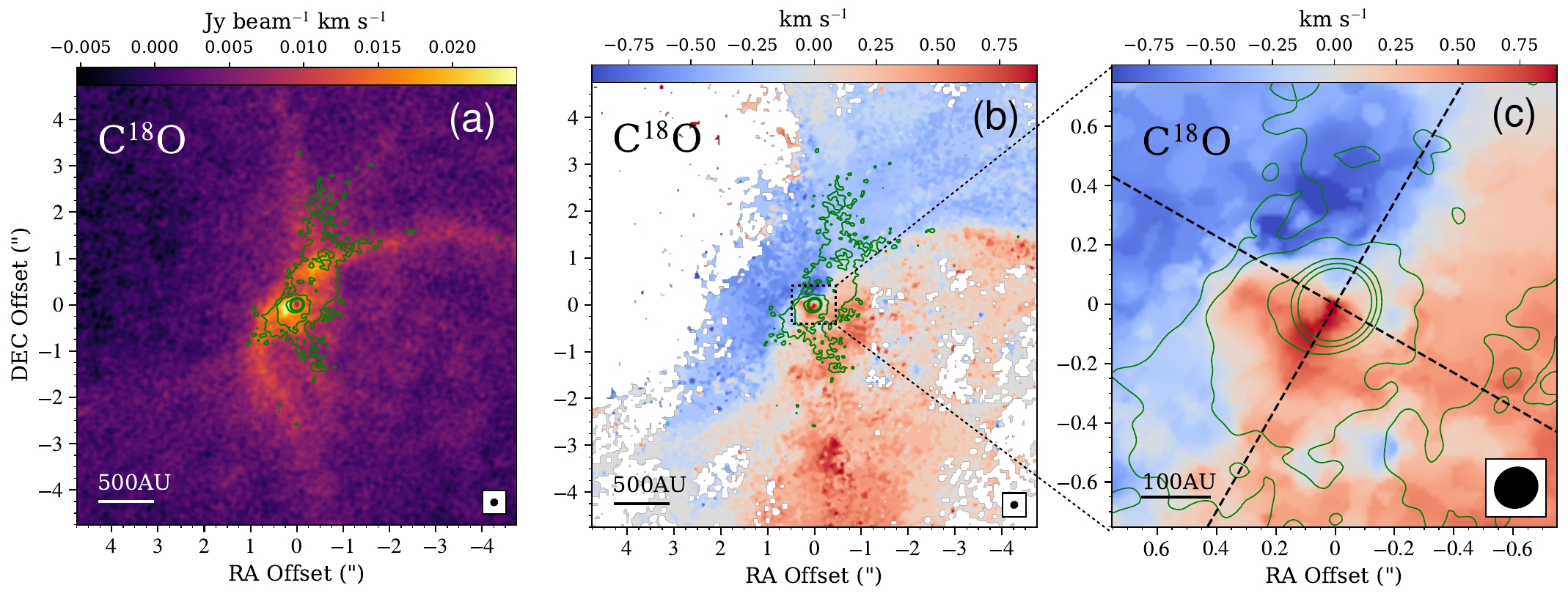}
\caption{From left to right: moment 0, moment 1, and a zoomed-in moment 1 map of C$^{18}$O. 
Green contours show the 1.3 mm continuum. 
The moment 0 map is integrated over $[-1.8, 1.8]~\kms$ relative to the systemic velocity. 
Dashed lines in the right panel mark the disk midplane and outflow axis (Figure~\ref{fig:Envelope_M0}). 
Black ellipses denote the synthesized beam.}
\label{fig:C18O_Moment}
\end{figure*}

\subsection{1.3 mm continuum}
\label{sec:continuum}

Figure~\ref{fig:Conti} presents the 1.3 mm continuum emission in comparison with the previous lower-resolution image. 
The continuum maps reveal a central compact source surrounded by fainter extended structures. 
The extended emission exhibits two arc-shaped features that coincide with the redshifted and blueshifted outflow cavities 
(traced by the blue curves; see \S\ref{sec:outflow_morphology}), particularly on the northern side. 
The peak position of the continuum emission is at 
$8^\mathrm{h}25^\mathrm{m}43^\mathrm{s}.55$, $-51^{\circ}00\arcmin33.66\arcsec$, 
consistent with previous measurements. 
The peak intensity in panel (b) and (c) of Figure~\ref{fig:Conti} is $21.5~\mJybeam$ (41 K).

A binary system with a projected separation of $0.23\arcsec$ ($\sim$100 au) 
has been identified from HST and JWST observations (\citealt[]{Reipurth2000,Nisini2024}). 
In the high-resolution 1.3 mm continuum image, 
we detect an asymmetric structure to the east of the central compact component. 
However, the peak of this structure is close to, but does not coincide with, 
the location of the companion (Figure~\ref{fig:Conti}c). 
Imaging with a Briggs robust parameter of $-0.5$ (Figure~\ref{fig:Conti}d) 
further reveals substructure around the companion. 
The companion (source B) is not associated with any continuum peak or compact condensation; 
instead, it lies within a dip between the termini of two spur-like features 
extending from the primary (source A). 
These features may trace substructures in the circumbinary disk induced by the companion (see \S\ref{sec:binary}).

We estimate the dust mass assuming optically thin, isothermal emission,
\begin{equation}
M_\mathrm{dust} = \frac{F_\nu D^2}{\kappa_\nu B_\nu(T_\mathrm{dust})},
\end{equation}
where $F_\nu$ is the flux density, $D$ is the source distance, $B_\nu$ is the Planck function, 
and $\kappa_\nu$ is the dust opacity at 225 GHz (1.3 mm). 
We adopt $\kappa = 0.899~\mathrm{cm}^2~\mathrm{g}^{-1}$ (\citealt[]{Ossenkopf1994}).

For the central compact component, a two-dimensional Gaussian fit yields flux densities of $24.88 \pm 0.18~\mJy$ 
from the $robust = -0.5$ image (Figure~\ref{fig:Conti}d) and $25.67 \pm 0.24~\mJy$ from the $robust = 0.5$ image (Figure~\ref{fig:Conti}c). 
The close agreement indicates that the central component is highly compact and only marginally resolved, 
such that the flux density is insensitive to the imaging weighting.
Assuming $T_\mathrm{dust} = 50~\mathrm{K}$ and a gas-to-dust ratio of $R_\mathrm{gd} = 100$, 
we derive gas masses of $(3.89 \pm 0.03)\times 10^{-2}~M_\odot$ and $(4.00 \pm 0.03)\times 10^{-2}~M_\odot$ 
from the $robust = -0.5$ and $robust = 0.5$ images, respectively. 
This mass can be interpreted as that of the circumstellar disk around source A. 
Here we adopt a dust temperature of 50 K, which is typical for the vicinity of protostars and 
consistent with the characteristic temperature range observed in Taurus disks (\citealt{Andrews2005}).

For the surrounding extended structure, integrating emission above $5\sigma$ and subtracting the compact component 
yields flux densities of $3.44 \pm 0.18~\mJy$ ($robust = -0.5$) and $11.31 \pm 0.24~\mJy$ ($robust = 0.5$). 
Assuming $T_\mathrm{dust} = 30~\mathrm{K}$ for the extended emission
-- lower than that adopted for the central component -- 
and $R_\mathrm{gd} = 100$, 
we derive corresponding gas masses of $(0.96 \pm 0.03)\times 10^{-2}~M_\odot$ and $(3.17 \pm 0.04)\times 10^{-2}~M_\odot$. 
As illustrated in Figures~\ref{fig:Conti}c and d, the central compact component likely traces the immediate circumbinary structure, 
while the surrounding extended structure includes additional contributions from more extended envelope emission.


\subsection{Envelope and Disk System}
\label{sec:disk}

Figure~\ref{fig:Envelope_M0} presents the moment 0 (integrated intensity) 
and moment 1 (intensity-weighted velocity) maps of the SO, CH$_3$OH, and H$_2$CO lines, 
with the images centered at the continuum position of source A. 
All three tracers probe compact structures around the central protostar. 
Among them, SO emission is the brightest and most extended, 
while CH$_3$OH and H$_2$CO are weaker and more compact. 
Clear velocity gradients are seen in all three lines, 
with redshifted emission toward the southeast and blueshifted emission toward the northwest. 
The velocity gradient is symmetric with respect to source A, 
while finer velocity structures around source B remain unresolved due to limited angular resolution.

Figure~\ref{fig:C18O_Moment} shows the moment 0 and moment 1 maps of the C$^{18}$O ($2-1$) line. 
In the moment 0 map(Figure~\ref{fig:C18O_Moment}a), the C$^{18}$O emission is significantly more extended than that of SO, CH$_3$OH, and H$_2$CO, 
tracing both the envelope and the base of the outflow cavity. 
In the central region(Figure~\ref{fig:C18O_Moment}c), C$^{18}$O also reveals a compact component; 
however, toward the continuum peak, the emission is suppressed due to absorption or self-absorption. 
On larger scales, the moment 1 map shows predominantly blueshifted emission to the northeast 
and redshifted emission to the southwest, consistent with outflow motions (Figure~\ref{fig:RGB}).

In the inner region, the C$^{18}$O moment 1 map exhibits a velocity gradient across source A 
that is aligned with those seen in SO, CH$_3$OH, and H$_2$CO (Figure~\ref{fig:Envelope_M0}). 
The consistency in both orientation and spatial extent among these tracers suggests a common kinematic origin. 
The gradient is approximately aligned with the major axis of the 1.3 mm continuum emission (Figure~\ref{fig:Conti}) and
perpendicular to the outflow axis (see below; also \citealt[]{Zhang2019}).
This is most naturally interpreted as rotation.
We therefore suggest that the C$^{18}$O, SO, CH$_3$OH, and H$_2$CO emission collectively traces a rotating inner envelope, 
with a possible contribution from a disk-like structure around source A. 
In \S\ref{sec:IRE}, we will employ a simple analytic model to fit emission from these lines 
and derive a dynamical mass of about $0.3M_{\odot}$ for the source A.


\subsection{Outflow}
\label{sec:outflow_morphology}

\begin{figure}[ht!]
\includegraphics[width=\columnwidth]{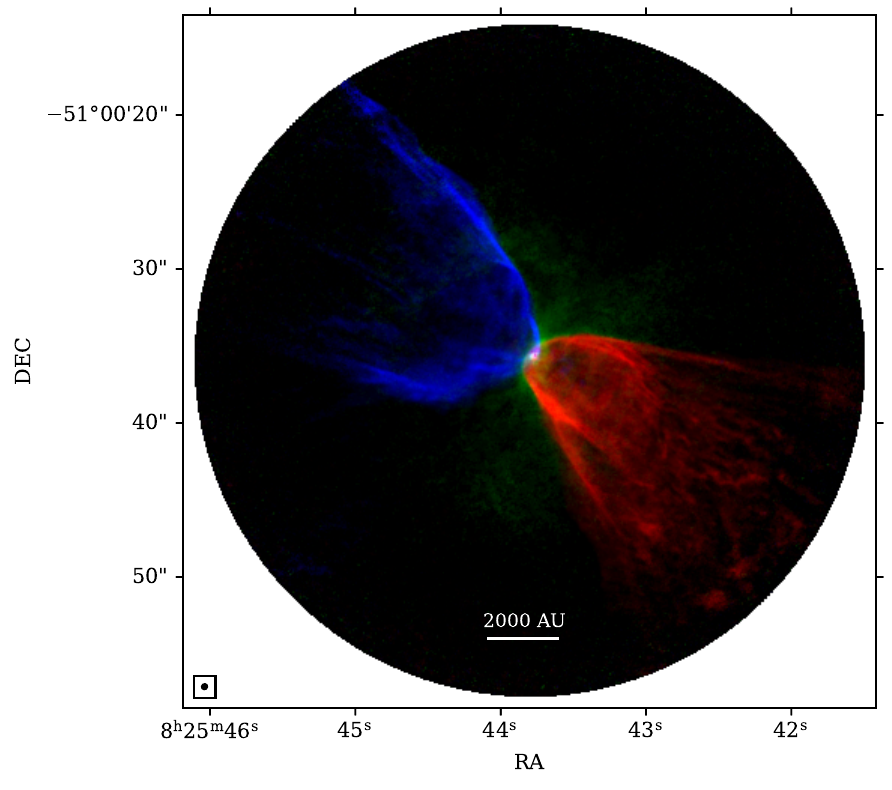}
\caption{$^{12}$CO integrated intensity maps of the HH 46/47 molecular outflow after primary beam correction. 
The minimum primary beam correction factor is 0.3.
The red and blue color scales show $^{12}$CO emission integrated 
over velocity ranges of $[+2, +52]~\kms$ and $[-35, -2]~\kms$, respectively. 
The green color scale shows C$^{18}$O emission integrated over $[-1.8, +1.8]~\kms$. 
All velocities are given relative to the systemic velocity of $+5.3~\kms$.}
\label{fig:RGB}
\end{figure}

\begin{figure*}[ht!]
\includegraphics[width=\textwidth]{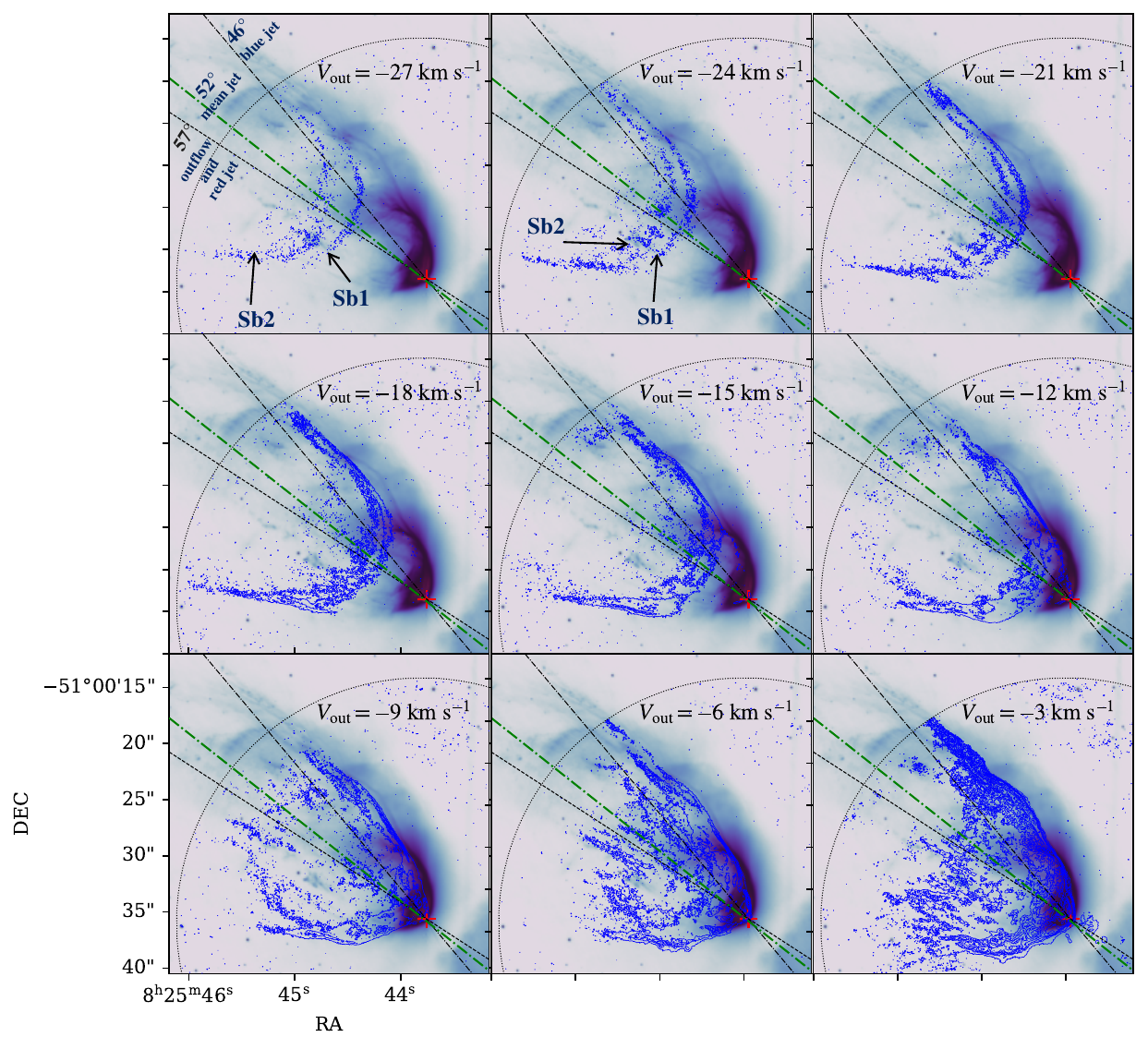}
\caption{Channel maps of the blueshifted $^{12}$CO outflow (contours) overlaid on the JWST NIRCam F200W image. The $^{12}$CO emission is after primary beam correction. The minimum primary beam response is set to be 0.3.
Contour levels start at $5\sigma$ and increase in steps of $5\sigma$ up to $45\sigma$ ($1\sigma = 0.6~\mJybeam$). 
Two shells structures (Sb1, Sb2) identified in \cite{Zhang2019} are labeled in $\vout = -27\kms$ and $-24\kms$ channels.
The red cross marks the position of the central source. 
The dashed circle indicates the field of view of the $^{12}$CO observations (primary beam response of 0.1). 
The dashed line marks the outflow axis determined in \S\ref{sec:vfield} 
($\pa = 57^\circ$), which also coincides with the position angle of the redshifted jet
measured from the 5.3 and 26~$\mu$m [FeII] emission (Figure~3 of \citealt{Nisini2024}). 
The black dot-dashed line shows the blueshifted [FeII] jet axis ($\pa = 46^\circ$). 
The green dot-dashed line marks the mean jet axis ($\pa = 52^\circ$), 
defined by the innermost blueshifted and redshifted 5.3~$\mu$m [FeII] jet knots
(Figure~3 of \citealt{Nisini2024}).
}
\label{fig:b_channel}
\end{figure*}

\begin{figure*}[ht!]
\includegraphics[width=\textwidth]{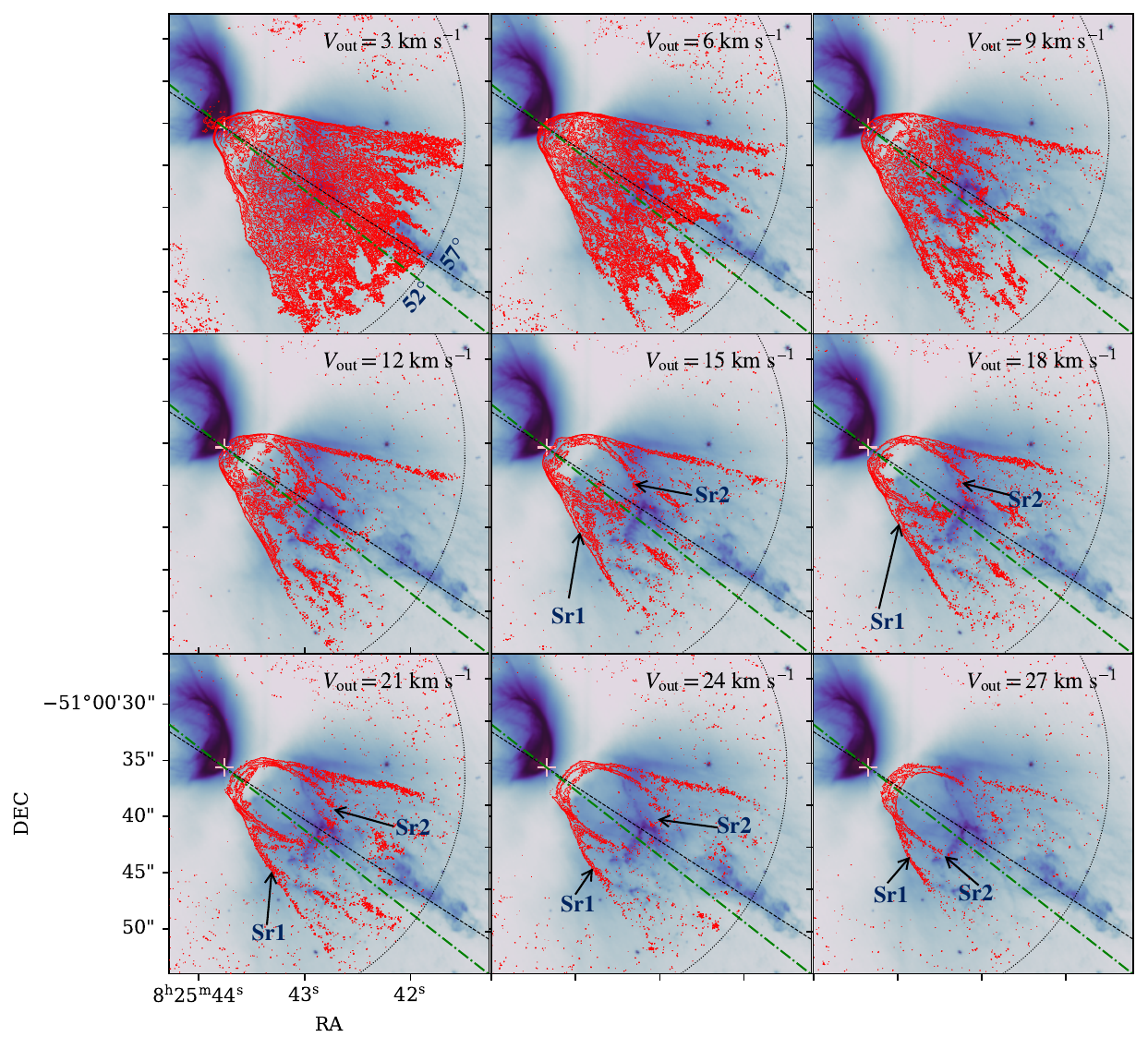}
\caption{Same as Figure~\ref{fig:b_channel}, but for the redshifted $^{12}$CO outflow. Shells Sr1 and Sr2 identified in \S\ref{sec:vfield} are labeled.}
\label{fig:r_channel}
\end{figure*}

\begin{figure*}[ht!]
\includegraphics[width=0.9\textwidth]{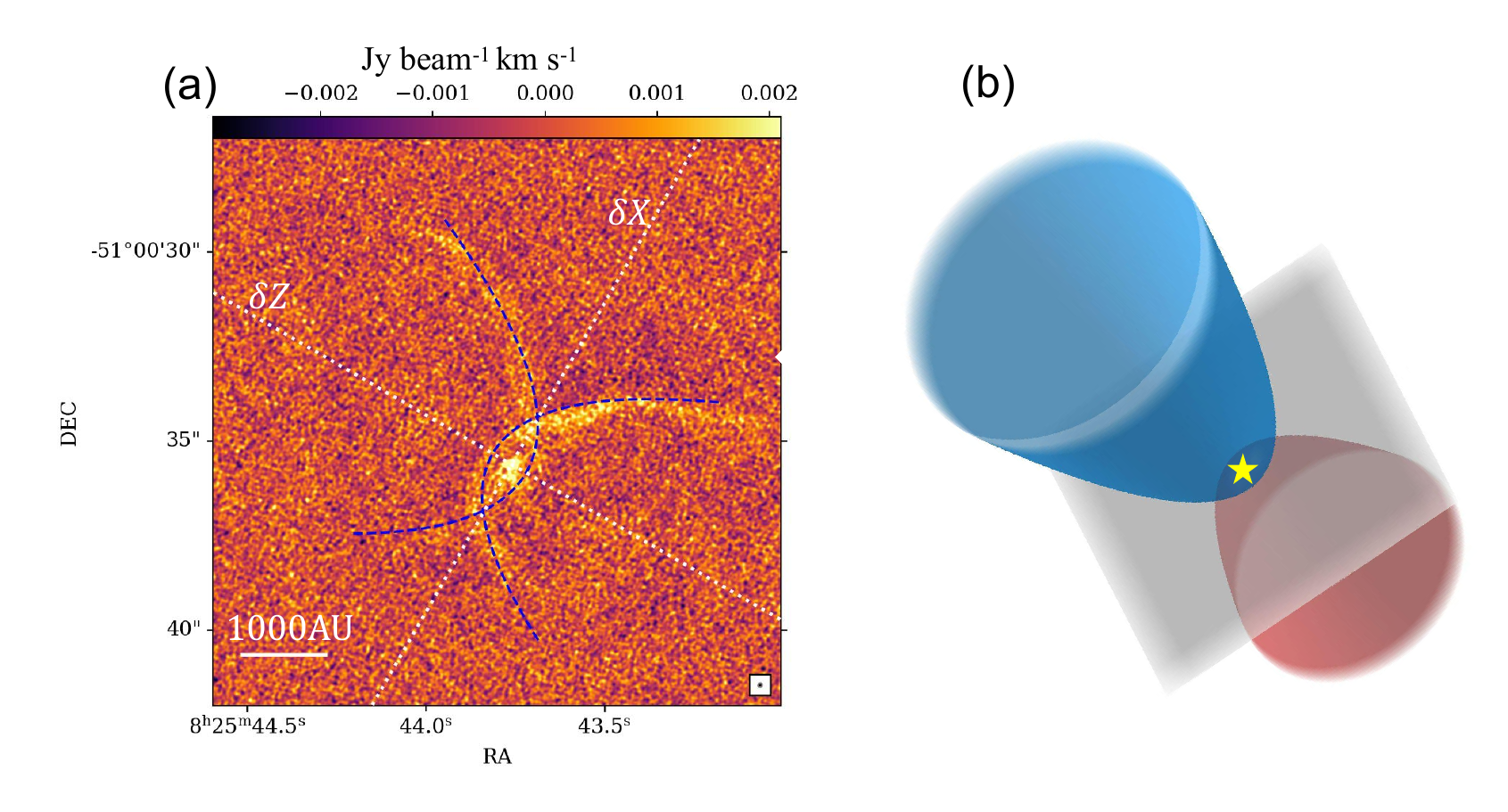}
\caption{{\bf (a):} H$_2$CO moment 0 map integrated over $[-1.5, 1.5]~\kms$ relative to the systemic velocity. 
The blue curves show the fitted projected shapes of the outflow cavities. This cavity is related to the blueshifted  $^{12}$CO shell and redshifted outer $^{12}$CO shell in Figure~\ref{fig:RGB}.
{\bf (b):} Schematic diagram illustrating two outflow cavities overlapping at their bases. 
The yellow star represents the position of the protostar.
}
\label{fig:eye}
\end{figure*}

The HH 46/47 molecular outflow has been extensively studied with ALMA 
(\citealt[]{Arce2013,Zhang2016,Zhang2019}). 
Our field of view is the same as in previous Band 6 observations (\citealt[]{Zhang2019}) 
and covers only the base of the outflow. 
Figure~\ref{fig:RGB} shows the integrated blueshifted and redshifted $^{12}$CO ($2-1$) emission. 
The observed morphologies are consistent with previous lower-resolution observations of the same line, 
exhibiting layered structures that are most prominent in the redshifted lobe. 
The outer boundaries of the $^{12}$CO emission coincide with the lower-velocity C$^{18}$O emission, 
which primarily traces the envelope material. 
This confirms that the arch-like structures seen in Figure~\ref{fig:C18O_Moment}a trace the outflow cavity walls. 
The extended structures seen in the continuum emission (Figure~\ref{fig:Conti}b) are also aligned with the outflow cavities, 
suggesting that outflowing material accumulates along the cavity walls.

Figure~\ref{fig:b_channel} shows channel maps of the blueshifted $^{12}$CO emission, 
compared with the JWST NIRCam F200W wide-band continuum image. 
On the blueshifted side, the high-resolution data confirm structures previously observed at lower resolution. 
At high velocities, at least two shells can be clearly distinguished, with their distances from the central source 
decreasing toward lower velocities. 
Following \cite{Zhang2019}, we label them as Sb1 and Sb2.
At lower velocities, as the shells approach the central source, a more complete elliptical morphology emerges. 
These features can be explained by two outflowing cones with similar opening angles but different velocities (\citealt[]{Zhang2019}).

The blueshifted outflow cavity is also visible in the JWST NIRCam image, particularly on the northern side. 
The near-infrared continuum emission is dominated by scattered light from the cavity walls. 
At high velocities, the $^{12}$CO emission lies interior to the NIR cavity. 
As the velocity decreases, the northern side of the $^{12}$CO shells approaches the NIR cavity walls 
(e.g., from $\vout=-21$ to $-6~\kms$). 
At $\vout=-3~\kms$, the outer boundary of the $^{12}$CO emission closely matches the NIR cavity. 
This further supports a scenario in which lower-velocity material accumulates along the cavity walls.

Most of the blueshifted $^{12}$CO emission is aligned with an axis (dashed line in Figure~\ref{fig:b_channel}, $\pa = 57^{\circ}$) 
that differs from the position angle of the blueshifted [FeII] jet seen by JWST 
(\citealt[]{Nisini2024}; dot-dashed line in Figure~\ref{fig:b_channel}, $\pa = 46^{\circ}$). 
No clear $^{12}$CO structures are directly associated with the jet. 
At low velocities (e.g., $\vout=-6~\kms$), the emission exhibits finger-like structures 
pointing away from the central source. 
Some align with the jet direction, but similar features are observed at a range of position angles, 
filling the outflow cavity.

Figures~\ref{fig:r_channel} show the channel maps of the redshifted $^{12}$CO emission. 
Similar to lower-resolution observations (\citealt[]{Zhang2019}), 
at least two shells are visible at higher velocities 
(labeled as Sr1 and Sr2, also see \S\ref{sec:vfield}).
In each channel, the wider shell(Sr1) is located farther from the central source than the narrower shell(Sr2). 
This morphology is consistent with two outflowing shells,
a wider, slower component and a narrower, faster component,
analogous to the blueshifted side.

The base of the redshifted outflow is not detected in the NIRCam image due to heavy extinction by the envelope. 
However, it is detected in MIRI observations of 
H$_2$ pure rotational lines, 
showing shell structures similar to the CO Sr2 shell (\citealt{Nisini2024,Navarro2025}),
although they lack the spectral resolution to distinguish morphological variations as a function of velocity.
Farther from the source, the cavity boundary becomes visible in the NIR 
and coincides with the boundary of the low-velocity $^{12}$CO emission. 
At higher velocities, the $^{12}$CO emission lies within the NIR cavity, 
consistent with the scenario proposed by \citet[]{Zhang2019} 
that the redshifted shells remain confined within the cavity and have not yet reached its boundary. 
The redshifted jet is detected in the 5.3 and 26~$\mu$m [FeII] emission 
and has a position angle of $57^\circ$ (\citealt{Nisini2024}), 
closely aligned with the axis of the CO outflow.
We will further analyze the morphology and kinematics of the redshifted outflow in \S\ref{sec:vfield}.

The boundaries of the outflow cavities are traced by arch-like structures seen in the continuum (Figure~\ref{fig:Conti}) 
and molecular lines such as C$^{18}$O (Figure~\ref{fig:C18O_Moment}). 
The cavity geometry is most clearly revealed in the large-scale H$_2$CO moment 0 map (Figure~\ref{fig:eye}). 
The two inclined outflow cavities overlap in projection near the central source, 
forming an eye-like morphology (see the schematic in Figure~\ref{fig:eye}b). 
Assuming the outflow cavity has an intrinsic parabolic shape $Z = A R^2$ 
and an inclination angle $\theta$ between the outflow axis and the plane of the sky, 
we fit the projected cavity boundaries using two free parameters, $A$ and $\theta$ 
(see Appendix~\ref{app:eye_equation} for details). 
We explored a range of plausible values and determined the best-fit parameters 
through visual comparison with the observed morphology.
The best-fit model yields an intrinsic cavity shape of $Z = 0.3~R^2$ 
(with $Z$ and $R$ in units of arcseconds) and an inclination angle of 
$\theta = 40^\circ$. The derived inclination is consistent with previous estimates 
based on jet proper motions ($37^\circ$; \citealt{Hartigan2005}).


\section{Analysis and Discussion}
\label{sec:discussion}

\begin{figure*}[ht!]
\centering
\includegraphics[width=0.9\textwidth]{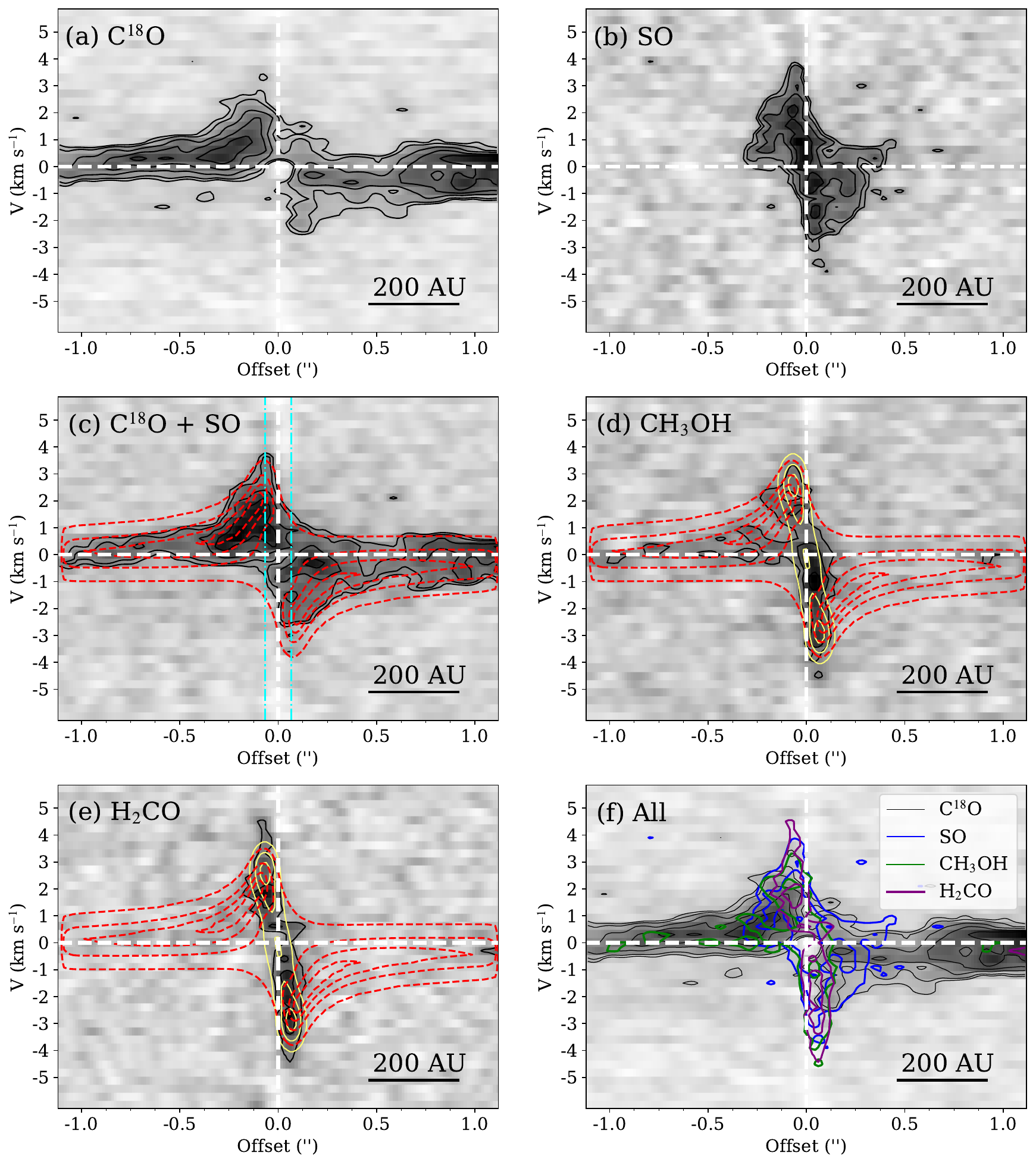}
\caption{
Transverse position–velocity (PV) diagrams for (a) C$^{18}$O, (b) SO, (c) C$^{18}$O+SO combined, (d) CH$_3$OH, (e) H$_2$CO, 
and (f) all lines combined. The PV diagrams are extracted along a cut perpendicular to the outflow axis, 
passing through the source center, with a width of one synthesized beam ($0.1\arcsec$; see Figure~\ref{fig:Conti}).
In panels (c)–(e), red dashed contours show the best-fit infalling–rotating envelope (IRE) model 
derived from the C$^{18}$O+SO emission ($M_* = 0.3~M_\odot$, $\rcb = 30~\au$). 
The centrifugal barrier radius $\rcb$ is indicated by the cyan dashed lines in panel (c). 
Yellow contours in panels (d) and (e) show the fitted ring models for CH$_3$OH and H$_2$CO (see text).
}
\label{fig:PV}
\end{figure*}

\begin{figure*}[ht!]
\centering
\includegraphics[width=0.9\textwidth]{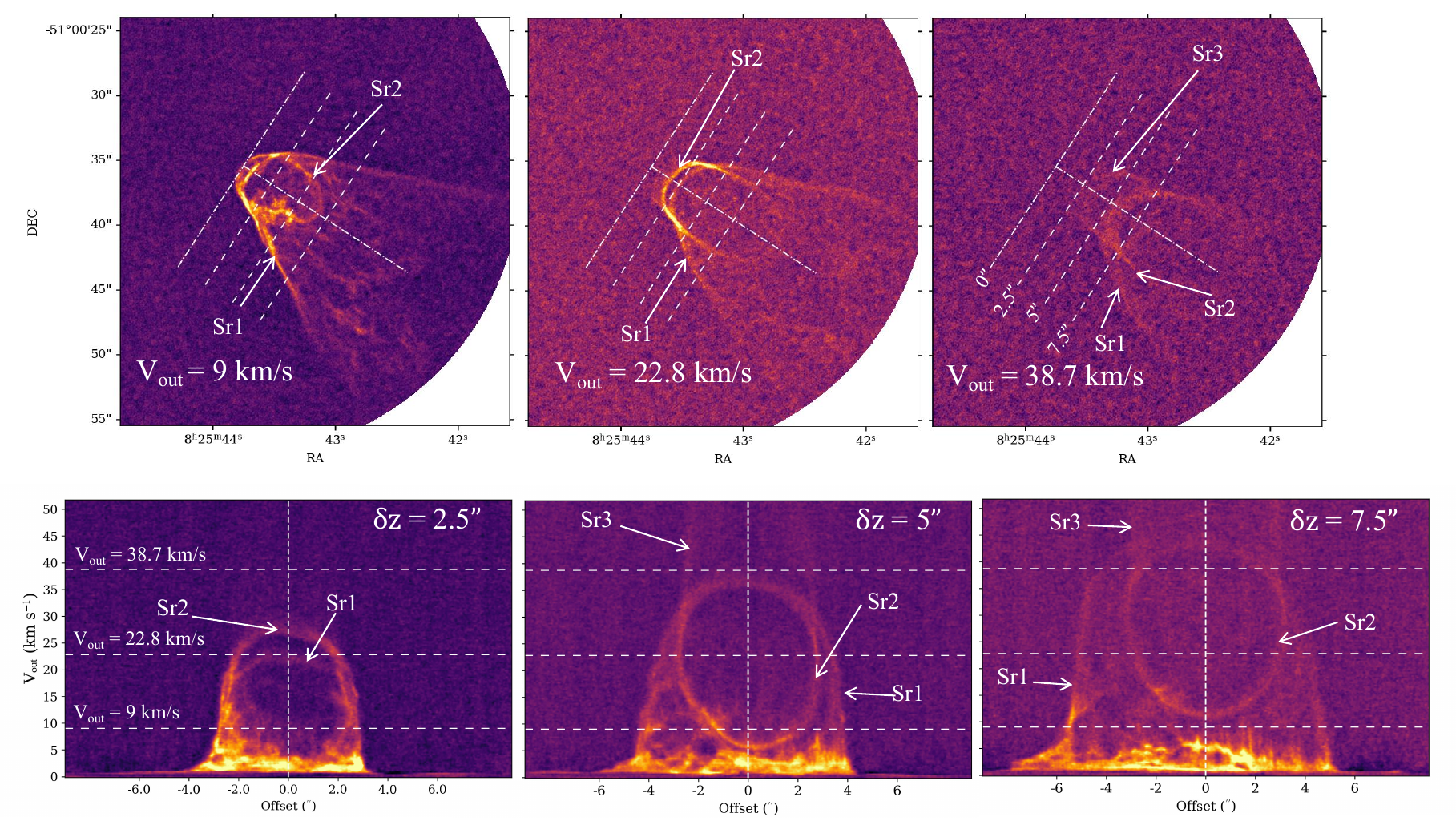}
\caption{
{\bf Upper:} Channel maps at $\vout = +9$, +22.8, and $+38.7~\kms$. 
The white dashed lines indicate the transverse PV cuts at $\delta z = 0\arcsec$, $2.5\arcsec$, $5\arcsec$, and $7.5\arcsec$. 
The three shells (Sr1, Sr2, and Sr3) are labeled by arrows. 
{\bf Bottom:} Transverse PV diagrams at $\delta z = 2.5\arcsec$, $5\arcsec$, and $7.5\arcsec$. 
The horizontal white dashed lines indicate the velocities corresponding to the channel maps shown in the upper panels.}
\label{fig:multi_shell}
\end{figure*}

\subsection{Envelope and Disk Kinematics}
\label{sec:IRE}

We use the C$^{18}$O, SO, CH$_3$OH, and H$_2$CO lines to investigate the dynamical structure 
of the envelope/disk system around the central source. 
Figure~\ref{fig:PV} presents position–velocity (PV) diagrams taken across the central source, 
perpendicular to the outflow axis 
(black dashed lines in Figures~\ref{fig:Conti} and \ref{fig:Envelope_M0}). 
All four tracers exhibit clear velocity gradients consistent with rotation; 
however, their distinct PV morphologies indicate that they probe different components of the system.

The transverse PV diagram of C$^{18}$O (Figure~\ref{fig:PV}a) 
shows a pattern commonly associated with an infalling-rotating envelope. 
At low velocities ($|v| < 1~\kms$), extended emission is seen on both sides of the source, 
with one side slightly blueshifted and the other slightly redshifted. 
The velocity increases toward smaller radii and reaches a maximum at an offset of $\sim0.1\arcsec$. 
In addition to the expected redshifted emission on one side and blueshifted emission on the other, 
emission is also detected in the upper-right and lower-left quadrants of the PV diagram. 
Such features cannot be explained by pure rotation but are naturally produced by infalling motion combined with rotation (e.g. \citealt{Sakai2014}). 
Toward the central position, absorption is present in the C$^{18}$O line, 
indicating that it predominantly traces cooler, outer material. 
Together, these characteristics are consistent with C$^{18}$O tracing a rotating, infalling envelope 
rather than a Keplerian disk.

The SO emission appears to trace a more compact region. 
In the transverse PV diagram (Figure~\ref{fig:PV}b), 
SO emission is confined to smaller radii than C$^{18}$O and reaches higher velocities, 
although the velocity peak remains offset from the central source. 
Emission is also detected in the upper-right and lower-left quadrants, 
again suggesting the presence of infall. 
We also find that the SO PV structure cannot be well reproduced by pure Keplerian rotation (Appendix~\ref{app:SO_PV}). 
These results suggest that SO traces the inner envelope and the transition region between the envelope and disk, 
while C$^{18}$O traces the outer envelope. 
The enhancement of SO may be associated with desorption from dust grains in shocked regions, 
possibly related to accretion shocks at the envelope–disk interface (e.g., \citealt[]{Zhang2023,Liu2025}).

To test the interpretation that C$^{18}$O and SO trace a rotating, infalling envelope, 
we combine their transverse PV diagrams (see Figure~\ref{fig:PV}c)
and compare them with an analytic model of an infalling–rotating envelope 
(IRE; e.g., \citealt[]{Ohashi1997,Sakai2014,Oya2016,Zhang2019massive,Zhang2022}).
The combined PV diagram is constructed by directly summing the emission 
from the two lines without normalization.
In this model, material undergoes ballistic infall under the gravitational potential of the central object, 
conserving both angular momentum and total energy. 
Assuming motion confined to the midplane, the rotational and infall velocities are given by
\begin{align}
\vrot &= \frac{j}{r}, \label{eq:vrot} \\
\vinf &= \sqrt{\frac{2GM_*}{r} - \left(\frac{j}{r}\right)^2}, \label{eq:vinf}
\end{align}
where $j$ is the specific angular momentum and $M_*$ is the central mass. 
The centrifugal barrier, defined as the radius where $\vinf = 0$, is given by
\begin{equation}
\rcb = \frac{j^2}{2GM_*}.
\end{equation}
At this radius, the rotational velocity reaches its maximum in the transverse PV diagram.
We note that, since the intensity ranges of the C$^{18}$O and SO PV diagrams are comparable, 
combining the two PV diagrams with or without normalization 
does not noticeably change the overall morphology of the combined PV diagram, 
particularly the location of the velocity maxima that primarily constrain the model. 
Consequently, the derived central mass, $M_*$, and centrifugal barrier radius, $r_\mathrm{CB}$, 
are not sensitive to whether the two tracers are normalized before being combined.

We generate model PV diagrams using the \texttt{FERIA} code (\citealt{Oya2022}) 
and compare them with the observed C$^{18}$O+SO PV diagram through visual inspection.
We adopt two free parameters: the central mass $M_*$ and the centrifugal barrier radius $\rcb$. 
The inclination angle $i$ is fixed to $40^\circ$ ($0^\circ$ for edge-on), as derived from the H$_2$CO cavity geometry 
(\S\ref{sec:outflow_morphology}) 
and consistent with jet measurements (\citealt{Hartigan2005}). 
The best-fit parameters are $M_* = 0.3~M_\odot$ and $\rcb = 30~\au$, 
corresponding to a specific angular momentum of $j = 126~\au~\kms$.
These values are comparable to those found in other low-mass Class 0 sources 
(e.g., \citealt{Sakai2014,Oya2016}). 
The derived $j$ is also consistent with the characteristic value of 
$\sim 6 \times 10^{-4}~\mathrm{pc~km~s^{-1}}$, 
or $123~\au~\kms$,
observed over radii of $\sim 50-1600$ au, 
where $j$ remains approximately constant, in a sample of Class 0 protostellar envelopes
(\citealt{Gaudel2020}).
A similar kinematic analysis was previously performed for this source by \cite{Zhang2016} 
using lower-resolution ($\sim 1\arcsec$, $\sim 450~\au$) $^{13}$CO and C$^{18}$O data. 
While the derived central mass ($M_* = 0.3~M_\odot$) is consistent with our result, 
the inferred centrifugal barrier radius ($\rcb \sim 380~\au$) and 
corresponding specific angular momentum ($j \sim 450~\au~\kms$) are significantly larger, 
likely reflecting the limited spatial resolution of the earlier observations.
Since the velocity gradient is centered on source A (Figure~\ref{fig:Envelope_M0}), 
the derived dynamical mass of $M_* = 0.3~M_\odot$ 
should be attributed to source A. 
This dynamical mass is significantly lower than the 
stellar mass of $\sim1.2~M_\odot$ estimated by \citet{Antoniucci2008} 
from infrared observations using 
pre-main-sequence evolutionary models \citep{Siess2000}. 
The latter estimate depends on the adopted 
stellar luminosity and evolutionary tracks, 
whereas our result is obtained directly from the gas kinematics and 
is therefore independent of such assumptions.

The transverse PV diagrams of CH$_3$OH and H$_2$CO (Figures~\ref{fig:PV}d and e) 
exhibit inclined linear features, characteristic of narrow rotating rings. 
The radius of such a ring corresponds to the offset at which the maximum velocity is reached, 
which coincides with the centrifugal barrier derived from the C$^{18}$O+SO analysis. 
We therefore model these tracers as emission from a narrow ring spanning $r = 30-35~\au$, 
whose motion follows equations (\ref{eq:vrot}) and (\ref{eq:vinf}),
with $M_* = 0.3~M_\odot$ and $\rcb = 30~\au$. 
The ring models are shown by the yellow contours in 
Figures \ref{fig:PV}(d) and (e).
The observed PV structures are well reproduced by this model.

These results suggest that CH$_3$OH and H$_2$CO predominantly trace a narrow region 
near the envelope–disk transition. 
Similar linear PV features have been reported in other sources (e.g., \citealt[]{Ohashi2014,Oya2016,Hsu2025}). 
The enhanced emission may arise from the release of molecules from dust grains 
due to elevated temperatures near the centrifugal barrier. 
Inside $\rcb$, a Keplerian disk is expected; 
however, the current angular resolution does not allow us to clearly resolve the disk kinematics. 
We note that, in reality, there is no sharp boundary between the envelope and the disk. 
Although the SO, CH$_3$OH, and H$_2$CO kinematics are well reproduced by the infalling-rotating envelope model extending inward to $\rcb$, 
these lines are likely to also trace the envelope-disk transition region. 
This simplification is not expected to significantly affect the derived stellar mass, 
since the kinematics remain dominated by the infalling-rotating envelope 
on the spatial scales probed by these tracers.

\begin{figure}[ht!]
\centering
\includegraphics[width=0.9\columnwidth]{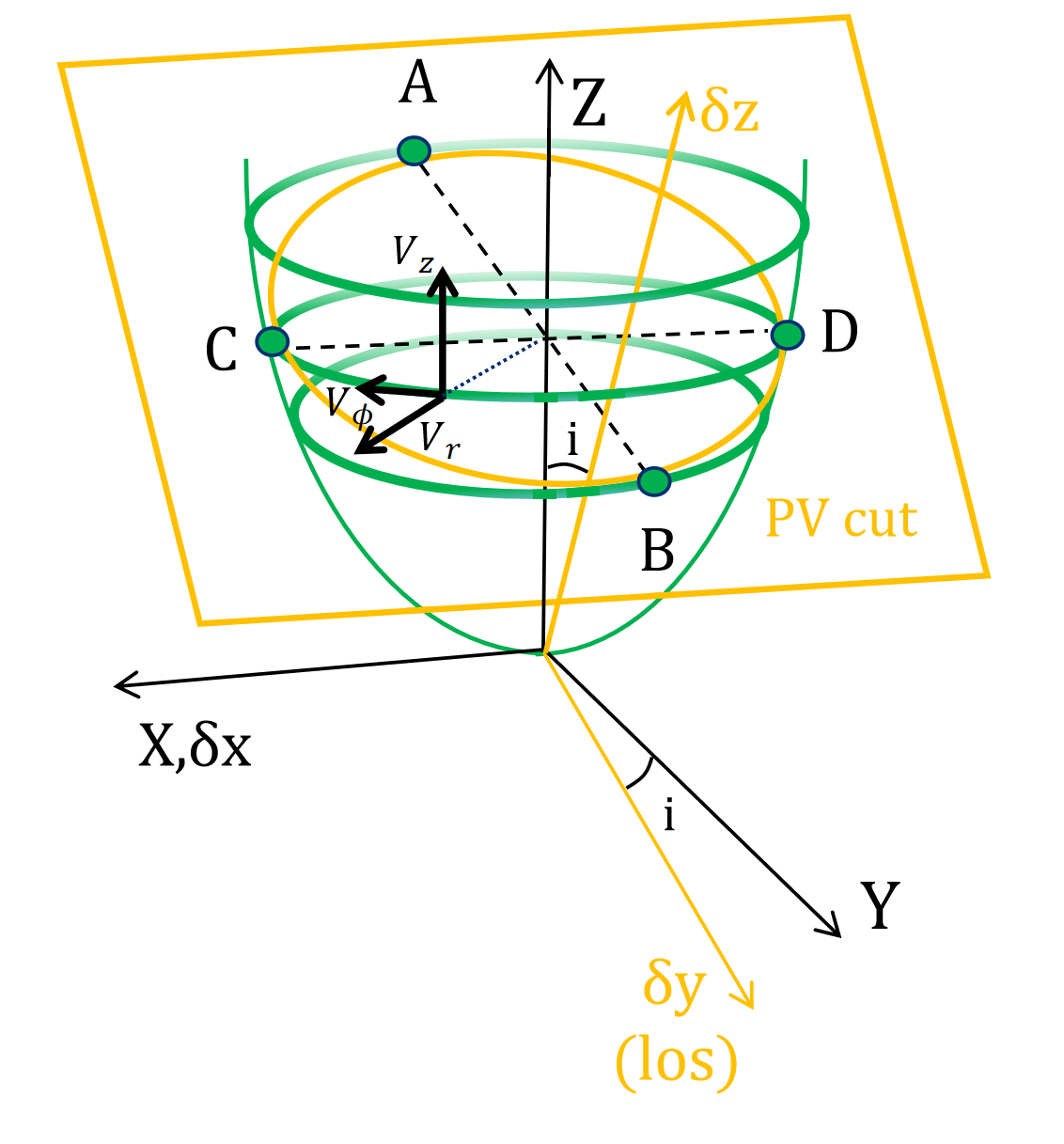}
\caption{Three-dimensional schematic illustrating the coordinate systems and velocity components in the outflow shell. The $(X, Y, Z)$ and $(\delta x, \delta y, \delta z)$ coordinates define the outflow frame and the observer’s frame, respectively (see \S\ref{sec:vfield} for details). A gas parcel on the shell has three velocity components: the axial velocity $V_z$, radial velocity $V_r$, and rotational velocity $V_{\phi}$. A position–velocity (PV) cut taken perpendicular to the outflow axis intersects the shell as a ring, illustrated by the yellow ring in the figure. The points A, B, C, and D correspond to the same locations shown in Figure~\ref{fig:Method_outflow_pv}.}
\label{fig:sketch}
\end{figure}

\begin{figure}[ht!]
    \centering
    \includegraphics[width=1\columnwidth]{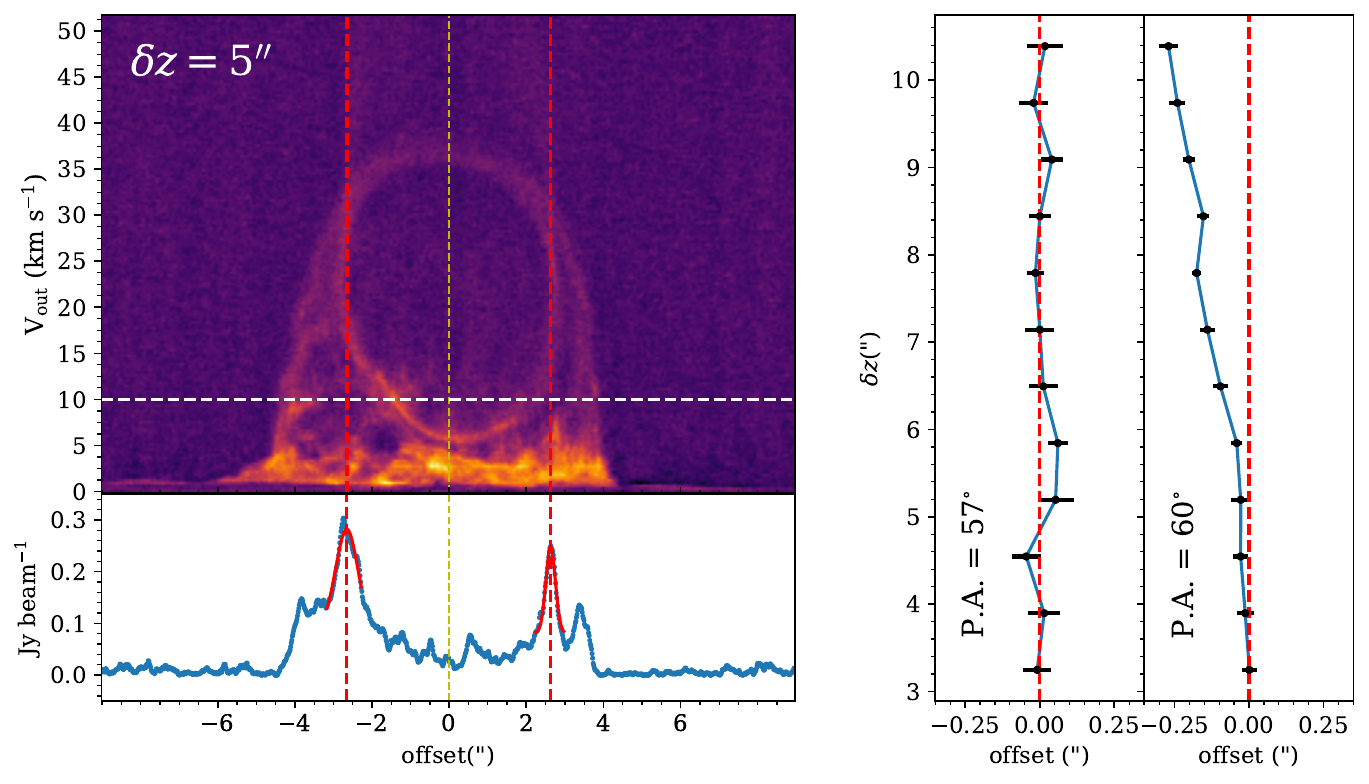}
    \caption{
    Method for determining the axis position angle of the outflow shell Sr2. 
    {\bf Upper-left:} 
    A typical transverse PV diagram assuming outflow axis $\pa = 57^{\circ}$ at $\delta z = 5\arcsec$. 
    The white dashed line marks $\vout = 10~\kms$. 
    {\bf Lower-left:} Intensity profile integrated from the PV diagram for $\vout > 10~\kms$. 
    The red curves show Gaussian fits to the two edges of the intensity profile, 
    and the red dashed lines mark the two peak positions (two edges).
    The yellow dashed line marks the midpoint between two edges.
    {\bf Right:} Offsets of the midpoint between the two edges of the outflow shell, 
    measured from transverse PV diagrams at different heights $\delta z$ and assuming different axis position angles. 
    The outflow axis position angle is determined to be $57^\circ$, for which the two edges are symmetric about the axis.
    }
    \label{fig:axis_method}
\end{figure}

\begin{figure}[ht!]
\centering
\includegraphics[width=0.9\columnwidth]{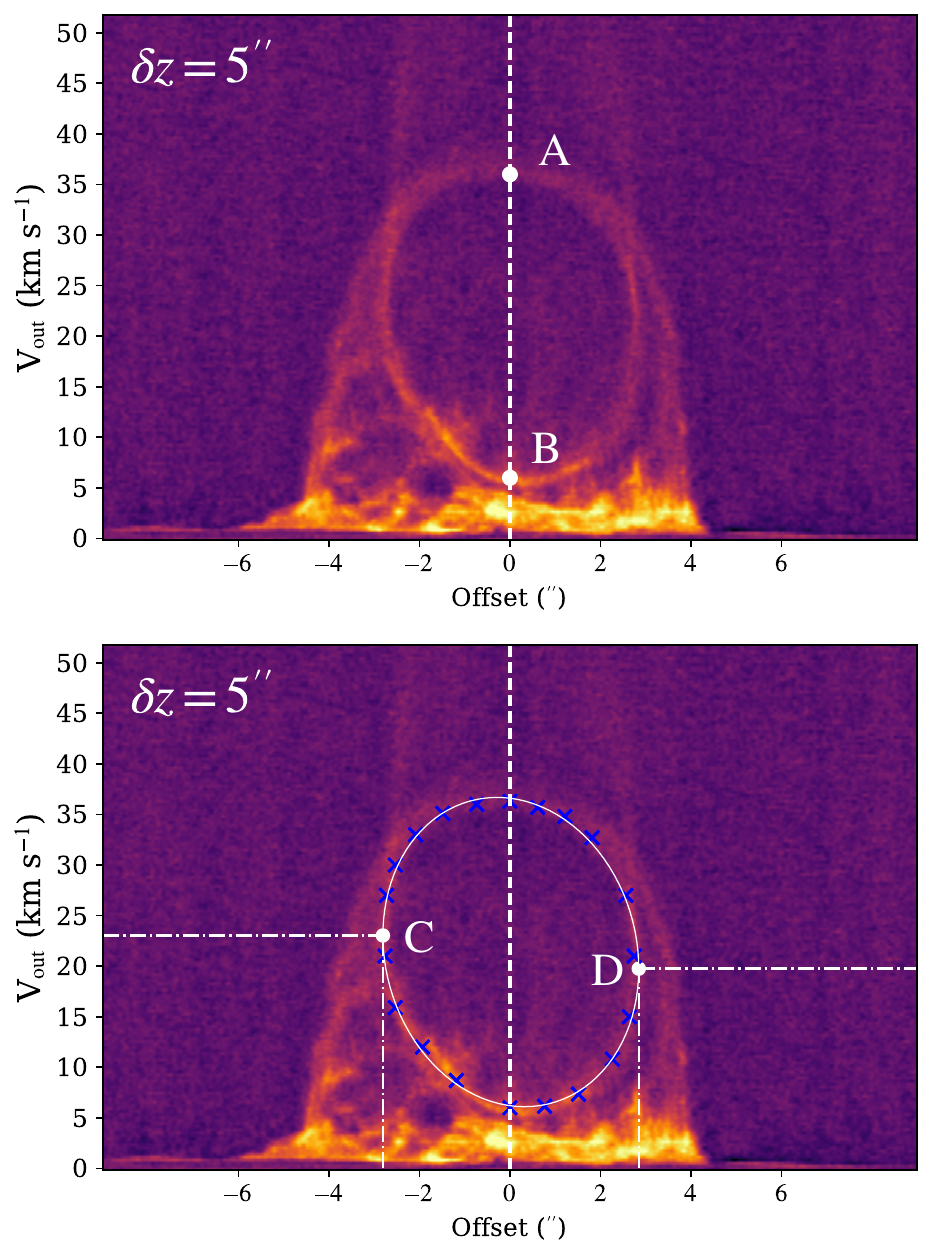}
\caption{{\bf Upper:} Transverse PV diagram at $\delta z = 5\arcsec$. 
Points A and B mark the velocities of the outflow shell Sr2 projected along the outflow axis.
{\bf Lower:} Ellipse fit to the outflow shell Sr2. 
The blue crosses indicate the selected data points used for the fit, 
and the white curve shows the best-fit ellipse. 
Points C and D mark the outermost points on the ellipse.}
\label{fig:Method_outflow_pv}
\end{figure}

\begin{figure*}
\centering
\includegraphics[width=0.8\textwidth]{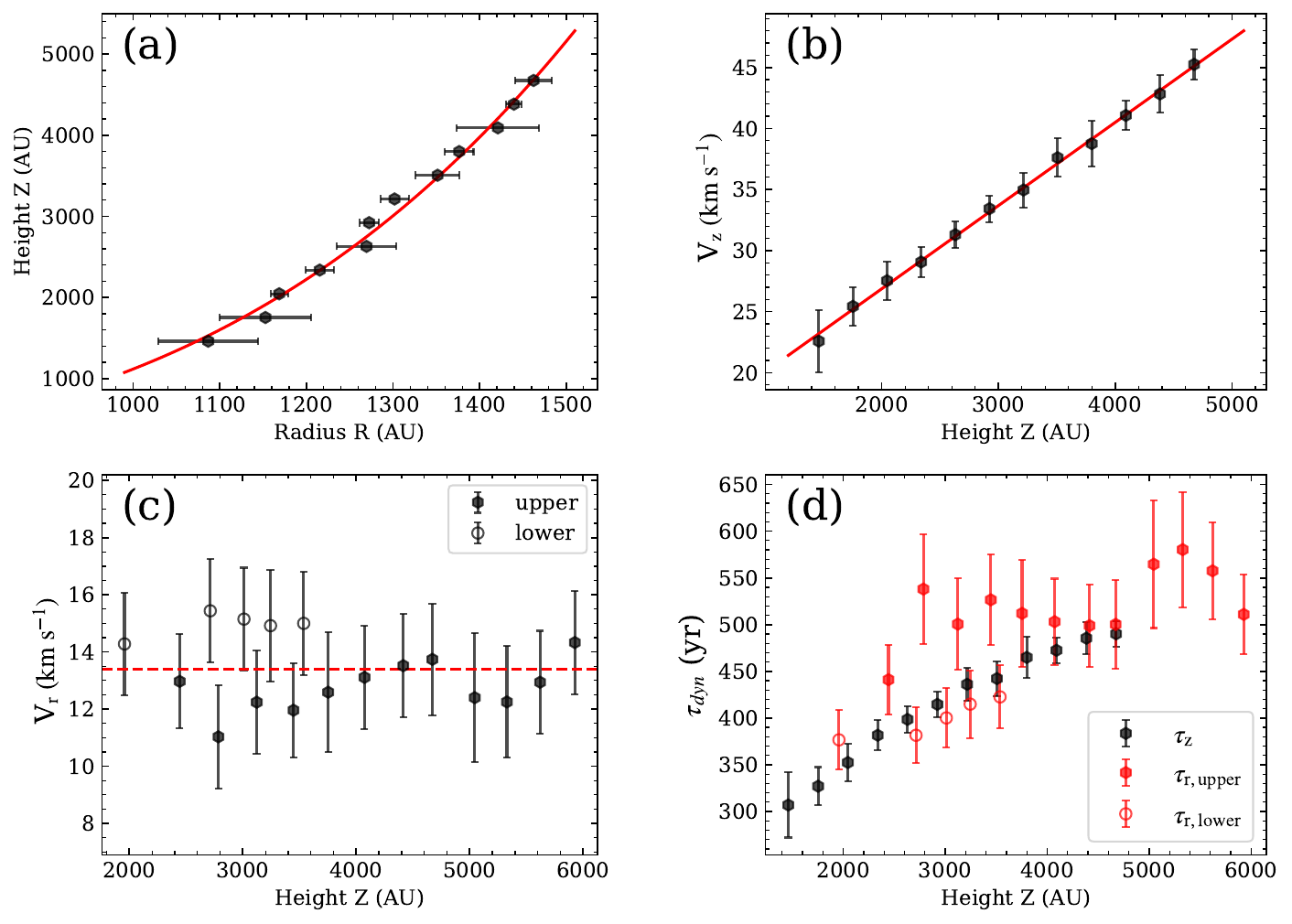}
\caption{Model-independent measurements of the morphology and velocity field of the outflow shell Sr2.
{\bf (a):} Radius of the Sr2 shell $R$ as a function of height $Z$. The red curve shows the best-fit relation, $Z \propto R^{3.6}$.
{\bf (b):} Axial velocity $V_z$ as a function of height $Z$. The red curve shows the best-fit relation.
{\bf (c):} Radial velocity $V_r$ as a function of height $Z$, showing two sets of measurements derived from the upper and lower on-axis points 
in the transverse PV diagrams (corresponding to the A and B points in Figures~\ref{fig:sketch} and \ref{fig:Method_outflow_pv}; 
see \S\ref{sec:vfield} for details).
The red dashed line indicates the average value of $13.5~\kms$.
{\bf (d):} Dynamical timescales $\tau_z = Z/V_z$ and $\tau_r = R/V_r$ as a function of height $Z$. $\tau_r$ is calculated from two groups of radial velocity $V_r$, same as (c). }
\label{fig:Velcoity_1}
\end{figure*}

\begin{figure}[ht!]
    \includegraphics[width=\columnwidth]{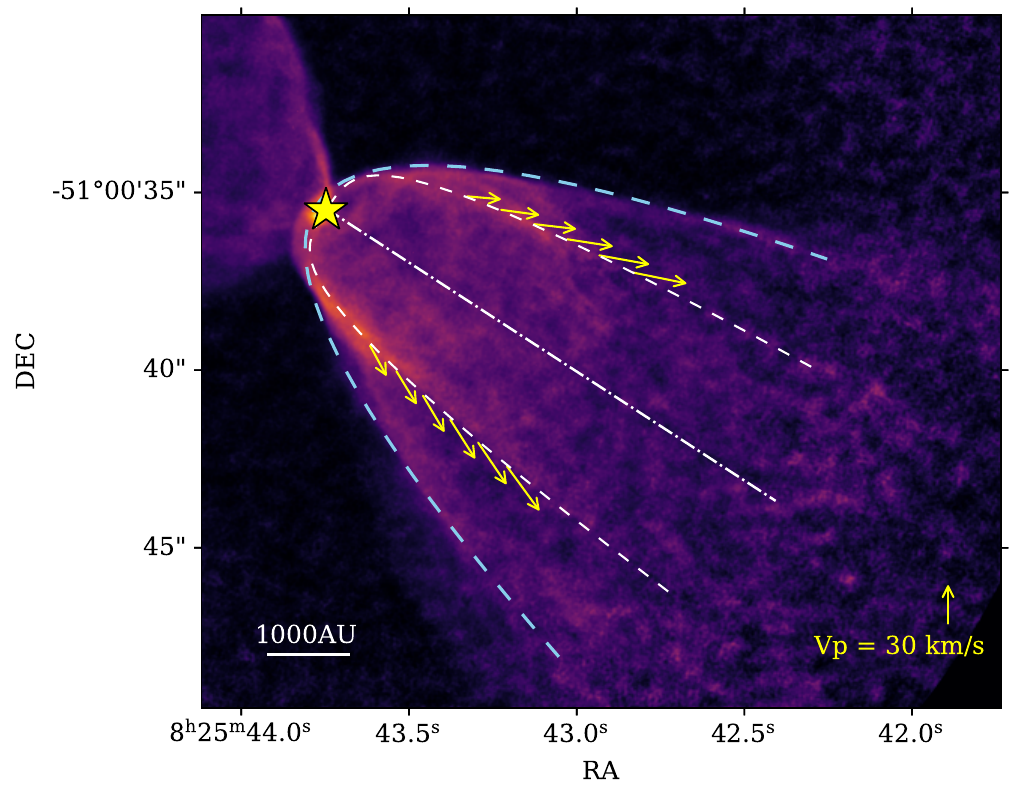}
    \caption{Projected poloidal velocities calculated from $V_r$ and $V_z$ for the Sr2 shell. 
    The background image shows the $^{12}$CO moment 0 map integrated over $[-33, 51]~\kms$. 
    The yellow arrows represent the projected poloidal velocity vectors $\mathbf{V}_p$ at $\delta z = 4\arcsec,5\arcsec,6\arcsec,7\arcsec,8\arcsec$, where $V_{r}$ equals its average value of $13.5~\kms$. 
    The reference arrow in the lower-right corner corresponds to $30~\kms$. 
    The blue dashed curve shows the boundary of redshifted outflow cavity ($Z \propto R^{2}$, see Figure \ref{fig:eye}).
    The white dashed curve shows the fitted Sr2 shell shape, $Z \propto R^{3.6}$, after projection, 
    and the white dash-dot line marks the outflow axis of the Sr2 shell.}
    \label{fig:Vp_image}
\end{figure}

\begin{figure*}[ht!]
\centering
\includegraphics[width=0.8\textwidth]{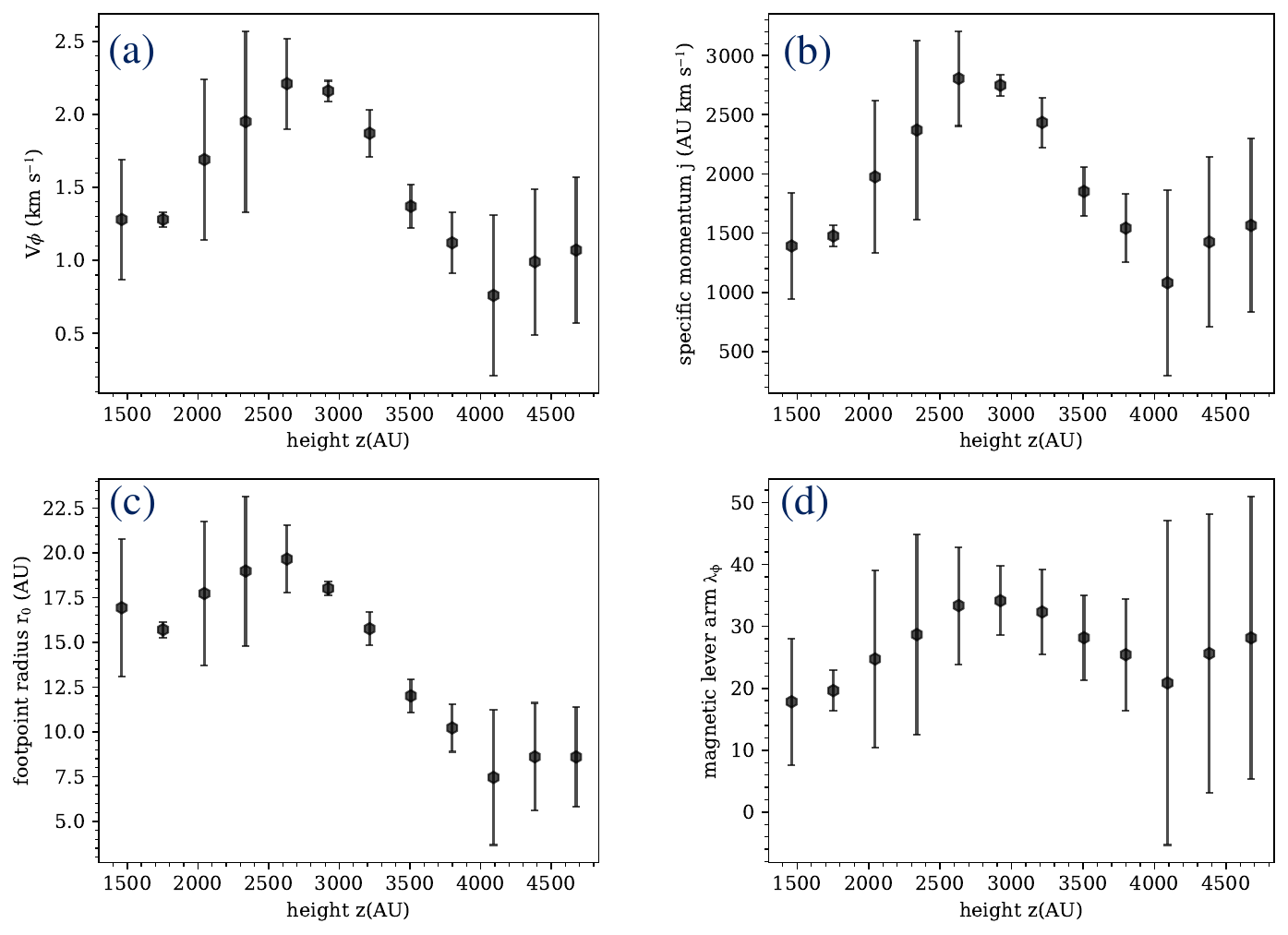}
\caption{
{\bf (a):} Rotation velocity $V_{\phi}$ of the Sr2 shell as a function of height $Z$, 
assuming that the measured transverse velocity gradient arises from rotation.
{\bf (b):} Specific angular momentum, $j = R V_{\phi}$, as a function of $Z$.
{\bf (c):} Derived disk-wind footpoint radius, $r_0$, as a function of $Z$.
{\bf (d):} Derived magnetic lever arm, $\lambda_{\phi}$, as a function of $Z$.
}
\label{fig:outflow_rotation}
\end{figure*}

\subsection{Velocity Field of the Redshifted Outflow Shells} 
\label{sec:vfield}

The HH 46/47 outflow exhibits multiple coherent, nested shells, as previously reported by \citet{Zhang2019} 
using lower-resolution data. Our high-resolution observations confirm these structures and reveal additional detail. 
In the blueshifted lobe, we do not identify new substructures from the $^{12}$CO channel maps or PV diagrams. 
In contrast, the redshifted lobe shows a more complex morphology, 
where we identify three distinct shells (Sr1, Sr2, Sr3) by searching feature correspondence between channel maps and transverse PV diagrams. 
Figure~\ref{fig:multi_shell} shows selected channel maps and transverse PV diagrams across outflow axis, where shells (Sr1, Sr2, Sr3) are labeled.
Shell Sr1 maintains a parabolic morphology in the channel maps and 
a reversed U-shaped structure in the transverse PV diagrams, 
indicating that it is a stacking of gas layers (e.g., \citealt{deValon2022}, Appendix D.2). 
In contrast, shell Sr2 consistently exhibits an elliptical morphology in the transverse PV diagrams at different heights, 
indicating that it can be viewed as a velocity-coherent thin shell (e.g., \citealt{Hirota2017}).
In the transverse PV diagram at $\delta z=7.5\arcsec$ (Figure~\ref{fig:multi_shell}), shells Sr1 and Sr2 are clearly separated.
A third shell, Sr3, is also identified in the channel maps. 
In the previous work (\citealt{Zhang2019}), Sr2 and Sr3 were interpreted as a single shell.
However, after carefully inspecting the continuous channel maps and transverse PV diagrams, 
we conclude that Sr2 and Sr3 are two distinct shells.
For example, in the channel map at $\vout=38.7~\kms$ (Figure~\ref{fig:multi_shell}), 
the base of the Sr2 shell has moved to close to that of the Sr1 shell, 
whereas Sr3 is clearly separated from them.
In the transverse PV diagram at $\delta z=5\arcsec$, 
the horizontal line marking $\vout=38.7~\kms$ intersects a triangle feature formed by shells Sr2 and Sr3.
Overall, although Sr2 and Sr3 have similar widths, 
they exhibit distinct kinematic properties, 
supporting their identification as two separate shells.

While previous studies \cite{Zhang2016} and \cite{Zhang2019} modeled the shells using wide-angle wind entrainment models proposed by \cite{Lee2000}, 
here we focus on the Sr2 shell to perform a model-independent analysis of its morphology and velocity structure 
including a search for possible rotation,
due to its velocity-coherent thin shell nature.

We define the coordinate systems following \citet{deValon2022}
(see Figure \ref{fig:sketch} for illustration).
In the outflow frame, the $Z$-axis is aligned with the outflow axis and the $X$-axis lies in the plane of the sky. 
In the observer’s frame, $\delta z$ denotes the projection of the outflow axis onto the plane of the sky, 
$\delta y$ corresponds to the line-of-sight direction, and $\delta x = X$ lies in the plane of the sky.

The determination of the outflow axis position angle is critical for deriving the velocity field. 
We follow a method similar to that of \citet{deValon2022}, 
using transverse PV diagrams at a series of projected heights $\delta z$. 
Assuming certain outflow position angle, 
at each height $\delta z$, we integrate the emission in the transverse PV diagram over 
velocities $\vout > 10~\kms$ to isolate the shell structure.
The emission profile at each $\delta z$ shows two intensity peaks 
corresponding to the shell edges (Figure \ref{fig:axis_method}).  
At some heights, more than two intensity peaks, arising from multiple shell structures, are present. 
In such cases, we select the pair of peaks corresponding to shell Sr2.
The midpoint between these two peaks, i.e. the geometry center of the outflow shell,
is expected to have zero offset relative to the assumed outflow axis if the assumed position angle is correct. 
We then perform a grid search over position angles around $60^{\circ}$ with a step size of $1^{\circ}$.
The best-fit outflow position angle is found to be $\pa = 57^\circ$, 
for which the measured midpoints at different heights are best aligned with the assumed outflow axis 
(Figure~\ref{fig:axis_method}). 
The alignment is evaluated by visual inspection.

Assuming the outflow shell is axisymmetric in both shape and velocity field, 
a transverse cut at projected height $\delta z$ intersects the shell Sr2 as a ring 
(see Figure \ref{fig:sketch}), 
which produces an ellipse-like structure in the PV diagram.
Figure~\ref{fig:Method_outflow_pv} shows an example of transverse PV diagram at $\delta z=5 \arcsec$. 
We first visually select points on this structure (blue crosses in the lower panel) 
and use an ellipse to fit them.
The outermost points of the fitted ellipse trace the shell edges
(e.g. points C and D in Figure~\ref{fig:Method_outflow_pv}  and Figure~\ref{fig:sketch}), 
where the line-of-sight velocity reflects the axial ($V_z$) and rotational ($V_\phi$) components, 
while the contribution from the cylindrical radial velocity ($V_r$) vanishes. 
Under these assumptions, the shell geometry and the axial and azimuthal velocity components at each $\delta z$ can be derived as follows:
\begin{align}
Z &= \delta z /\cos i, \\
R &= \frac{|\delta x_C| + |\delta x_D|}{2}, \\
V_z &= \frac{v_{\rm out,C} + v_{\rm out,D}}{2\sin i}, \\
V_\phi &= \frac{v_{\rm out,C} - v_{\rm out,D}}{2\cos i}, \label{eq:V_phi}
\end{align}
where $i$ is the inclination angle between outflow axis and plane of sky (Figure \ref{fig:sketch}).
We adopt $i$ = 40$^{\circ}$, same as the inclination angle value used in the envelope PV diagram fitting 
(Figure \ref{fig:PV}). 

We construct transverse PV diagrams at projected heights $\delta z$ ranging from $2.5\arcsec$ (900 au) to $8\arcsec$ (3600 au), with intervals of $0.5\arcsec$ (225 au). For each $\delta z$, we apply the analysis procedure described above to derive the morphology and velocity structure of shell Sr2, including $Z(R)$, $V_z(Z)$, and $V_{\phi}(Z)$. The results are shown in Figures~\ref{fig:Velcoity_1}a,b and \ref{fig:outflow_rotation}a.

We fit the shell morphology $Z(R)$ (Figure~ \ref{fig:Velcoity_1}a) with a power-law relation and best fit result is  
\begin{equation}
Z = 0.13\,R^{3.6} \label{eq:Z_R}
\end{equation}
(see the white dashed curve in Figure~\ref{fig:Vp_image}).
The axial velocity $V_z$ increases approximately linearly with height $Z$ (Figure~\ref{fig:Velcoity_1}b), and the best-fit relation is
\begin{equation}
V_z = 2.94\,Z + 13.23. \label{eq:VZ_R}
\end{equation}
Here $Z$ and $R$ are in arcseconds and $V_z$ is in $\kms$. 
The velocity component $V_{\phi}$ shows a non-monotonic relationship with height $Z$, 
ranging from 0.8 $\kms$ to 2.3 $\kms$. 
The measured $V_{\phi}$ may arise from true outflow rotation, 
asymmetries in the ambient medium, 
or deviations from axisymmetry in the outflow shell. 
Assuming that $V_{\phi}$ reflects rotation, we further derive the 
specific angular momentum profiles ($j = R V_{\phi}$; Figure~\ref{fig:outflow_rotation}b). 
We discuss the implications of these results in the context of disk-wind scenarios in \S\ref{sec:Disk Wind scenario}.

After having the profile $Z(R)$ and $V_z(Z)$,
we can use the upper and lower on-axis points (A and B points in Figure \ref{fig:Method_outflow_pv} and \ref{fig:sketch})
in transverse PV diagrams to estimate the radial velocity $V_r$.
The upper on-axis points (A points) correspond to the far side of the outflow shell, 
while the lower on-axis points (B points) correspond to the near side.
The velocity decomposition is given by
\begin{align}
v_{\rm out}(A) &= V_z(A)\sin i + V_r(A)\cos i, \\
v_{\rm out}(B) &= V_z(B)\sin i - V_r(B)\cos i.
\end{align}
At a given projected height $\delta z$, points A and B correspond to different intrinsic heights $Z$ 
due to projection effects (Figure~\ref{fig:sketch}). 
Therefore, the relations $R(Z)$ and $V_z(Z)$ (Equations~\ref{eq:Z_R} and \ref{eq:VZ_R}) 
are required to determine the axial velocity components $V_z(A)$ and $V_z(B)$ for each $\delta z$.

Repeating this procedure over a range of projected heights, 
we derive the $V_r$ profiles as a function of $Z$ for both the A and B point sets (Figure~\ref{fig:Velcoity_1}c). 
The number of $V_r$ measurements from the B points is smaller, 
as the lower edges of the PV ellipses are more easily blended with other structures 
and are less well defined (e.g., the $\delta z = 2.5\arcsec$ slice in Figure~\ref{fig:multi_shell}). 
The $V_r$ profiles derived from the near and far sides are broadly consistent and 
show little variation with $Z$, with an average value of $\sim 13.5~\kms$.
However, the $V_r$ values derived from the near-side (lower) points are systematically higher than those derived from the far-side (upper) points, although they remain consistent within the uncertainties. This offset may arise from the indirect and relatively complicated procedure used to derive $V_r$.

Combining $V_r$ and $V_z$, we derive the poloidal velocity vector, $\mathbf{V}_p = \mathbf{V}_r + \mathbf{V}_z$ (Figure~\ref{fig:Vp_image}).
Because $V_r$ is not measured at exactly the same $Z$ as $V_z$, we adopt the average value of $V_r = 13.5~\kms$ 
when computing $\mathbf{V}_p$. 
The resulting poloidal velocity vectors are not tangent to the shell surface but instead are nearly radial, 
indicating that the shell material is expanding rather than flowing along the shell. 
The implications of this result are discussed in \S\ref{sec:Entrainment scenario}.



\subsection{Origin of the CO Outflow}

\subsubsection{Disk Wind scenario}
\label{sec:Disk Wind scenario}

A natural interpretation of the observed transverse velocity gradient in the Sr2 shell 
is outflow rotation, as expected if the CO outflow itself is 
a magneto-centrifugally launched disk wind 
(e.g., \citealt{Bjerkeli2016,Hirota2017,Lee2017,Zhang2018,Tabone2020,deValon2022,Bacciotti2025}). 
To test this scenario, we infer the launching radius and magnetic lever arm 
from the measured rotational and poloidal velocities.

Assuming that the Sr2 shell represents a steady, axisymmetric MHD wind 
with minimal interaction with the surrounding medium, 
the footpoint radius $r_0$ can be estimated following \citet{Anderson2003}:
\begin{align}
RV_{\phi}(GM_*)^{1/2}r_0^{-3/2} - \frac{3}{2}(GM_*)r_0^{-1} - \frac{V_p^2 + V_{\phi}^2}{2} = 0,
\end{align}
where 
$R$ is the shell radius, 
$V_p$ is the poloidal velocity, 
and $V_{\phi}$ is the rotational velocity. 
This relation follows from conservation of energy and angular momentum along a streamline, 
neglecting enthalpy and gravitational potential at large distances.

The derived launching radius is $r_0 \sim 7.5 - 20~\au$ (Figure~\ref{fig:outflow_rotation}), 
comparable to values reported in wide-angle molecular outflows 
(e.g. \citealt{Bjerkeli2016,Tabone2017,Zhang2018,Louvet2018,deValon2022,
Bacciotti2025}),
but higher than values derived from class 0/I jets 
seen in millimeter (e.g. \citealt{Lee2017})
or optical (e.g. \citealt{Bacciotti2002,Coffey2007}).
From \S\ref{sec:IRE}, the centrifugal barrier is $\rcb \sim 30$ au, 
within which a Keplerian disk is expected. 
The inferred launching radius is therefore broadly consistent 
with the scenario of a disk wind.

We further estimate the magnetic lever arm parameter using $\lambda_{\phi} = RV_{\phi}/(\Omega_0 r_0^2)$, 
where $\Omega_0$ is the Keplerian angular velocity at $r_0$. 
The derived values(Figure~\ref{fig:outflow_rotation}d), $\lambda_{\phi} \sim 19-32$, however, 
are significantly larger than those reported in other systems 
(typically $\lambda \lesssim 10$; e.g., HH 30, DG Tau B, HL Tau, and NGC 1333 IRAS 4C; \citealt{Louvet2018,Vazquez2024,deValon2022,Bacciotti2025,Zhang2018}). 
Although such large magnetic lever arms are formally allowed in classical disk-wind models 
(e.g., \citealt{Blandford1982}), models with typical mass-loading efficiencies 
($\dot{M}_{out}/\dot{M}_{acc}\sim 0.1$) 
generally favor smaller values of $\lambda \lesssim 10$ 
(e.g., \citealt{Ferreira1997AA}). 

In the self-similar disk-wind model of \citet{Ferreira1997AA}, the maximum mass-loading efficiency is
$\dot{M}_{out}/\dot{M}_{acc} \sim (r_e/r_i)^\xi - 1$,
where $\xi=1/[2(\lambda-1)]$, and $r_e$ and $r_i$ are the outer and inner radii of 
the wind-launching region. 
For $\lambda=20$, corresponding to $\xi\approx0.026$, 
and adopting an extreme value of $r_e/r_i\sim1000$ 
(i.e., assuming that the wind-launching region extends all the way to the stellar surface), 
we obtain $\dot{M}_{out}/\dot{M}_{acc}\sim0.2$. 
Given the observed outflow rate of $\dot{M}_{out}\sim10^{-5}~M_\odot~{\rm yr}^{-1}$ (\S\ref{sec:out_mass}), such a model would require accretion rates of order 
$10^{-5}-10^{-4}~M_\odot~{\rm yr}^{-1}$, 
substantially higher than the typical accretion rates expected for low-mass forming stars like HH~46/47. 
In fact, an accretion rate of $2.2 \times10^{-7} M_{\odot}~\rm yr^{-1}$ was reported for HH~46~IRS by \cite{Antoniucci2008}. 

This discrepancy suggests that a large fraction of the observed outflowing material is not directly launched as a disk wind.

Furthermore, the inferred launching radii of $r_0 \sim 7.5 - 20~\au$
lie in the weakly ionized outer disk, 
where modern magneto-thermal wind models predict significantly smaller lever-arm parameters, 
typically $\lambda\lesssim3$ \citep{Bai2016,Lesur2021AA}. 
The combination of large inferred $\lambda_{\phi}$ values, the implied low mass-loading efficiencies, and the large launching radii therefore argues against interpreting the observed CO shell rotation as the direct signature of a launched disk wind.
This conclusion is consistent with the results of \citet{Birney2024}, 
who reported similarly large
specific angular momentum
($\sim3000~\au~\kms$) in the blueshifted outflow lobe 
based on infrared H$_2$ observations.
They also found that the transverse velocity gradient in the blueshifted lobe 
(redshifted to the northwest and blueshifted to the southeast) 
is opposite to that of the envelope/disk system, 
and therefore argued against an interpretation in terms of wind rotation.
In contrast, the transverse velocity gradient that we observe in the redshifted outflow is in the same sense as the envelope/disk rotation.
The opposite senses of the transverse velocity gradients observed in the two outflow lobes 
are difficult to reconcile with a rotating disk wind, 
and instead suggest that wind rotation is unlikely to be the primary origin of the observed velocity gradients.

\subsubsection{Entrainment scenario}
\label{sec:Entrainment scenario}

An alternative explanation is that the CO emission traces entrained material instead of directly launched material. 
In this scenario, the observed outflow arises from the interaction between a jet/wind and the ambient medium. 
Simulations have shown that most of the wide angle outflows are due to entrainment, 
with a mass ratio $\sim$1:4 between the launched material to the total outflowing (launched and entrained) material
(e.g. \citealt[]{Offner2017}).
Entrainment can occur through jet-driven bow shocks (e.g., \citealt{Raga1993,Rabenanahary2022}) or 
wide-angle winds sweeping up surrounding gas (e.g., \citealt{Shu1991,Lee2001,Shang2023}). 

In entrainment scenario, apparent velocity gradients across the outflow axis 
may arise from asymmetries in the ambient medium rather than intrinsic outflow rotation.
In our analysis, the derivation of the rotational velocity assumes that the outflow shell is axisymmetric.
However, an asymmetric ambient medium may distort the shell morphology away from perfect axisymmetry.
Combined with projection effects, such distortions can cause the transverse PV diagram to deviate from the idealized elliptical shape
(Figure~\ref{fig:Method_outflow_pv}),
thereby introducing an apparent transverse velocity gradient or altering the magnitude of the measured gradient.
In this case, the inferred large specific angular momentum does not necessarily result from 
differential interaction between the wind and the ambient material, 
but may instead reflect a geometric effect caused by the distorted shell morphology.


Several lines of evidence support the entrainment scenario. 
First, the derived poloidal velocity $\mathbf{V}_p$ is not tangent to the shell surface (Figure~\ref{fig:Vp_image}), 
indicating that the shell is expanding rather than representing a steady flow along streamlines. 
The dynamical timescales $\tau_r = R/V_r$ and $\tau_z = Z/V_z$ are comparable at different heights 
(Figure~\ref{fig:Velcoity_1}d), 
implying that the expansion is approximately radial. 
Here, the two groups of $\tau_r$ shown in the figure 
correspond to values derived from $V_r$ measured on the near side 
(lower on-axis points in the transverse PV diagrams) and the far side (upper on-axis points) of the outflow shell.
This radial expansion behavior is consistent with numerical simulations of 
entrainment (e.g., \citealt{Lee2001,Shang2023}).

Second, while in principle, a disk wind could also produce expanding shell-like structures 
if the emission traces localized density enhancements rather a group of streamlines, 
such models generally predict velocity stratification, 
with higher-velocity material confined closer to the axis (e.g., \citealt{deValon2022}).
This, however, is not observed in our data.
Instead, the Sr2 shell moves outward with increasing velocity (Figure~\ref{fig:r_channel}), 
inconsistent with this expectation.
Moreover, both $\tau_r$ and $\tau_z$ increase with height $Z$, 
indicating that the shell is decelerating as it interacts with and accumulates ambient material. 
This behavior further supports the entrainment scenario.

Finally, JWST H$_2$ observations of the HH 46/47 outflow (\citealt{Navarro2025}) 
show that shock-excited gas lies interior to the CO Sr2 shell and arises from interactions between jets or wide-angle winds 
and the surrounding medium. 
The radial velocity of H$_2$ emission within 1500 $\au$ of the central source reaches up to 50 $\kms$,
while the CO radial velocity only reaches $\sim30~\kms$ in the same region (see \citealt{Navarro2025}).
These spatial and kinematic relationship suggest that the CO shell traces slower swept-up ambient material 
rather than directly launched wind material.
Taken together, these results strongly favor an entrainment origin for the CO shells.

However, the mechanism responsible for entrainment remains uncertain. 
The observed molecular outflow could be driven by a wide-angle wind, 
a jet-driven bow shock, or a combination of both. 
\citet{Zhang2019} successfully reproduced the multiple shell structures in HH~46/47 
using an episodic wide-angle wind model similar to that proposed by \citet{Lee2000}. 
Our model-independent analysis of the Sr2 shell confirms its overall expanding nature, 
consistent with wide-angle wind entrainment, 
but also reveals deviations from the simple analytic prescriptions 
adopted in previous studies.

An alternative to the wide-angle wind entrainment scenario is jet entrainment.
Recent JWST observations provide new constraints on this scenario.
The blueshifted [FeII] jet has a position angle that differs from the CO outflow axis 
by approximately $11^\circ$ (see Figure~\ref{fig:b_channel}), 
which at first glance appears inconsistent with a simple jet-entrainment scenario. 
However, the redshifted [FeII] jet seen in MIR 
is closely aligned with the derived CO outflow axis. 
Moreover, the mean jet axis, defined by the innermost blueshifted and redshifted 
5.3 $\mu$m [FeII] knots (Figure~3 of \citealt{Nisini2024}), 
is also consistent with the CO outflow orientation. 
These observations suggest that temporal variations in the jet direction 
could produce an entrained molecular outflow whose time-averaged axis 
matches the observed CO outflow.

If jet entrainment is indeed responsible for driving the CO outflow,
it is still unclear whether the CO outflow is driven solely by the source A jet,
or by a combination of jets from both sources A and B.
Although a jet from source B has not been directly detected in the infrared,
an H$_2$ knot that might be associated with this jet (knot A1) has been identified
(\citealt{Nisini2024,Navarro2025}), implying a position angle of about $70^\circ$.
It is worth noting that the blueshifted [FeII] jet from source A
and the inferred source B jet lie on opposite sides of the CO outflow axis,
with similar position angle offsets.
Therefore, the source B jet may also contribute
to driving the CO molecular outflow.
However, no clear signature in the CO emission corresponding to the
source B outflow is seen in our CO data
(see Figures~\ref{fig:RGB} and \ref{fig:b_channel}).
A CO component extending toward the southeast was
seen at blueshifted low velocities of $\sim -1~\kms$,
and was suggested to be associated with the secondary outflow (\citealt{Arce2013}).
However, this component is largely misaligned with the
inferred source B jet direction.
As this feature is only seen at very low velocities,
it is not clearly distinguished in our high-resolution data.

Multiple nested molecular shells are also predicted by the unified outflow models 
of \citet{Shang2023} and \citet{ai2024}, 
which include both a collimated jet and a surrounding wide-angle wind. 
In these models, a broad compressed mixing layer develops between the CO shells
and the ambient envelope. 
Interestingly, the observed redshifted CO shell follows a significantly narrower 
profile ($Z\propto R^{3.6}$; \S\ref{sec:vfield}) 
than the outer cavity boundary ($Z\propto R^2$; \S\ref{sec:outflow_morphology}), 
as illustrated in Figure~\ref{fig:Vp_image}. 
The relatively slow outflowing material located between these two structures 
is consistent with the compressed mixing layer predicted by the unified models. 
Overall, the current observations do not uniquely distinguish 
between wide-angle wind entrainment, 
jet-driven entrainment, 
and the unified outflow scenario.

\begin{table*}[ht!]
    \caption{Derived Properties of the CO Outflow} 
    \begin{tabular}{ccccccc} 
         \hline\hline
         Lobe & $M_\mathrm{out}$ & $P_\mathrm{out}$ & $E_\mathrm{out}$ & 
         $\dot{M}_\mathrm{out}$ & $\dot{P}_\mathrm{out}$ & 
         $\dot{E}_\mathrm{out}$ \\ 
         & ($10^{-2} M_{\odot}$) & ($10^{-2}~M_\odot~\kms$) & ($10^{42}$ erg) & 
         ($M_\odot~\mathrm{Myr}^{-1}$) & ($M_\odot~\kms~\mathrm{Myr}^{-1}$) & 
         ($10^{46}~\mathrm{erg}~\mathrm{Myr}^{-1}$)\\ 
         \hline
         blue& 9.8& 40.6 & 22.0 &10& 50 &0.9\\
         red& 21.1& 92.7& 56.6 &19& 103&2.1\\
         \hline
   \end{tabular}
    \label{tab:outflow}
\end{table*}

\begin{figure*}[ht!]
\centering
\includegraphics[width=0.9\textwidth]{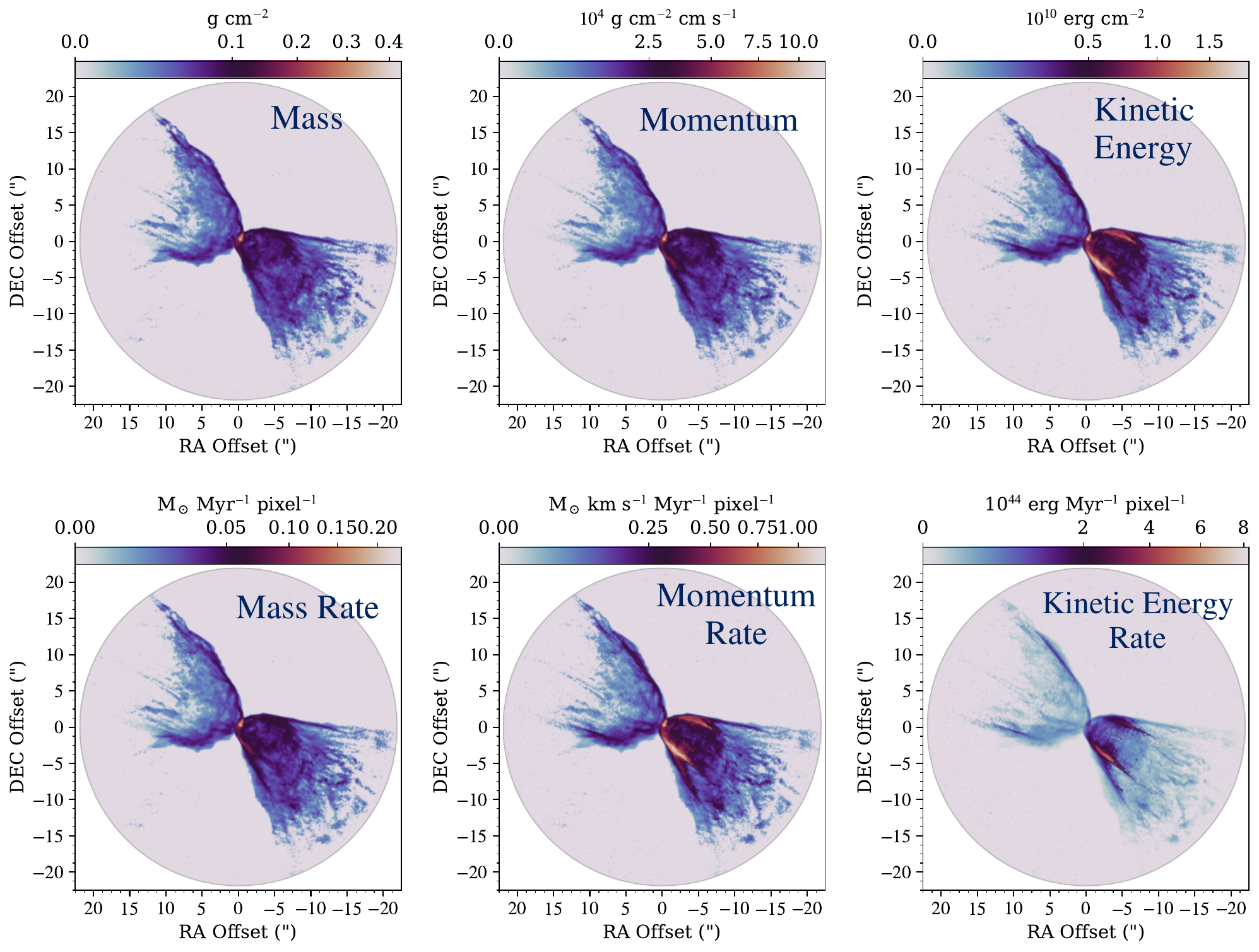}
\caption{{\bf Upper:} Surface density maps of outflow mass, momentum, and kinetic energy, 
including only pixels with intensities per channel $>4\sigma$.
{\bf Lower:} Maps of the outflow mass, momentum, and kinetic energy rates,including only pixels with intensities per channel $>4\sigma$. Both the upper and lower panels have been corrected for the primary beam response. The minimum primary beam correction factor is 0.3.}
\label{fig:quantity_map}
\end{figure*}

\begin{figure}[ht!]
\centering
\includegraphics[width=\columnwidth]{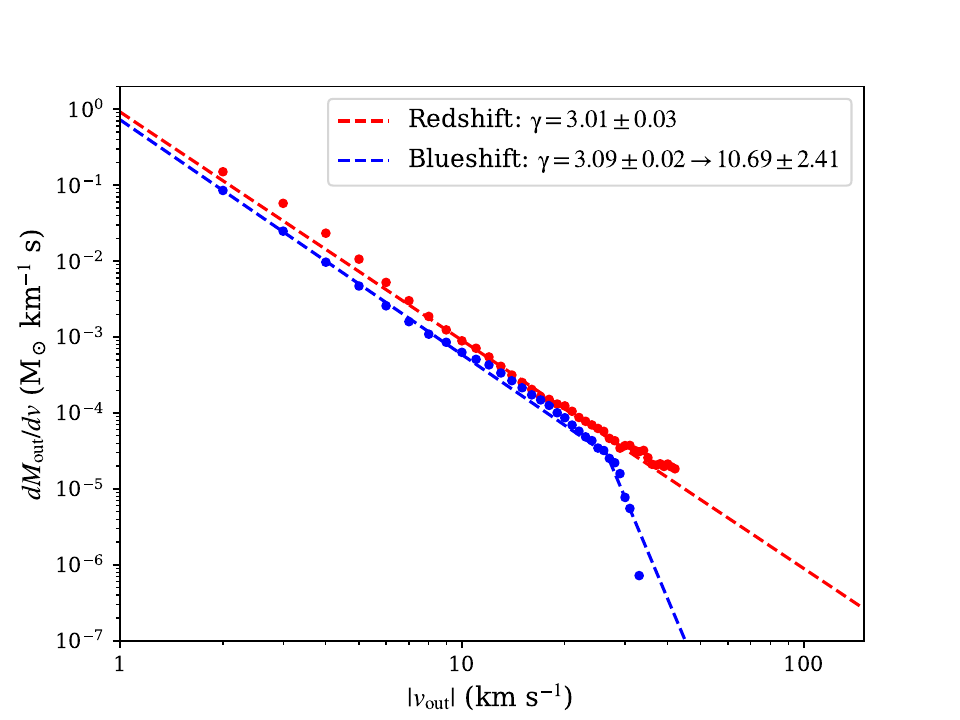}
\caption{Mass–velocity relation of the HH 46/47 outflow. The outflow mass is calculated after primary beam correction. Only the area where the primary beam correction factor $>0.3$ is included. The blue and red points represent the blueshifted and redshifted lobes, respectively. The blue and red dashed lines show the power-law fits to the mass spectra.}
\label{fig:mass_spec}
\end{figure}

\begin{figure}[ht!]
\centering
\includegraphics[width=\columnwidth]{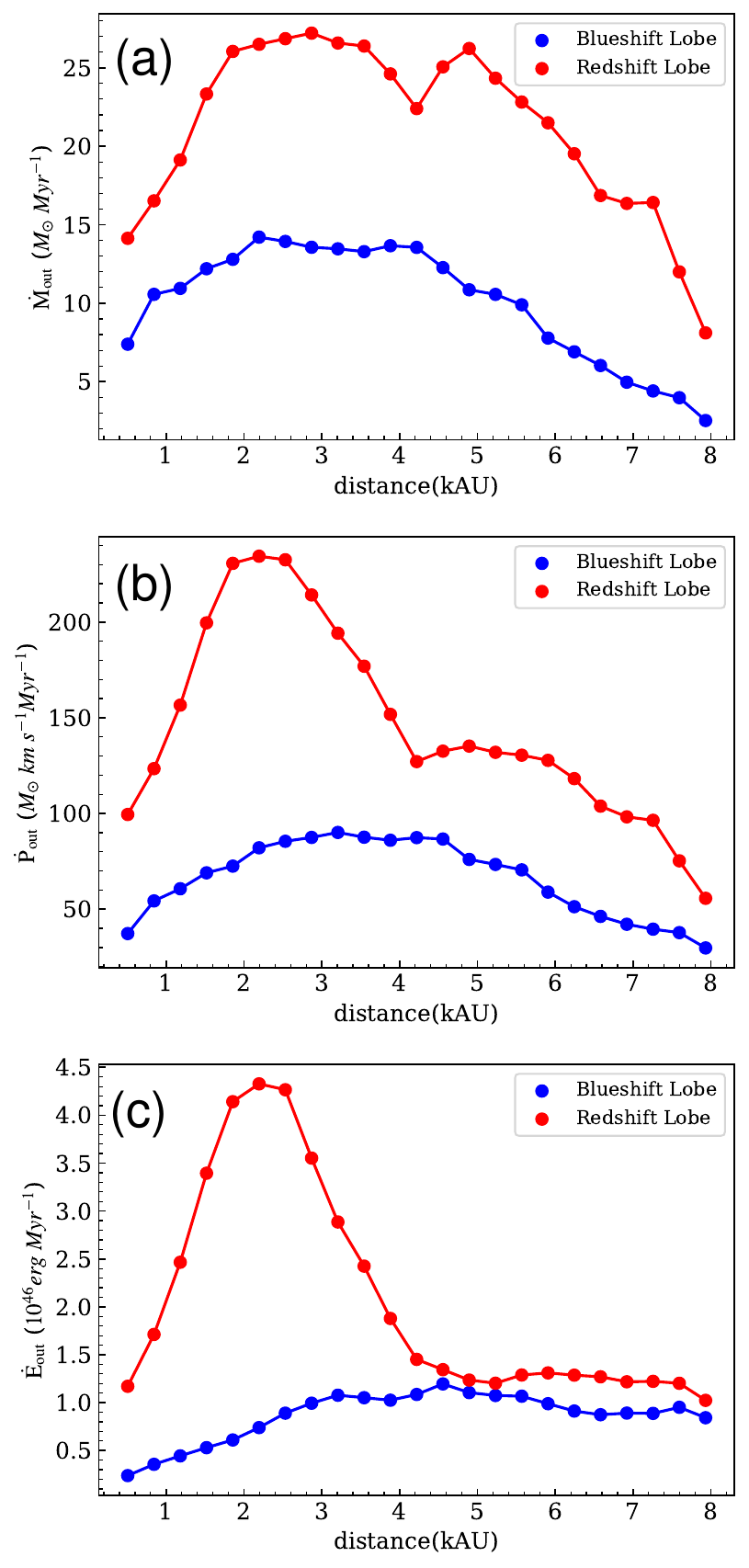}
\caption{Radial profiles of the outflow mass rate (panel~a), momentum rate (panel~b), and kinetic energy rate (panel~c). The profile is after primary beam correction, and only the area where the primary beam correction factor $>0.3$ is included. }
\label{fig:RM}
\end{figure}

\subsection{Mass, Momentum and Kinetic Energy Distributions of the CO Outflow}
\label{sec:out_mass}

In this section, we use $^{12}$CO (2-1), $^{13}$CO  (2-1), and C$^{18}$O (2-1) to derive the mass, momentum, 
and kinetic energy distributions near the outflow base. 
Optical depth corrections for $^{12}$CO are performed using $^{13}$CO and C$^{18}$O 
following \citet{Dunham2014,Zhang2016}, with details given in Appendix~\ref{app:column densities}.
Although \citet{Zhang2016} carried out a similar analysis using large-scale, lower-resolution CO (1-0) observations, 
the present CO (2-1) data probe the inner outflow at much higher angular resolution and are sensitive to warmer and higher-velocity gas
($|v_\mathrm{out,max}| \sim 50~\kms$, compared to $\sim20~\kms$ in \citealt{Zhang2016}).
Consequently, the derived mass-velocity relation and the estimated mass, momentum, and kinetic energy rates 
differ from those in the previous study, as discussed below.

We first compute the spatial distributions of the outflow mass, momentum, and kinetic energy surface densities 
(Figure~\ref{fig:quantity_map}) as
\begin{align}
\Sigma_M(x,y) & = \sum_{\vout} \Sigma_M (x,y,\vout), \\
\Sigma_P(x,y) & = \sum_{\vout} \Sigma_M (x,y,\vout) \voutc, \\
\Sigma_E(x,y) & = \sum_{\vout} \frac{1}{2} \Sigma_M(x,y,\vout) \voutc^2,
\end{align}
where $\Sigma_M(x,y,\vout)$ is the mass surface density in each velocity channel, 
and $\voutc = \vout/\sin i$ is the inclination-corrected velocity, with $i = 40^{\circ}$ (see \S\ref{sec:vfield}).
Only pixels with intensities $>4\sigma$ in each channel are included. 
Primary beam correction has been applied before, with the minimum correction factor set to be 0.3. 
We also only include emission with $|\vout|>2~\kms$ 
to exclude the contribution of the cloud material.
Compared to the distribution maps presented in \citet{Zhang2016}, 
our new results reveal layered shell structures in the redshifted outflow lobe, particularly in the momentum and energy maps, 
indicating that this region is where the bulk of the momentum and kinetic energy are injected.


Integrating over the emission area yields the total outflow mass, 
momentum, and kinetic energy (Table~\ref{tab:outflow}). 
These values should be regarded as lower limits, 
as the lack of short-spacing data filters out extended emission,
and our field of view only covers the inner region of the outflow.
Assuming $T_{ex} = 15~K$, the derived outflow mass is lower than the result in \cite{Zhang2016}.
However, the momentum and kinetic energy in both blueshifted and redshifted lobes are all higher than that of \cite{Zhang2016}, 
likely reflecting the detected higher outflow velocity.
Compared to averaged values measured for Class I outflows in Orion A cloud (\citealt{Hsieh2023}),
our derived total outflow mass $M$, momentum $P$ and kinetic energy $E$ are about $7, 6, 5$ times lower, respectively.


Figure~\ref{fig:mass_spec} shows the outflow mass spectra ($dM_{\rm out}/dv$), 
which follow power-law relations ($dM_{\rm out}/dv \propto v^{-\gamma}$). 
The redshifted lobe is well fitted by a single power law with $\gamma = 3.01$ over $2 < |\vout| < 42~\kms$. 
The blueshifted lobe exhibits a broken power law, with $\gamma = 3.09$ for $2 < |\vout| < 30~\kms$ 
and $\gamma = 10.69$ for $|\vout| > 30~\kms$. 
Broken power-law behavior is commonly observed in molecular outflows 
from both low- and high-mass protostars (e.g., \citealt{Arce2007,Su2004,Yang2024}), 
with typical break velocities of $|\vout| \sim 6-12~\kms$ (\citealt{Arce2007}), 
and can be reproduced by theoretical models and simulations (\citealt{Matzner1999,Rabenanahary2022,Li2017}).
Previous studies of the HH 46/47 outflow (\citealt{Zhang2016}) showed that, 
after optical depth correction, the mass spectra follow a single power law with index $\sim 3$ up to $|\vout| \sim 15~\kms$, 
limited by sensitivity. 
Our measurements are consistent at low velocities but extend to higher velocities, revealing a continuous power-law behavior 
with $\gamma \sim 3$ up to $|\vout| \sim 42~\kms$.
At the highest velocities, the blueshifted emission partially moves out of the field of view, 
leading to an apparent steepening of the spectrum for $|\vout| \gtrsim 30~\kms$. 
Taking this effect into account
and since there are only four velocity bins supporting a broken law when $|\vout| \gtrsim 30~\kms$, our results are consistent with a single power-law mass spectrum 
with $\gamma \sim 3$ over the full observed velocity range.

Based on the outflow mass, momentum, and kinetic energy maps, 
we derive the corresponding rate distributions. 
A common approach is to estimate the rates by dividing the outflow quantities ($M$, $P$, $E$) 
by a dynamical timescale $t_{\rm dyn} = R_{\rm lobe}/v_{\rm max}$ 
(e.g. \citealt{vanderMarel2013,Zhang2016}). 
However, this method is highly uncertain, 
as both $R_{\rm lobe}$ and $v_{\rm max}$ depend on observational sensitivity and field of view. 
Moreover, it assumes that the outflowing material is directly launched from the central source and 
moves to their current positions at nearly constant velocity, 
which is unlikely for outflows mostly composed of entrained material (e.g., \citealt{Offner2017}).
To obtain more reliable rate estimates
with respect to the analysis presented in \citet{Zhang2016} 
who used low-resolution data,
we adopt the ring method (\citealt{Hsieh2023}), 
which divides the outflow into concentric rings and computes the rates within each ring
(See Figure 6 in their paper for schematic illustration):
\begin{align}
\dot{M} &= \sum_{v_{\rm chan}} \frac{\frac{dm}{dv}}{\Delta R_{\rm corr}/v_{\rm out,corr}}\Delta v_{\rm chan}, \\
\dot{P} &= \sum_{v_{\rm chan}} \frac{\frac{dm}{dv} v_{\rm out,corr}}{\Delta R_{\rm corr}/v_{\rm out,corr}}\Delta v_{\rm chan}, \\
\dot{E} &= \sum_{v_{\rm chan}} \frac{\frac{1}{2}\frac{dm}{dv} v_{\rm out,corr}^2}{\Delta R_{\rm corr}/v_{\rm out,corr}}\Delta v_{\rm chan},
\end{align}
where $dm/dv$ is the mass per unit velocity for each channel within a ring, 
$\Delta R_{\rm corr}$ is the inclination-corrected ring width (adopted as $0.75\arcsec$, 340 au), 
and $v_{\rm out,corr}$ is the inclination-corrected velocity.
Applying this method over a series of radii, 
we derive the radial profiles of the mass, momentum, and kinetic energy rates (Figure~\ref{fig:RM}). 
The redshifted lobe exhibits systematically higher rates than the blueshifted lobe, 
likely reflecting environmental asymmetry, 
as the blueshifted lobe lies near the edge of the parent cloud and may entrain less material. 
The energy rate peaks at $\sim 2500$ au ($5.6\arcsec$), 
coinciding with the high-velocity inner shell (Figure~\ref{fig:quantity_map}).
By averaging over all radial bins, we obtain the total outflow mass, momentum, and kinetic energy rates, which are listed in Table~\ref{tab:outflow}. 
Compared to the mass and momentum rates derived by \citet{Zhang2016},
the new results based on the ring method are slightly higher in value, 
yet remain comparable in order of magnitude.
In contrast to the total outflow mass ($M$), momentum ($P$), and kinetic energy ($E$), 
the derived mass-loss rate ($\dot{M}$), momentum rate ($\dot{P}$), and kinetic energy injection rate ($\dot{E}$) 
are approximately 5, 8, and 30 times higher, respectively, 
than the average values reported for Class I outflows in the Orion A cloud (\citealt{Hsieh2023}). 
This enhancement is likely attributable to the higher outflow velocities traced by our observations, 
which reach $|v_\mathrm{out,max}|\sim50~\kms$.

\begin{figure*}[ht!]
    \centering
    \includegraphics[width=1\textwidth]{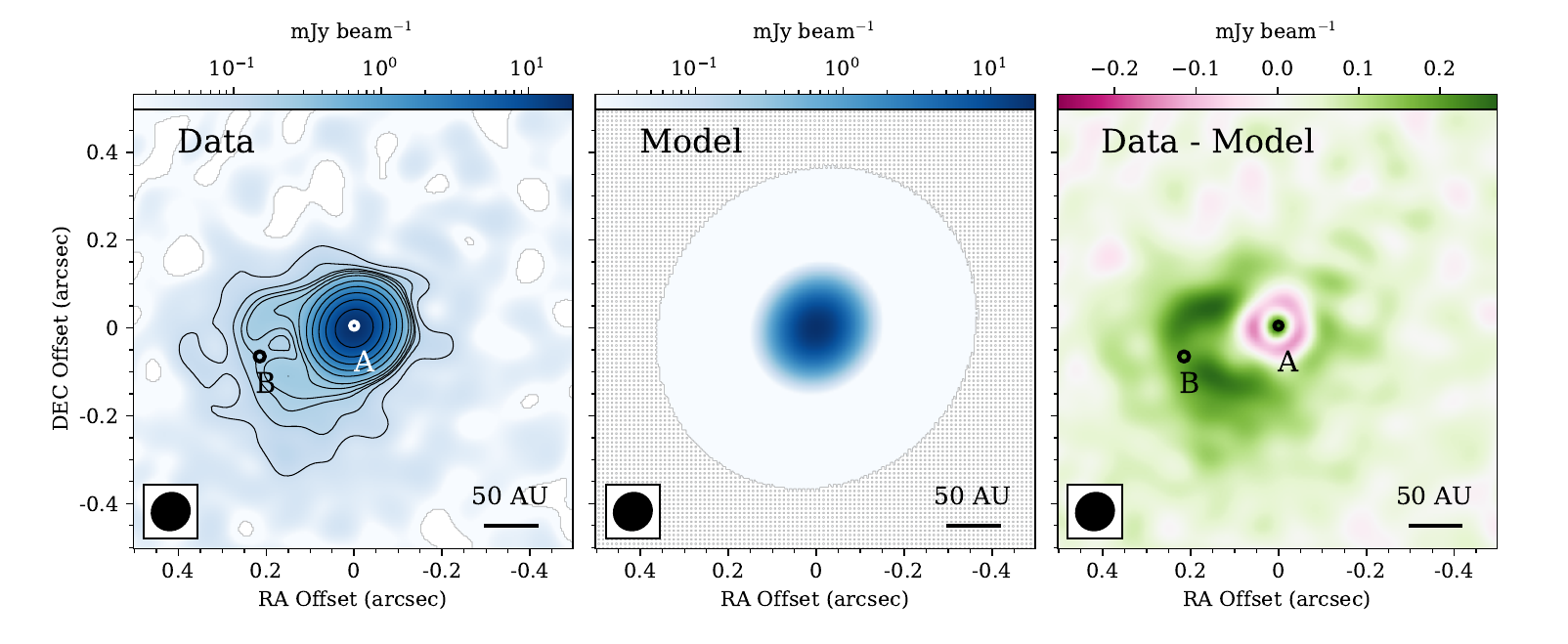}
    \caption{\textit{From left to right}: 1.3 mm continuum image (robust = $-0.5$), 2D Gaussian model, and residual map (data $-$ model).}
    \label{fig:2D_Gauss}
\end{figure*}

From the measured outflow mass-loss rate of $\dot{M}_{\rm outflow} = 29~M_\odot~\mathrm{Myr^{-1}}$, 
we estimate the timescale for dispersing the core. 
By fitting the sub-mm continuum emission, \cite{vanKempen09} 
derived a remaining core mass of $5.1~M_\odot$ within 0.1 pc ($46\arcsec$, $2.08\times10^4$ au). 
The corresponding 
remaining 
core-dispersal timescale is therefore
\begin{equation}
\tau_{\rm des} \approx \frac{M_{\rm core}}{\dot{M}_{\rm outflow}} \approx 0.17~\mathrm{Myr}.
\label{eq:tau_core}
\end{equation}
Adopting a typical age of $\sim 0.13-0.26$ Myr for an early Class I source as a proxy 
for the outflow age (\citealt{Dunham2015}), 
the total core-dispersal timescale becomes $\sim 0.3-0.43$ Myr. 
This is comparable to the typical combined Class 0+I lifetime of $\sim 0.40-0.78$ Myr 
(\citealt{Dunham2015}).
\footnote{
Note the typical lifetime used above from \cite{Dunham2015} assumes that the Class II stage has a lifetime of $\sim2$ Myr. 
However, recent work suggests that the Class II lifetime could be approximately twice as long 
($5.8 \pm 0.3$ Myr instead of the canonical $2 - 3$ Myr; \citealt{Polnitzky2026}). 
If this is the case, the total core-dispersal timescale would become $\sim0.43 - 0.68$ Myr, 
and the Class 0+I lifetime would become $\sim0.80 - 1.56$ Myr.
} 
This timescale is also shorter than the previously estimated total core-dispersal time of 0.52–1 Myr reported by \citet{Zhang2016},
whose estimate relied on the simple assumption of a constant solid-angle growth rate due to the lack of a reliable mass outflow rate.
In contrast, our newly derived value of $\sim0.3 - 0.43$ Myr,
based on a more direct method (equation \ref{eq:tau_core}), seems more reliable.
However,
we also note that our observations primarily probe the inner outflow, 
and extended emission is likely missing; 
therefore, the derived mass-loss rate should be considered a lower limit. 
Consequently, the true dispersal timescale may be shorter. 
These results suggest that the outflow is capable of dispersing the core 
within the Class 0+I evolutionary timescale.

To further examine spatial variations, we apply the pixel flux tracing (PFT) method (\citealt{Hsieh2023}), 
which treats each pixel as a ring. The crossing time is $\tau_{\rm pix} = \frac{\Delta R_{\rm pix}/\cos i}{\vout/\sin i}$
and the corresponding rates per pixel are
\begin{align}
\dot{M}(x,y) & = \sum \frac{M(x,y,\vout)}{\tau_{\rm pix}},\\
\dot{P}(x,y) & = \sum \frac{M(x,y,\vout)\,v_{\rm out,corr}}{\tau_{\rm pix}},\\
\dot{E}(x,y) & = \frac{1}{2}\sum \frac{M(x,y,\vout)\,v_{\rm out,corr}^2}{\tau_{\rm pix}},
\end{align}
where $\dot{M}(x,y)$, $\dot{P}(x,y)$, and $\dot{E}(x,y)$ are the mass, momentum, and energy rates per pixel,
and $M(x,y,\vout)$ is the mass in each pixel and each channel. 
The resulting rate distribution maps are shown in Figure~\ref{fig:quantity_map}.
The mass rate map traces the mass entrained by the outflow,
the momentum rate map shows the force acting on the entrained gas, and
the kinetic energy rate map shows the mechanical luminosity in each pixel 
and highlight the high-velocity shell component.
In \cite{Hsieh2023}, kinetic energy rate maps often highlight the protostellar jets, which is unseen in our result.
This fact further supports that the jets in HH46/47 is not molecular. 


\subsection{Substructure in 1.3 mm continuum}
\label{sec:binary}

As shown in \S\ref{sec:continuum}, the continuum emission is dominated by a bright central component surrounded by fainter, structured emission. 
To better highlight the extended features, Figure~\ref{fig:2D_Gauss} presents a Gaussian fit to the central component, 
with the residual map revealing the surrounding structures. 
The central component is well described by a single Gaussian profile, 
while two spur-like features extending from the central source toward the position of source B are clearly visible in the residual map.

\begin{figure*}[ht!]
\centering
\includegraphics[width=0.8\textwidth]{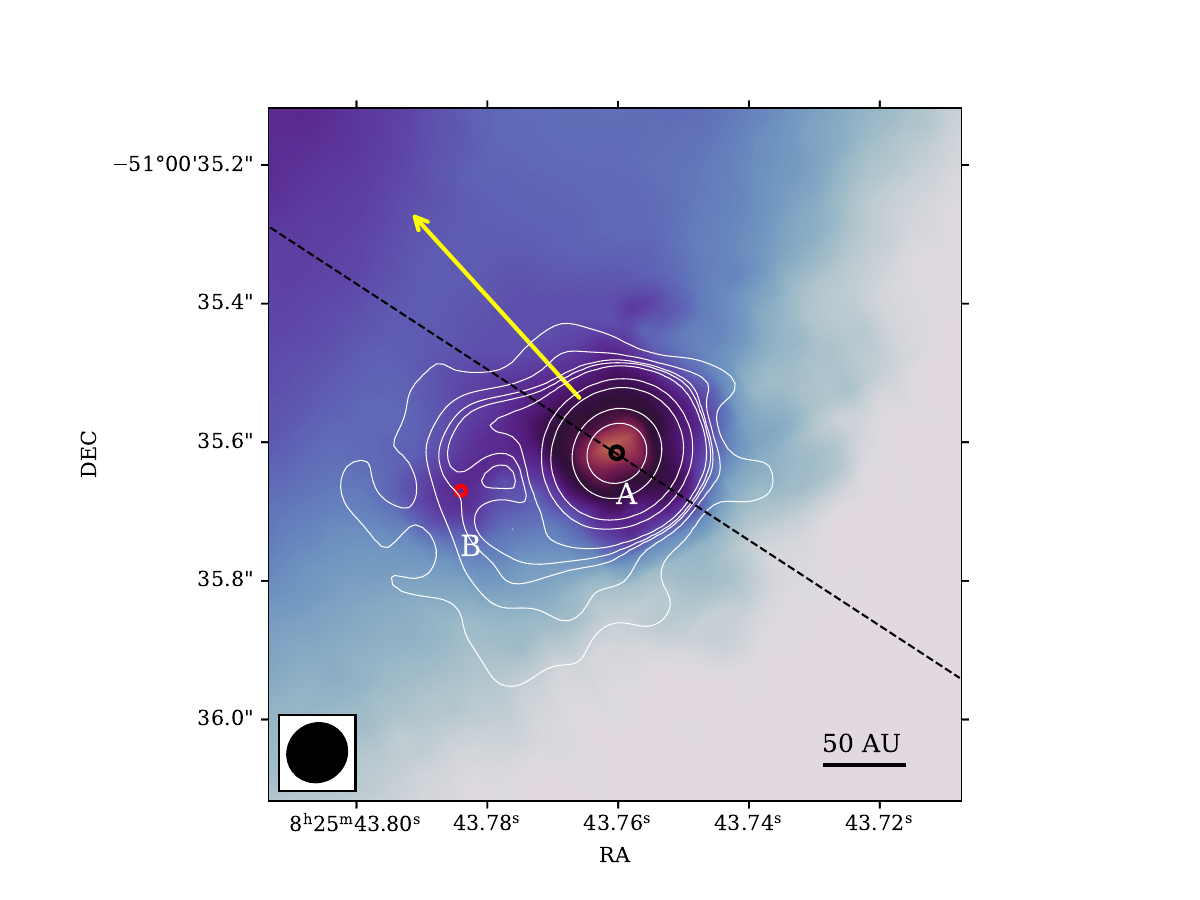}
\caption{JWST NIRCam F187N (1.87~$\mu$m) image of the binary system. The locations of sources A and B are marked with two circles. 
White contours show the 1.3 mm continuum emission (robust = $-0.5$). 
The yellow arrow indicates the direction of the
blueshifted [FeII] jet seen by JWST 
(\citealt[]{Nisini2024}; $\pa = 46^{\circ}$).
The black dashed line marks the outflow axis ($\pa = 57^{\circ}$), derived in \S\ref{sec:vfield}. 
Black ellipses denote the synthesized beam of the 1.3 mm continuum. 
}
\label{fig:JWST_F187N}
\end{figure*}

Figure~\ref{fig:JWST_F187N} overlays the continuum contours on the JWST NIRCam 1.87~$\mu$m narrow-band image. 
The position of source B coincides with a local minimum in the continuum emission 
and lies at the endpoints of the two spurs.
We interpret these spurs as substructures in the circumbinary disk, 
potentially induced by the gravitational influence of the companion.
Such perturbations inevitably affect the kinematic structure of the disk-envelope system
(e.g., producing a quadrupolar velocity pattern in the moment-1 map; \citealt{Jorgensen2022}).
However, the molecular-line kinematics discussed in \S\ref{sec:IRE} are highly symmetric about source A
(Figure~\ref{fig:Envelope_M0}),
suggesting that the gravitational perturbation induced by source B 
has only a minor influence on the circumstellar disk around source A.


An intriguing aspect of this system is that 
the companion is clearly detected in the infrared 
but not in the millimeter continuum. 
The projected separation of the two components, $\sim$100 au, 
is close to the peak of the separation distribution of low-mass binary systems \citep{Tobin16,Tobin22}, 
making a chance alignment due to projection effects unlikely.
\citet[]{Nisini2024} reported a position angle change of $13^\circ$ counterclockwise of source B with respect to source A between HST observation and JWST observation over 25 years.
This shift is also compatible with a binary orbital motion with a period of several hundred years.
Moreover, although source B is not detected as a distinct millimeter continuum peak, 
the continuum emission exhibits substructures near its infrared position 
that may be physically associated with the companion. 
Taken together, these observations strongly suggest that 
the two infrared sources constitute a genuine binary system.

One possible explanation for the absence of a millimeter counterpart is that 
source B remains deeply embedded within the circumbinary disk and 
therefore does not appear as a distinct continuum peak. 
However, this scenario would require the disk to be optically thick at 1.3 mm, 
which appears inconsistent with both the detection of continuum substructures 
and the local intensity dip observed at the position of source B. 
Furthermore, such an embedded configuration would likely produce stronger infrared extinction, 
contrary to the prominent appearance of source B in the near-infrared.

A more plausible explanation is that source B is relatively unobscured, 
either because it has partially cleared its surrounding material or 
because it is viewed at a more face-on inclination. 
The latter possibility is supported by the bubble-like H$_2$ feature 
identified in infrared observations (knot A1 in \citealt{Nisini2024,Navarro2025}), 
which closely resembles the bow-shock structures observed in other nearly face-on 
protostellar outflows \citep{Guillermo2026,Hodapp2014}. 
In such a geometry, the source could remain bright in the infrared while 
exhibiting only weak millimeter continuum emission.
Future observations with higher angular resolution and sensitivity will be essential for resolving the circumbinary and circumstellar structures and clarifying the nature and evolutionary state of this binary system.



\section{Conclusions}
\label{sec:conclusion}

We present ALMA $0.1\arcsec$ ($\sim 45~\au$) resolution observations at 1.3 mm of the low-mass protostellar outflow source HH 46/47. 
The observations include sensitive continuum data and spectral coverage of CO and its isotopologues for outflow studies, 
as well as multiple molecular lines (SO, CH$_3$OH, and H$_2$CO) tracing the disk–envelope system. Our main conclusions are as follows:

\begin{enumerate}

\item The 1.3 mm continuum emission consists of a bright central component associated with the primary source 
and extended emission with substructures. 
The companion source, identified in optical and infrared observations, 
is not detected in the 1.3 mm continuum despite sufficient angular resolution to resolve the binary system. 
Instead, it coincides with a local minimum in the continuum emission.

\item Two substructures (``spurs'') extending from the primary source toward the companion are revealed in the continuum image. 
These features likely trace perturbations in the circumbinary disk induced by the gravitational influence of the companion.
The absence of millimeter emission toward the companion, combined with its infrared brightness, 
suggests that it is in a relatively more evolved (exposed) 
stage with little dense material in its immediate vicinity,
or it has a more face-on inclination. 

\item Multiple molecular tracers -- $\chem{C^{18}O}$, SO, $\chem{H_2CO}$, and $\chem{CH_3OH}$ -- 
reveal a rotating and infalling envelope transitioning to an inner disk at a radius of $\sim 30$ au around a $0.3~M_\odot$ protostar. 
The different tracers probe distinct regions, with $\chem{C^{18}O}$ tracing the outer envelope, SO tracing the inner envelope, 
and $\chem{H_2CO}$ and $\chem{CH_3OH}$ tracing density enhancements near the centrifugal barrier.

\item The high-resolution $\chem{^{12}CO}$ emission confirms the presence of multiple shell structures in the outflow. 
Using a model-independent analysis of a well-defined shell in the redshifted lobe, 
we derive its three-dimensional morphology and velocity field. 
The shell material is found to move predominantly radially rather than along the shell surface, 
indicating an expanding structure.

\item A transverse velocity gradient is detected across the outflow axis. 
If interpreted as rotation within a magneto-centrifugal disk-wind framework, 
the inferred magnetic lever arm ($\lambda \sim 20-30$) is significantly larger than expected ($\lambda<10$). 
Together with the radial expansion, this strongly disfavors the scenario that the observed CO shell is directly launched as a disk wind.
Instead, the observed shell kinematics are consistent with an entrainment scenario.


\item Using CO and its isotopologues, we derive the outflow mass, momentum, and kinetic energy,
as well as their spatial distributions and radial profiles of the corresponding rates. 
The redshifted lobe exhibits systematically higher rates than the blueshifted lobe, likely reflecting environmental asymmetry.


\end{enumerate}

\begin{acknowledgments}
We thank the anonymous referee for their constructive suggestions. 
YZ acknowledges support from Yangyang Development Fund.
HGA acknowledges support from NSF grant AST-2407116.
This paper makes use of the following ALMA data: ADS/JAO.ALMA \#2018.1.01625.S, ADS/JAO.ALMA \#2015.1.01068.S. 
ALMA is a partnership of ESO (representing its member states), NSF (USA) and NINS (Japan), 
together with NRC (Canada), NSTC and ASIAA (Taiwan), and KASI (Republic of Korea), 
in cooperation with the Republic of Chile. 
The Joint ALMA Observatory is operated by ESO, AUI/NRAO and NAOJ. 
The National Radio Astronomy Observatory is a facility of the National Science Foundation 
operated under cooperative agreement by Associated Universities, Inc.
This work is based in part on observations made with the NASA/ESA/CSA James Webb Space Telescope. 
The data were obtained from the Mikulski Archive for Space Telescopes at the Space Telescope Science Institute, 
which is operated by the Association of Universities for Research in Astronomy, Inc., 
under NASA contract NAS 5-03127 for JWST. 
These observations are associated with program \#4441.
\end{acknowledgments}

\appendix

\section{Fitting the Outflow Cavity Geometry Using H$_2$CO Emission}
\label{app:eye_equation}

As discussed in \S\ref{sec:outflow_morphology}, molecular lines such as H$_2$CO reveal two arch-like structures 
that trace the outflow cavities and intersect near their base. 
We define an outflow-centered coordinate system $O-XYZ$, 
in which the $Z$-axis corresponds to the outflow axis, 
and an observer’s coordinate system $O-\delta x \delta y \delta z$, 
in which the line of sight is along the $\delta y$-axis, the projected outflow axis lies along the $\delta z$-axis,
and $\delta x =X$.
The angle $\theta$ between the $Z$- and $z$-axes is the inclination of the outflow axis relative to the plane of the sky. We let $\delta z$-axis has $\pa=57^{\circ}$, which is derived in \S\ref{sec:vfield}.

Assuming that the outflow cavity wall follows a parabolic shape (Figure~\ref{fig:eye}b),
\begin{equation}
Z = A(X^2 + Y^2),
\end{equation}
the equation of the cavity wall in the observer’s frame can be obtained via a coordinate rotation, yielding
\begin{equation}
A\cos^2\theta \delta y^2 - (2A\delta z\sin\theta\cos\theta + \sin\theta)\delta y + A \delta z^2 \sin^2\theta + A \delta x^2 - \delta z\cos\theta = 0,
\label{eq:cavity_fit}
\end{equation}
which is a quadratic equation in $\delta y$.

For an observer, the projected outflow cavity on the plane of the sky ($\delta x \delta z$-plane) consists of all $(\delta x, \delta z)$ positions 
for which Equation~\ref{eq:cavity_fit} has real solutions. This condition leads to
\begin{equation}
|\delta z| \geq A \delta x^2 \cos\theta - \frac{\sin^2\theta}{4A\cos\theta}.
\end{equation}
The boundary of the projected cavity is therefore given by
\begin{equation}
|\delta z| = A \delta x^2 \cos\theta - \frac{\sin^2\theta}{4A\cos\theta},
\label{eq:H2CO_fit}
\end{equation}
which describes a parabolic relation between the projected transverse distance $\delta x$ 
and the projected distance along the outflow axis $\delta z$. 
We use this relation to fit the H$_2$CO moment 0 map shown in Figure~\ref{fig:eye}.

\section{Comparison of SO Transverse PV with Keplerian Rotation}
\label{app:SO_PV}

\begin{figure*}[ht!]
    \centering
    \includegraphics[width=0.5\linewidth]{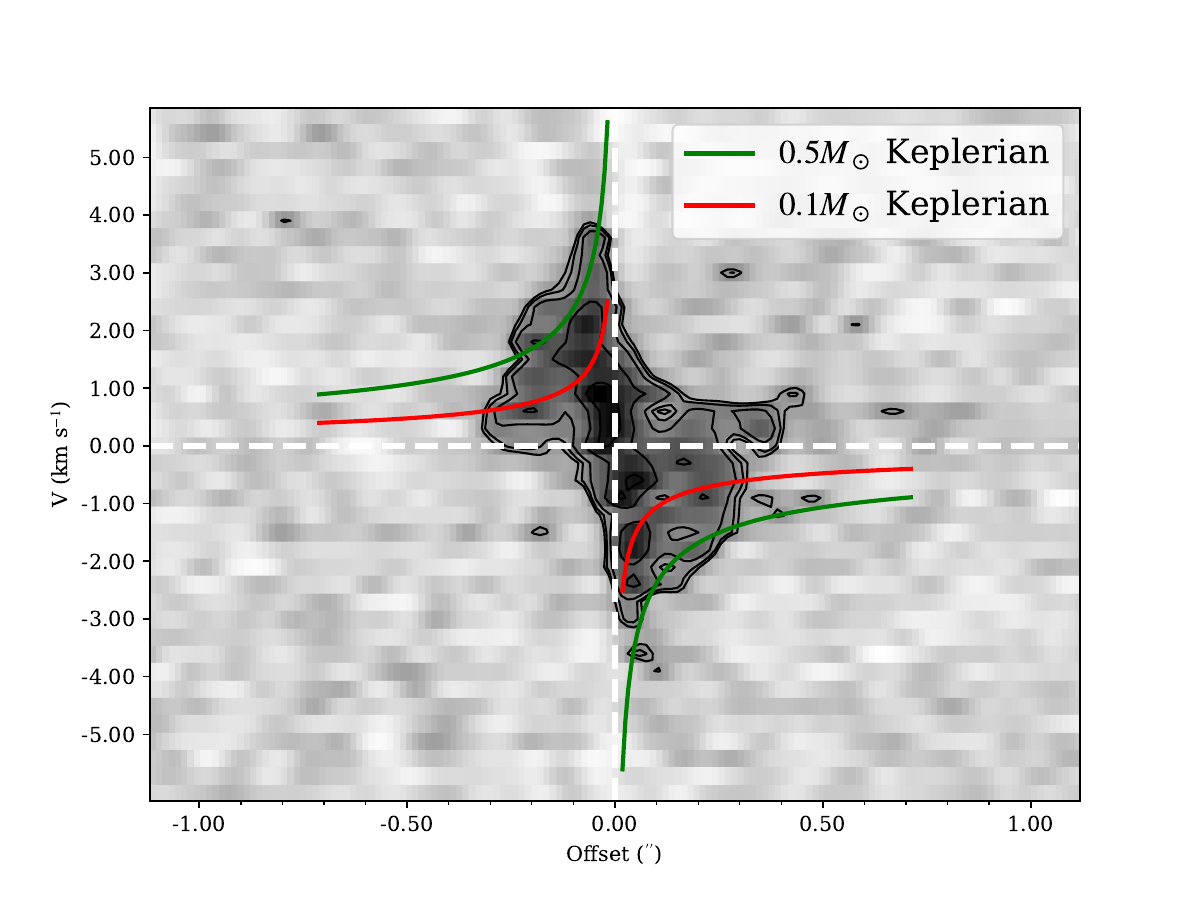}
    \caption{Transverse SO position–velocity (PV) diagram compared with Keplerian rotation curves. 
    The green and red curves represent Keplerian rotation for central masses of $0.5~M_\odot$ and $0.1~M_\odot$, respectively.}
    \label{fig:SO_N0_Keplerian}
\end{figure*}

Figure~\ref{fig:SO_N0_Keplerian} compares the transverse SO position–velocity (PV) diagram 
with Keplerian rotation curves. 
Two cases with central masses of $0.1$ and $0.5~M_\odot$ are shown. 
Compared with the infalling–rotating envelope model discussed in the main text, 
a pure Keplerian rotation model does not adequately reproduce the observed emission distribution.

\section{Deriving Column Densities of the CO Outflow}
\label{app:column densities}

\begin{figure*}[ht!]
\includegraphics[width=0.48\textwidth]{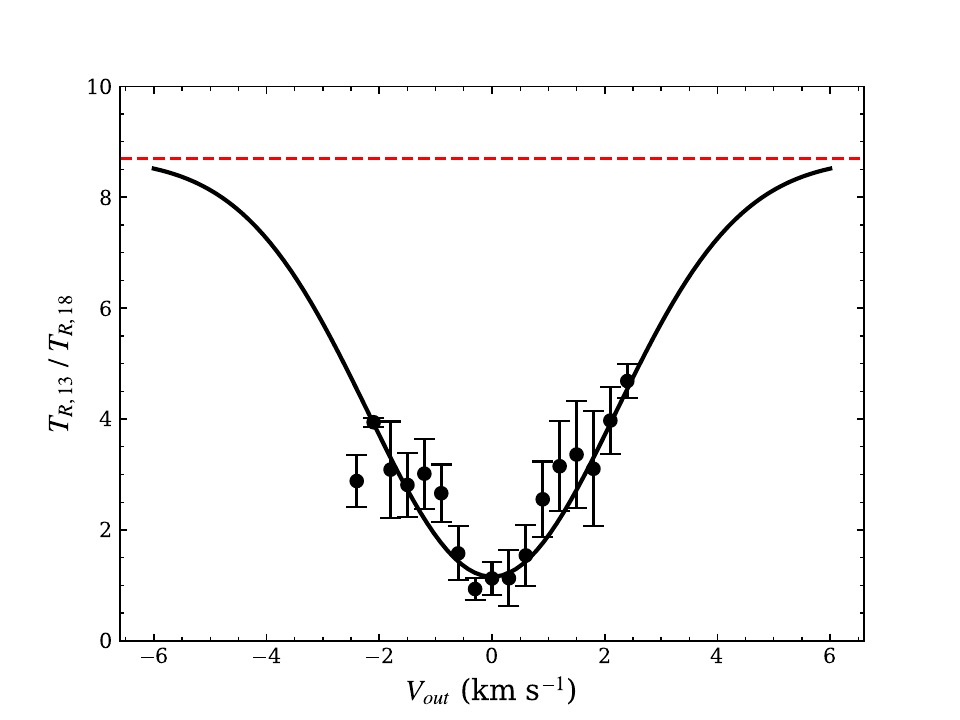}
\includegraphics[width=0.48\textwidth]{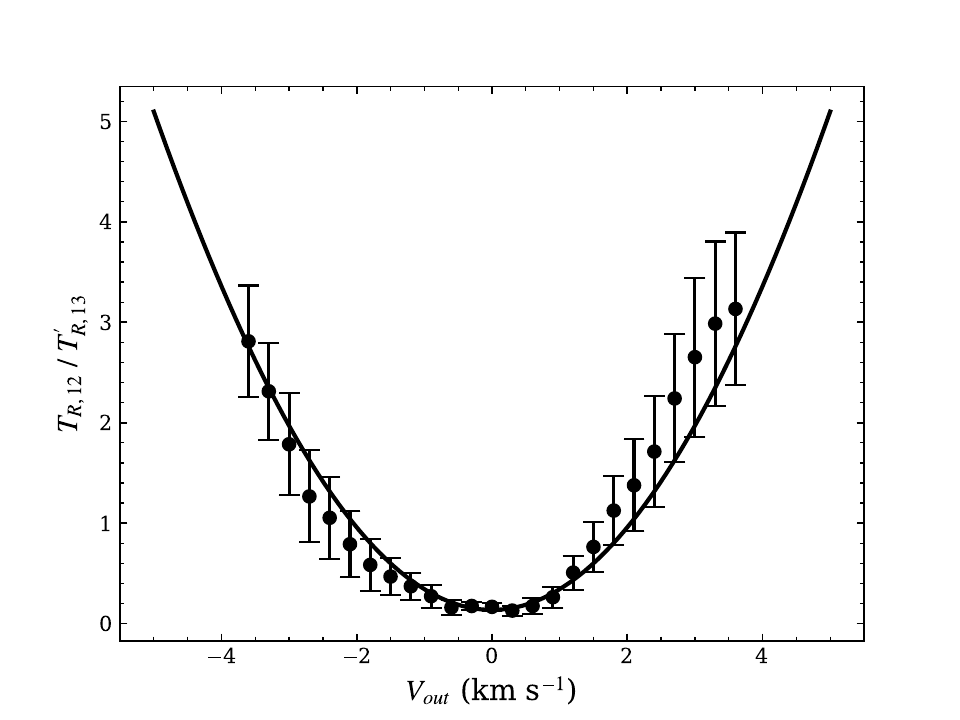}
\caption{Left: average radiation temperature ratio $T_{R,13}(v)/T_{R,18}(v)$ between $^{13}$CO and C$^{18}$O in each velocity channel. The solid curve is the Gaussian fit to the ratio profile. 
The horizontal line marks the abundance ratio between $^{13}$CO and C$^{18}$O of $R_{13,18}=8.7$.
Right: Similar to the left panel, but for the radiation temperature ratio between $^{12}$CO and $^{13}$CO
where $^{13}$CO emission has been optical depth corrected.}
\label{fig:factor}
\end{figure*}

We outline the formalism used to derive the outflow column densities from CO and its isotopologues, 
which are subsequently used to calculate the mass, momentum, and kinetic energy distributions. 
The method generally follows \citet{Dunham2014,Zhang2016,Hsieh2023}.

Under the assumption of local thermodynamic equilibrium (LTE), the CO column density per unit velocity interval is given by
\begin{equation}
\frac{dN}{dv} = \left(\frac{dN}{dv}\right)_{\rm thin} F_{\tau}(v),
\label{eq:dNdv}
\end{equation}
\begin{equation}
\left(\frac{dN}{dv}\right)_{\rm thin} = \left(\frac{8\pi k \nu_{ul}^2}{h c^3 A_{ul} g_u}\right) Q_{\rm rot}(T_{\rm ex})
\exp\left(\frac{E_u}{k T_{\rm ex}}\right) \frac{T_R(v)}{f},
\end{equation}
\begin{equation}
F_{\tau}(v) = \frac{\tau_v}{1 - \exp(-\tau_v)},
\end{equation}
where $F_{\tau}(v)$ is the optical-depth correction factor, 
$\nu_{ul}$ is the transition frequency, 
$A_{ul}$ is the Einstein coefficient (Table~\ref{tab:lines}), 
$E_u$ and $g_u$ are the upper-level energy and degeneracy, 
$T_{\rm ex}$ is the excitation temperature, 
$Q_{\rm rot}$ is the partition function, 
$f$ is the beam-filling factor, 
and $T_R(v)$ is the background-subtracted radiation temperature.

Following \citet{Zhang2016}, we first use C$^{18}$O emission to correct for the optical depth of $^{13}$CO, 
and then apply a further correction to $^{12}$CO. 
The brightness temperature ratio between $^{13}$CO and C$^{18}$O is given by
\begin{equation}
\frac{T_{R,13}(v)}{T_{R,18}(v)} = \frac{1 - \exp(-\tau_{v,13})}{1 - \exp(-\tau_{v,18})}.
\end{equation}
Since C$^{18}$O is generally optically thin (e.g., \citealt{Kong2018}), this simplifies to
\begin{equation}
\frac{T_{R,13}(v)}{T_{R,18}(v)} \approx \frac{1 - \exp(-\tau_{v,13})}{\tau_{v,13}} R_{13,18} = \frac{R_{13,18}}{F_{\tau,13}(v)},
\end{equation}
where $R_{13,18} \approx \tau_{v,13}/\tau_{v,18}$ is the abundance ratio.

The ratio $T_{R,13}/T_{R,18}$ is derived from the data (Figure~\ref{fig:factor}a). 
For each velocity channel, we compute pixel-by-pixel ratios where both lines are detected above $5\sigma$, 
and then calculate a weighted mean using $w = 1/\sigma^2_{\rm pixel}$. 
Since $^{13}$CO and C$^{18}$O are detected only at low velocities ($|\vout| \le 5~\kms$), 
we fit the ratio with a Gaussian function of $\vout$ to extrapolate to higher velocities. 
The fitted relation is
\begin{equation}
\frac{T_{R,13}(v)}{T_{R,18}(v)} = R_{13,18} - (7.55 \pm 0.14)\exp\left(-\frac{\vout^2}{2(2.20 \pm 0.06)^2}\right),
\end{equation}
where we adopt $R_{13,18} = 8.7$ \citep{Wilson1992}. 
This ratio reaches its minimum near the systemic velocity (optically thick) 
and approaches $R_{13,18}$ at higher velocities (optically thin).
The corresponding correction factor is
\begin{equation}
F_{\tau,13}(v) = R_{13,18}\frac{T_{R,18}(v)}{T_{R,13}(v)}.
\end{equation}

For $^{12}$CO, the optical-depth correction factor is
\begin{equation}
\begin{split}
F_{\tau,12}(v) &= \frac{\tau_{v,12}}{1 - \exp(-\tau_{v,12})} \\
&= R_{12,13} F_{\tau,13}(v) \frac{T_{R,13}(v)}{T_{R,12}(v)} \\
&= R_{12,13} \frac{T'_{R,13}(v)}{T_{R,12}(v)},
\end{split}
\end{equation}
where $T'_{R,13}(v) \equiv F_{\tau,13}(v) T_{R,13}(v)$ is the corrected $^{13}$CO radiation temperature.

We then fit the ratio $T_{R,12}(v)/T'_{R,13}(v)$ with a Gaussian profile to extrapolate to higher velocities (Figure \ref{fig:factor}b):
\begin{equation}
\frac{T_{R,12}(v)}{T'_{R,13}(v)} = R_{12,13} - (61.86 \pm 0.02)\exp\left(-\frac{\vout^2}{2(12.23 \pm 0.53)^2}\right),
\end{equation}
where $R_{12,13} = 62$ \citep{Langer1993}.

Finally, the H$_2$ column density is derived from the $^{12}$CO column density as
\begin{equation}
N_{\mathrm{H}_2} = X_{^{12}\mathrm{CO}} \sum \left(\frac{dN_{^{12}\mathrm{CO}}}{dv}\right) \Delta v,
\end{equation}
where $X_{^{12}\mathrm{CO}} = 10^4$ is the abundance ratio between H$_2$ and $^{12}$CO, and $\Delta v$ is the channel width.
The summation is performed over velocity channels with $|\vout| > 2~\kms$ to exclude ambient cloud emission. 
We adopt an excitation temperature of $T_{\rm ex} = 15~\K$ for the CO ($2-1$) transition.

\bibliography{reference}{}

@article{Zhang2016,
  title = {ALMA CYCLE 1 OBSERVATIONS OF THE HH46/47 MOLECULAR OUTFLOW: STRUCTURE,  ENTRAINMENT,  AND CORE IMPACT},
  volume = {832},
  ISSN = {1538-4357},
  url = {http://dx.doi.org/10.3847/0004-637X/832/2/158},
  DOI = {10.3847/0004-637x/832/2/158},
  number = {2},
  journal = {\apj},
  publisher = {American Astronomical Society},
  author = {Zhang,  Yichen and Arce,  Héctor G. and Mardones,  Diego and Cabrit,  Sylvie and Dunham,  Michael M. and Garay,  Guido and Noriega-Crespo,  Alberto and Offner,  Stella S. R. and Raga,  Alejandro C. and Corder,  Stuartt A.},
  year = {2016},
  month = nov,
  pages = {158}
}

@article{Zhang2019,
  title = {An Episodic Wide-angle Outflow in HH 46/47},
  volume = {883},
  ISSN = {1538-4357},
  url = {http://dx.doi.org/10.3847/1538-4357/ab3850},
  DOI = {10.3847/1538-4357/ab3850},
  number = {1},
  journal = {\apj},
  publisher = {American Astronomical Society},
  author = {Zhang,  Yichen and Arce,  Héctor G. and Mardones,  Diego and Cabrit,  Sylvie and Dunham,  Michael M. and Garay,  Guido and Noriega-Crespo,  Alberto and Offner,  Stella S. R. and Raga,  Alejandro C. and Corder,  Stuartt A.},
  year = {2019},
  month = sep,
  pages = {1}
}

@article{Dunham2014,
  title = {MOLECULAR OUTFLOWS DRIVEN BY LOW-MASS PROTOSTARS. I. CORRECTING FOR UNDERESTIMATES WHEN MEASURING OUTFLOW MASSES AND DYNAMICAL PROPERTIES},
  volume = {783},
  ISSN = {1538-4357},
  url = {http://dx.doi.org/10.1088/0004-637X/783/1/29},
  DOI = {10.1088/0004-637x/783/1/29},
  number = {1},
  journal = {\apj},
  publisher = {American Astronomical Society},
  author = {Dunham,  Michael M. and Arce,  Héctor G. and Mardones,  Diego and Lee,  Jeong-Eun and Matthews,  Brenda C. and Stutz,  Amelia M. and Williams,  Jonathan P.},
  year = {2014},
  month = feb,
  pages = {29}
}

@article{Kong2018,
  title = {The CARMA-NRO Orion Survey},
  volume = {236},
  ISSN = {1538-4365},
  url = {http://dx.doi.org/10.3847/1538-4365/aabafc},
  DOI = {10.3847/1538-4365/aabafc},
  number = {2},
  journal = {\apjs},
  publisher = {American Astronomical Society},
  author = {Kong,  Shuo and Arce,  Héctor G. and Feddersen,  Jesse R. and Carpenter,  John M. and Nakamura,  Fumitaka and Shimajiri,  Yoshito and Isella,  Andrea and Ossenkopf-Okada,  Volker and Sargent,  Anneila I. and Sánchez-Monge,  Álvaro and Suri,  S\"{u}meyye T. and Kauffmann,  Jens and Pillai,  Thushara and Pineda,  Jaime E. and Koda,  Jin and Bally,  John and Lis,  Dariusz C. and Padoan,  Paolo and Klessen,  Ralf and Mairs,  Steve and Goodman,  Alyssa and Goldsmith,  Paul and McGehee,  Peregrine and Schilke,  Peter and Teuben,  Peter J. and Maureira,  María José and Hara,  Chihomi and Ginsburg,  Adam and Burkhart,  Blakesley and Smith,  Rowan J. and Schmiedeke,  Anika and Pineda,  Jorge L. and Ishii,  Shun and Sasaki,  Kazushige and Kawabe,  Ryohei and Urasawa,  Yumiko and Oyamada,  Shuri and Tanabe,  Yoshihiro},
  year = {2018},
  month = may,
  pages = {25}
}

@article{Hsieh2023,
  title = {The Evolution of Protostellar Outflow Cavities,  Kinematics,  and Angular Distribution of Momentum and Energy in Orion A: Evidence for Dynamical Cores},
  volume = {947},
  ISSN = {1538-4357},
  url = {http://dx.doi.org/10.3847/1538-4357/acba13},
  DOI = {10.3847/1538-4357/acba13},
  number = {1},
  journal = {\apj},
  publisher = {American Astronomical Society},
  author = {Hsieh,  Cheng-Han  and Arce,  Héctor G. and Li,  Zhi-Yun and Dunham,  Michael and Offner,  Stella and Stephens,  Ian W. and Stutz,  Amelia and Megeath,  Tom and Kong,  Shuo and Plunkett,  Adele and Tobin,  John J. and Zhang,  Yichen and Mardones,  Diego and Pineda,  Jaime E. and Stanke,  Thomas and Carpenter,  John},
  year = {2023},
  month = apr,
  pages = {25}
}

@article{Wilson1992,
  title = {Abundances in the interstellar medium},
  volume = {4},
  ISSN = {1432-0754},
  url = {http://dx.doi.org/10.1007/BF00873568},
  DOI = {10.1007/bf00873568},
  number = {1},
  journal = {The Astronomy and Astrophysics Review},
  publisher = {Springer Science and Business Media LLC},
  author = {Wilson,  T. L. and Matteucci,  F.},
  year = {1992},
  pages = {1–33}
}

@article{Langer1993,
  title = {(C-12)/(C-13) isotope ratio in the local interstellar medium from observations of (C-13)(O-18) in molecular clouds},
  volume = {408},
  ISSN = {1538-4357},
  url = {http://dx.doi.org/10.1086/172611},
  DOI = {10.1086/172611},
  journal = {\apj},
  publisher = {American Astronomical Society},
  author = {Langer,  William D. and Penzias,  Arno A.},
  year = {1993},
  month = may,
  pages = {539}
}

@article{vanderMarel2013,
  title = {Outflow forces of low-mass embedded objects in Ophiuchus: a quantitative comparison of analysis methods},
  volume = {556},
  ISSN = {1432-0746},
  url = {http://dx.doi.org/10.1051/0004-6361/201220717},
  DOI = {10.1051/0004-6361/201220717},
  journal = {\aap},
  publisher = {EDP Sciences},
  author = {van der Marel,  N. and Kristensen,  L. E. and Visser,  R. and Mottram,  J. C. and Yıldız,  U. A. and van Dishoeck,  E. F.},
  year = {2013},
  month = jul,
  pages = {A76}
}

@article{Offner2017,
  title = {Impact of Protostellar Outflows on Turbulence and Star Formation Efficiency in Magnetized Dense Cores},
  volume = {847},
  ISSN = {1538-4357},
  url = {http://dx.doi.org/10.3847/1538-4357/aa8996},
  DOI = {10.3847/1538-4357/aa8996},
  number = {2},
  journal = {\apj},
  publisher = {American Astronomical Society},
  author = {Offner,  Stella S. R. and Chaban,  Jonah},
  year = {2017},
  month = sep,
  pages = {104}
}

@article{Ohashi1997,
  title = {Interferometric Imaging of IRAS 04368+2557 in the L1527 Molecular Cloud Core: A Dynamically Infalling Envelope with Rotation},
  volume = {475},
  ISSN = {1538-4357},
  url = {http://dx.doi.org/10.1086/303533},
  DOI = {10.1086/303533},
  number = {1},
  journal = {\apj},
  publisher = {American Astronomical Society},
  author = {Ohashi,  Nagayoshi and Hayashi,  Masahiko and Ho,  Paul T. P. and Momose,  Munetake},
  year = {1997},
  month = jan,
  pages = {211–223}
}

@article{Sakai2014,
  title = {Change in the chemical composition of infalling gas forming a disk around a protostar},
  volume = {507},
  ISSN = {1476-4687},
  url = {http://dx.doi.org/10.1038/nature13000},
  DOI = {10.1038/nature13000},
  number = {7490},
  journal = {Nature},
  publisher = {Springer Science and Business Media LLC},
  author = {Sakai,  Nami and Sakai,  Takeshi and Hirota,  Tomoya and Watanabe,  Yoshimasa and Ceccarelli,  Cecilia and Kahane,  Claudine and Bottinelli,  Sandrine and Caux,  Emmanuel and Demyk,  Karine and Vastel,  Charlotte and Coutens,  Audrey and Taquet,  Vianney and Ohashi,  Nagayoshi and Takakuwa,  Shigehisa and Yen,  Hsi-Wei and Aikawa,  Yuri and Yamamoto,  Satoshi},
  year = {2014},
  month = feb,
  pages = {78–80}
}

@article{Oya2022,
  title = {FERIA: Flat Envelope Model with Rotation and Infall under Angular Momentum Conservation},
  volume = {134},
  ISSN = {1538-3873},
  url = {http://dx.doi.org/10.1088/1538-3873/ac8839},
  DOI = {10.1088/1538-3873/ac8839},
  number = {1039},
  journal = {\pasp},
  publisher = {IOP Publishing},
  author = {Oya,  Yoko and Kibukawa,  Hirofumi and Miyake,  Shota and Yamamoto,  Satoshi},
  year = {2022},
  month = sep,
  pages = {094301}
}

@article{Nisini2024,
  title = {PROJECT-J: JWST Observations of HH46 IRS and Its Outflow. Overview and First Results},
  volume = {967},
  ISSN = {1538-4357},
  url = {http://dx.doi.org/10.3847/1538-4357/ad3d5a},
  DOI = {10.3847/1538-4357/ad3d5a},
  number = {2},
  journal = {\apj},
  publisher = {American Astronomical Society},
  author = {Nisini,  Brunella and Navarro,  Maria Gabriela and Giannini,  Teresa and Antoniucci,  Simone and Kavanagh,  Patrick,  J. and Hartigan,  Patrick and Bacciotti,  Francesca and Caratti o Garatti,  Alessio and Noriega-Crespo,  Alberto and van Dishoeck,  Ewine F. and Whelan,  Emma T. and Arce,  Hector G. and Cabrit,  Sylvie and Coffey,  Deirdre and Fedele,  Davide and Eisl\"{o}ffel,  Jochen and Palumbo,  Maria Elisabetta and Podio,  Linda and Ray,  Tom P. and Schultze,  Megan and Urso,  Riccardo G. and Alcalá,  Juan M. and Bautista,  Manuel A. and Codella,  Claudio and Greene,  Thomas P. and Manara,  Carlo F.},
  year = {2024},
  month = may,
  pages = {168}
}

@article{Su2004,
  title = {Bipolar Molecular Outflows from High‐Mass Protostars},
  volume = {604},
  ISSN = {1538-4357},
  url = {http://dx.doi.org/10.1086/381880},
  DOI = {10.1086/381880},
  number = {1},
  journal = {\apj},
  publisher = {American Astronomical Society},
  author = {Su,  Yu‐Nung and Zhang,  Qizhou and Lim,  Jeremy},
  year = {2004},
  month = mar,
  pages = {258–271}
}

@INPROCEEDINGS{Arce2007,
       author = {{Arce}, H.~G. and {Shepherd}, D. and {Gueth}, F. and {Lee}, C.-F. and {Bachiller}, R. and {Rosen}, A. and {Beuther}, H.},
        title = "{Molecular Outflows in Low- and High-Mass Star-forming Regions}",
    booktitle = {Protostars and Planets V},
         year = 2007,
       editor = {{Reipurth}, Bo and {Jewitt}, David and {Keil}, Klaus},
        month = jan,
        pages = {245},
          doi = {10.48550/arXiv.astro-ph/0603071},
archivePrefix = {arXiv},
       eprint = {astro-ph/0603071},
 primaryClass = {astro-ph},
       adsurl = {https://ui.adsabs.harvard.edu/abs/2007prpl.conf..245A}
}

@article{Yang2024,
  title = {Surveys of clumps,  cores,  and condensations in Cygnus-X: SMA observations of SiO (5−4)},
  volume = {684},
  ISSN = {1432-0746},
  url = {http://dx.doi.org/10.1051/0004-6361/202346873},
  DOI = {10.1051/0004-6361/202346873},
  journal = {\aap},
  publisher = {EDP Sciences},
  author = {Yang,  Kai and Qiu,  Keping and Pan,  Xing},
  year = {2024},
  month = apr,
  pages = {A140}
}

@article{Rabenanahary2022,
  title = {Wide-angle protostellar outflows driven by narrow jets in stratified cores},
  volume = {664},
  ISSN = {1432-0746},
  url = {http://dx.doi.org/10.1051/0004-6361/202243139},
  DOI = {10.1051/0004-6361/202243139},
  journal = {\aap},
  publisher = {EDP Sciences},
  author = {Rabenanahary,  M. and Cabrit,  S. and Meliani,  Z. and Pineau des For\^ets,  G.},
  year = {2022},
  month = aug,
  pages = {A118}
}

@article{Li2017,
  title = {Formation of stellar clusters in magnetized,  filamentary infrared dark clouds},
  volume = {473},
  ISSN = {1365-2966},
  url = {http://dx.doi.org/10.1093/mnras/stx2611},
  DOI = {10.1093/mnras/stx2611},
  number = {3},
  journal = {\mnras},
  publisher = {Oxford University Press (OUP)},
  author = {Li,  Pak Shing and Klein,  Richard I. and McKee,  Christopher F.},
  year = {2017},
  month = nov,
  pages = {4220–4241}
}

@article{Hartigan2005,
  title = {Proper Motions of the HH 47 Jet Observed with theHubble Space Telescope},
  volume = {130},
  ISSN = {1538-3881},
  url = {http://dx.doi.org/10.1086/491673},
  DOI = {10.1086/491673},
  number = {5},
  journal = {\apj},
  publisher = {American Astronomical Society},
  author = {Hartigan,  Patrick and Heathcote,  Steve and Morse,  Jon A. and Reipurth,  Bo and Bally,  John},
  year = {2005},
  month = nov,
  pages = {2197–2205}
}

@article{deValon2022,
  title = {Modeling the CO outflow in DG Tauri B: Swept-up shells versus perturbed MHD disk wind},
  volume = {668},
  ISSN = {1432-0746},
  url = {http://dx.doi.org/10.1051/0004-6361/202141316},
  DOI = {10.1051/0004-6361/202141316},
  journal = {\aap},
  publisher = {EDP Sciences},
  author = {de Valon,  A. and Dougados,  C. and Cabrit,  S. and Louvet,  F. and Zapata,  L. A. and Mardones,  D.},
  year = {2022},
  month = dec,
  pages = {A78}
}

@article{Louvet2018,
  title = {The HH30 edge-on T Tauri star: A rotating and precessing monopolar outflow scrutinized by ALMA},
  volume = {618},
  ISSN = {1432-0746},
  url = {http://dx.doi.org/10.1051/0004-6361/201731733},
  DOI = {10.1051/0004-6361/201731733},
  journal = {\aap},
  publisher = {EDP Sciences},
  author = {Louvet,  F. and Dougados,  C. and Cabrit,  S. and Mardones,  D. and Ménard,  F. and Tabone,  B. and Pinte,  C. and Dent,  W. R. F.},
  year = {2018},
  month = oct,
  pages = {A120}
}

@ARTICLE{Bacciotti2025,
       author = {{Bacciotti}, F. and {Nony}, T. and {Podio}, L. and {Dougados}, C. and {Garufi}, A. and {Cabrit}, S. and {Codella}, C. and {Zimniak}, N. and {Ferreira}, J.},
        title = "{ALMA chemical survey of disk-outflow sources in Taurus (ALMA-DOT): VII. The layered molecular outflow from HL Tau and its relationship with the ringed disk}",
      journal = {\aap},
         year = 2025,
        month = dec,
       volume = {704},
          eid = {A157},
        pages = {A157},
          doi = {10.1051/0004-6361/202453649},
archivePrefix = {arXiv},
       eprint = {2501.03920},
 primaryClass = {astro-ph.GA},
       adsurl = {https://ui.adsabs.harvard.edu/abs/2025A&A...704A.157B}
}

@ARTICLE{Ossenkopf1994,
       author = {{Ossenkopf}, V. and {Henning}, Th.},
        title = "{Dust opacities for protostellar cores.}",
      journal = {\aap},
         year = 1994,
        month = nov,
       volume = {291},
        pages = {943-959},
       adsurl = {https://ui.adsabs.harvard.edu/abs/1994A&A...291..943O}
}

@ARTICLE{Ohashi2014,
       author = {{Ohashi}, Nagayoshi and {Saigo}, Kazuya and {Aso}, Yusuke and {Aikawa}, Yuri and {Koyamatsu}, Shin and {Machida}, Masahiro N. and {Saito}, Masao and {Takahashi}, Sanemichi Z. and {Takakuwa}, Shigehisa and {Tomida}, Kengo and {Tomisaka}, Kohji and {Yen}, Hsi-Wei},
        title = "{Formation of a Keplerian Disk in the Infalling Envelope around L1527 IRS: Transformation from Infalling Motions to Kepler Motions}",
      journal = {\apj},
         year = 2014,
        month = dec,
       volume = {796},
       number = {2},
          eid = {131},
        pages = {131},
          doi = {10.1088/0004-637X/796/2/131},
archivePrefix = {arXiv},
       eprint = {1410.0172},
 primaryClass = {astro-ph.GA},
       adsurl = {https://ui.adsabs.harvard.edu/abs/2014ApJ...796..131O}
}

@ARTICLE{Hsu2025,
       author = {{Hsu}, Shih-Ying and {Lee}, Chin-Fei and {Johnstone}, Doug and {Liu}, Sheng-Yuan and {Liu}, Tie and {Bronfman}, Leonardo and {Chen}, Huei-Ru Vivien and {Dutta}, Somnath and {Eden}, David J. and {Hirano}, Naomi and {Juvela}, Mika and {Kim}, Kee-Tae and {Kuan}, Yi-Jehng and {Kwon}, Woojin and {Lee}, Chang Won and {Lee}, Jeong-Eun and {Li}, Shanghuo and {Lin}, Sheng-Jun and {Liu}, Chun-Fan and {Liu}, Xunchuan and {L{\'o}pez-V{\'a}zquez}, J.~A. and {Luo}, Qiuyi and {Rawlings}, Mark G. and {Sahu}, Dipen and {Sanhueza}, Patricio and {Shang}, Hsien and {Tatematsu}, Ken'ichi and {Yang}, Yao-Lun},
        title = "{ALMASOP. A Rotating Feature Rich in Complex Organic Molecules in a Protostellar Core}",
      journal = {\apj},
         year = 2025,
        month = aug,
       volume = {989},
       number = {1},
          eid = {56},
        pages = {56},
          doi = {10.3847/1538-4357/ade7fc},
archivePrefix = {arXiv},
       eprint = {2506.15140},
 primaryClass = {astro-ph.GA},
       adsurl = {https://ui.adsabs.harvard.edu/abs/2025ApJ...989...56H}
}

@ARTICLE{Oya2016,
       author = {{Oya}, Yoko and {Sakai}, Nami and {L{\'o}pez-Sepulcre}, Ana and {Watanabe}, Yoshimasa and {Ceccarelli}, Cecilia and {Lefloch}, Bertrand and {Favre}, C{\'e}cile and {Yamamoto}, Satoshi},
        title = "{Infalling-Rotating Motion and Associated Chemical Change in the Envelope of IRAS 16293-2422 Source A Studied with ALMA}",
      journal = {\apj},
         year = 2016,
        month = jun,
       volume = {824},
       number = {2},
          eid = {88},
        pages = {88},
          doi = {10.3847/0004-637X/824/2/88},
archivePrefix = {arXiv},
       eprint = {1605.00340},
 primaryClass = {astro-ph.SR},
       adsurl = {https://ui.adsabs.harvard.edu/abs/2016ApJ...824...88O}
}

@ARTICLE{Vazquez2024,
       author = {{L{\'o}pez-V{\'a}zquez}, J.~A. and {Lee}, Chin-Fei and {Fern{\'a}ndez-L{\'o}pez}, M. and {Louvet}, Fabien and {Guerra-Alvarado}, O. and {Zapata}, Luis A.},
        title = "{Multiple Shells Driven by Disk Winds: ALMA Observations in the HH 30 Outflow}",
      journal = {\apj},
         year = 2024,
        month = feb,
       volume = {962},
       number = {1},
          eid = {28},
        pages = {28},
          doi = {10.3847/1538-4357/ad132a},
archivePrefix = {arXiv},
       eprint = {2312.03272},
 primaryClass = {astro-ph.SR},
       adsurl = {https://ui.adsabs.harvard.edu/abs/2024ApJ...962...28L}
}

@ARTICLE{Muller05,
       author = {{M{\"u}ller}, Holger S.~P. and {Schl{\"o}der}, Frank and {Stutzki}, J{\"u}rgen and {Winnewisser}, Gisbert},
        title = "{The Cologne Database for Molecular Spectroscopy, CDMS: a useful tool for astronomers and spectroscopists}",
      journal = {Journal of Molecular Structure},
         year = 2005,
        month = may,
       volume = {742},
       number = {1-3},
        pages = {215-227},
          doi = {10.1016/j.molstruc.2005.01.027},
       adsurl = {https://ui.adsabs.harvard.edu/abs/2005JMoSt.742..215M}
}

@ARTICLE{vanKempen09,
       author = {{van Kempen}, T.~A. and {van Dishoeck}, E.~F. and {G{\"u}sten}, R. and {Kristensen}, L.~E. and {Schilke}, P. and {Hogerheijde}, M.~R. and {Boland}, W. and {Nefs}, B. and {Menten}, K.~M. and {Baryshev}, A. and {Wyrowski}, F.},
        title = "{APEX-CHAMP$^{+}$ high-J CO observations of low-mass young stellar objects. I. The HH 46 envelope and outflow}",
      journal = {\aap},
         year = 2009,
        month = jul,
       volume = {501},
       number = {2},
        pages = {633-646},
          doi = {10.1051/0004-6361/200912013},
archivePrefix = {arXiv},
       eprint = {0905.2878},
 primaryClass = {astro-ph.SR},
       adsurl = {https://ui.adsabs.harvard.edu/abs/2009A&A...501..633V}
}

@ARTICLE{Zhang2019massive,
       author = {{Zhang}, Yichen and {Tan}, Jonathan C. and {Sakai}, Nami and {Tanaka}, Kei E.~I. and {De Buizer}, James M. and {Liu}, Mengyao and {Beltr{\'a}n}, Maria T. and {Kratter}, Kaitlin and {Mardones}, Diego and {Garay}, Guido},
        title = "{An Ordered Envelope-Disk Transition in the Massive Protostellar Source G339.88-1.26}",
      journal = {\apj},
         year = 2019,
        month = mar,
       volume = {873},
       number = {1},
          eid = {73},
        pages = {73},
          doi = {10.3847/1538-4357/ab0553},
archivePrefix = {arXiv},
       eprint = {1811.04381},
 primaryClass = {astro-ph.GA},
       adsurl = {https://ui.adsabs.harvard.edu/abs/2019ApJ...873...73Z}
}

@ARTICLE{Zhang2022,
       author = {{Zhang}, Yichen and {Tanaka}, Kei E.~I. and {Tan}, Jonathan C. and {Yang}, Yao-Lun and {Greco}, Eva and {Beltran}, Maria T. and {Sakai}, Nami and {De Buizer}, James M. and {Rosero}, Viviana and {Fedriani}, Rub{\'e}n and {Garay}, Guido},
        title = "{Massive Protostars in a Protocluster - A Multi-scale ALMA View of G35.20-0.74N.}",
      journal = {\apj},
         year = 2022,
        month = sep,
       volume = {936},
       number = {1},
          eid = {68},
        pages = {68},
          doi = {10.3847/1538-4357/ac847f},
archivePrefix = {arXiv},
       eprint = {2207.11320},
 primaryClass = {astro-ph.GA},
       adsurl = {https://ui.adsabs.harvard.edu/abs/2022ApJ...936...68Z}
}

@article{Anderson2003,
  title = {Locating the Launching Region of T Tauri Winds: The Case of DG Tauri},
  volume = {590},
  ISSN = {1538-4357},
  url = {http://dx.doi.org/10.1086/376824},
  DOI = {10.1086/376824},
  number = {2},
  journal = {\apj},
  publisher = {American Astronomical Society},
  author = {Anderson,  Jeffrey M. and Li,  Zhi-Yun and Krasnopolsky,  Ruben and Blandford,  Roger D.},
  year = {2003},
  month = may,
  pages = {L107–L110}
}

@ARTICLE{Blandford1982,
       author = {{Blandford}, R.~D. and {Payne}, D.~G.},
        title = "{Hydromagnetic flows from accretion disks and the production of radio jets.}",
      journal = {\mnras},
         year = 1982,
        month = jun,
       volume = {199},
        pages = {883-903},
          doi = {10.1093/mnras/199.4.883},
       adsurl = {https://ui.adsabs.harvard.edu/abs/1982MNRAS.199..883B}
}

@article{Zhang2018,
  title = {Rotation in the NGC 1333 IRAS 4C Outflow},
  volume = {864},
  ISSN = {1538-4357},
  url = {http://dx.doi.org/10.3847/1538-4357/aad7ba},
  DOI = {10.3847/1538-4357/aad7ba},
  number = {1},
  journal = {\apj},
  publisher = {American Astronomical Society},
  author = {Zhang,  Yichen and Higuchi,  Aya E. and Sakai,  Nami and Oya,  Yoko and López-Sepulcre,  Ana and Imai,  Muneaki and Sakai,  Takeshi and Watanabe,  Yoshimasa and Ceccarelli,  Cecilia and Lefloch,  Bertrand and Yamamoto,  Satoshi},
  year = {2018},
  month = aug,
  pages = {76}
}

@ARTICLE{Lee2001,
       author = {{Lee}, Chin-Fei and {Stone}, James M. and {Ostriker}, Eve C. and {Mundy}, Lee G.},
        title = "{Hydrodynamic Simulations of Jet- and Wind-driven Protostellar Outflows}",
      journal = {\apj},
         year = 2001,
        month = aug,
       volume = {557},
       number = {1},
        pages = {429-442},
          doi = {10.1086/321648},
archivePrefix = {arXiv},
       eprint = {astro-ph/0104373},
 primaryClass = {astro-ph},
       adsurl = {https://ui.adsabs.harvard.edu/abs/2001ApJ...557..429L}
}

@ARTICLE{Arce2013,
       author = {{Arce}, H{\'e}ctor G. and {Mardones}, Diego and {Corder}, Stuartt A. and {Garay}, Guido and {Noriega-Crespo}, Alberto and {Raga}, Alejandro C.},
        title = "{ALMA Observations of the HH 46/47 Molecular Outflow}",
      journal = {\apj},
         year = 2013,
        month = sep,
       volume = {774},
       number = {1},
          eid = {39},
        pages = {39},
          doi = {10.1088/0004-637X/774/1/39},
archivePrefix = {arXiv},
       eprint = {1304.0674},
 primaryClass = {astro-ph.SR},
       adsurl = {https://ui.adsabs.harvard.edu/abs/2013ApJ...774...39A}
}

@ARTICLE{Shu1991,
       author = {{Shu}, Frank H. and {Ruden}, Steven P. and {Lada}, Charles J. and {Lizano}, Susana},
        title = "{Star Formation and the Nature of Bipolar Outflows}",
      journal = {\apjl},
         year = 1991,
        month = mar,
       volume = {370},
        pages = {L31},
          doi = {10.1086/185970},
       adsurl = {https://ui.adsabs.harvard.edu/abs/1991ApJ...370L..31S}
}

@article{Shang2023,
  title = {A Unified Model for Bipolar Outflows from Young Stars: Kinematic Signatures of Jets,  Winds,  and Their Magnetic Interplay with the Ambient Toroids},
  volume = {944},
  ISSN = {1538-4357},
  url = {http://dx.doi.org/10.3847/1538-4357/aca763},
  DOI = {10.3847/1538-4357/aca763},
  number = {2},
  journal = {\apj},
  publisher = {American Astronomical Society},
  author = {Shang ,  Hsien  and Liu ,  Chun-Fan  and Krasnopolsky,  Ruben and Wang ,  Liang-Yao},
  year = {2023},
  month = feb,
  pages = {230}
}

@ARTICLE{Raga1993,
       author = {{Raga}, A. and {Cabrit}, S.},
        title = "{Molecular outflows entrained by jet bowshocks.}",
      journal = {\aap},
         year = 1993,
        month = oct,
       volume = {278},
        pages = {267-278},
       adsurl = {https://ui.adsabs.harvard.edu/abs/1993A&A...278..267R}
}

@article{Zhang2023,
  title = {The Perseus ALMA Chemistry Survey (PEACHES). II. Sulfur-bearing Species and Dust Polarization Revealing Shocked Regions in Protostars in the Perseus Molecular Cloud},
  volume = {946},
  ISSN = {1538-4357},
  url = {http://dx.doi.org/10.3847/1538-4357/acbdf7},
  DOI = {10.3847/1538-4357/acbdf7},
  number = {2},
  journal = {\apj},
  publisher = {American Astronomical Society},
  author = {Zhang,  Ziwei E. and Yang,  Yao-lun and Zhang,  Yichen and Cox,  Erin G. and Zeng,  Shaoshan and Murillo,  Nadia M. and Ohashi,  Satoshi and Sakai,  Nami},
  year = {2023},
  month = apr,
  pages = {113}
}

@ARTICLE{Reipurth2000,
       author = {{Reipurth}, Bo and {Yu}, Ka Chun and {Heathcote}, Steve and {Bally}, John and {Rodr{\'\i}guez}, Luis F.},
        title = "{Hubble Space Telescope NICMOS Images of Herbig-Haro Energy Sources: [Fe II] Jets, Binarity, and Envelope Cavities}",
      journal = {\aj},
         year = 2000,
        month = sep,
       volume = {120},
       number = {3},
        pages = {1449-1466},
          doi = {10.1086/301510},
       adsurl = {https://ui.adsabs.harvard.edu/abs/2000AJ....120.1449R}
}

@ARTICLE{Schwartz1977,
       author = {{Schwartz}, R.~D.},
        title = "{Evidence of star formation triggered by expansion of the Gum Nebula.}",
      journal = {\apjl},
         year = 1977,
        month = feb,
       volume = {212},
        pages = {L25-L26},
          doi = {10.1086/182367},
       adsurl = {https://ui.adsabs.harvard.edu/abs/1977ApJ...212L..25S}
}

@ARTICLE{Noriega-Crespo2004,
       author = {{Noriega-Crespo}, Alberto and {Moro-Martin}, Amaya and {Carey}, Sean and {Morris}, Patrick W. and {Padgett}, Deborah L. and {Latter}, William B. and {Muzerolle}, James},
        title = "{Is the Cepheus E Outflow Driven by a Class 0 Protostar?}",
      journal = {\apjs},
         year = 2004,
        month = sep,
       volume = {154},
       number = {1},
        pages = {402-407},
          doi = {10.1086/423136},
archivePrefix = {arXiv},
       eprint = {astro-ph/0506710},
 primaryClass = {astro-ph},
       adsurl = {https://ui.adsabs.harvard.edu/abs/2004ApJS..154..402N}
}

@ARTICLE{Birney2024,
       author = {{Birney}, M. and {Dougados}, C. and {Whelan}, E.~T. and {Nisini}, B. and {Cabrit}, S. and {Zhang}, Y.},
        title = "{A kinematical study of the launching region of the blueshifted HH 46/47 outflow with SINFONI K-band observations}",
      journal = {\aap},
         year = 2024,
        month = dec,
       volume = {692},
          eid = {A143},
        pages = {A143},
          doi = {10.1051/0004-6361/202451433},
archivePrefix = {arXiv},
       eprint = {2407.07233},
 primaryClass = {astro-ph.SR},
       adsurl = {https://ui.adsabs.harvard.edu/abs/2024A&A...692A.143B}
}

@article{Navarro2025,
  title = {PROJECT-J: The Shocking H
                    2
                    Outflow from HH 46},
  volume = {995},
  ISSN = {1538-4357},
  url = {http://dx.doi.org/10.3847/1538-4357/ae1f8f},
  DOI = {10.3847/1538-4357/ae1f8f},
  number = {2},
  journal = {\apj},
  publisher = {American Astronomical Society},
  author = {Navarro,  Maria Gabriela and Nisini,  Brunella and Giannini,  Teresa and Kavanagh,  Patrick J. and Caratti o Garatti,  Alessio and Antoniucci,  Simone and Arce,  Hector G. and Bacciotti,  Francesca and Cabrit,  Sylvie and Coffey,  Deirdre and Dougados,  Catherine and Eisl\"{o}ffel,  Jochen and Hartigan,  Patrick and Crespo,  Alberto Noriega- and Podio,  Linda and van Dishoeck,  Ewine F. and Whelan,  Emma T.},
  year = {2025},
  month = dec,
  pages = {199}
}

@INPROCEEDINGS{Frank2014,
       author = {{Frank}, A. and {Ray}, T.~P. and {Cabrit}, S. and {Hartigan}, P. and {Arce}, H.~G. and {Bacciotti}, F. and {Bally}, J. and {Benisty}, M. and {Eisl{\"o}ffel}, J. and {G{\"u}del}, M. and {Lebedev}, S. and {Nisini}, B. and {Raga}, A.},
        title = "{Jets and Outflows from Star to Cloud: Observations Confront Theory}",
    booktitle = {Protostars and Planets VI},
         year = 2014,
       editor = {{Beuther}, Henrik and {Klessen}, Ralf S. and {Dullemond}, Cornelis P. and {Henning}, Thomas},
        month = jan,
        pages = {451-474},
          doi = {10.2458/azu_uapress_9780816531240-ch020},
archivePrefix = {arXiv},
       eprint = {1402.3553},
 primaryClass = {astro-ph.SR},
       adsurl = {https://ui.adsabs.harvard.edu/abs/2014prpl.conf..451F}
}

@ARTICLE{Hirota2017,
       author = {{Hirota}, Tomoya and {Machida}, Masahiro N. and {Matsushita}, Yuko and {Motogi}, Kazuhito and {Matsumoto}, Naoko and {Kim}, Mi Kyoung and {Burns}, Ross A. and {Honma}, Mareki},
        title = "{Disk-driven rotating bipolar outflow in Orion Source I}",
      journal = {Nature Astronomy},
         year = 2017,
        month = jul,
       volume = {1},
          eid = {0146},
        pages = {0146},
          doi = {10.1038/s41550-017-0146},
archivePrefix = {arXiv},
       eprint = {1712.04606},
 primaryClass = {astro-ph.SR},
       adsurl = {https://ui.adsabs.harvard.edu/abs/2017NatAs...1E.146H}
}

@article{Matzner1999,
     author = {{Matzner}, Christopher D. and {McKee}, Christopher F.},
        title = "{Bipolar Molecular Outflows Driven by Hydromagnetic Protostellar Winds}",
      journal = {\apjl},
         year = 1999,
        month = dec,
       volume = {526},
       number = {2},
        pages = {L109-L112},
          doi = {10.1086/312376},
archivePrefix = {arXiv},
       eprint = {astro-ph/9909479},
 primaryClass = {astro-ph},
       adsurl = {https://ui.adsabs.harvard.edu/abs/1999ApJ...526L.109M}
}

@ARTICLE{Ulrich1976,
       author = {{Ulrich}, R.~K.},
        title = "{An infall model for the T Tauri phenomenon.}",
      journal = {\apj},
         year = 1976,
        month = dec,
       volume = {210},
        pages = {377-391},
          doi = {10.1086/154840},
       adsurl = {https://ui.adsabs.harvard.edu/abs/1976ApJ...210..377U}
}

@ARTICLE{Liu2025,
       author = {{Liu}, X.-C. and {van Dishoeck}, E.~F. and {Hogerheijde}, M.~R. and {van Gelder}, M.~L. and {Chen}, Y. and {Liu}, T. and {van't Hoff}, M. and {Drozdovskaya}, M.~N. and {Artur de la Villarmois}, E. and {Mai}, X.-F. and {Tychoniec}, {\L}.},
        title = "{Sulfur oxides tracing streamers and shocks at low-mass protostellar disk─envelope interfaces}",
      journal = {\aap},
         year = 2025,
        month = sep,
       volume = {701},
          eid = {A141},
        pages = {A141},
          doi = {10.1051/0004-6361/202554186},
archivePrefix = {arXiv},
       eprint = {2507.22870},
 primaryClass = {astro-ph.GA},
       adsurl = {https://ui.adsabs.harvard.edu/abs/2025A&A...701A.141L}
}

@ARTICLE{Lee2000,
       author = {{Lee}, Chin-Fei and {Mundy}, Lee G. and {Reipurth}, Bo and {Ostriker}, Eve C. and {Stone}, James M.},
        title = "{CO Outflows from Young Stars: Confronting the Jet and Wind Models}",
      journal = {\apj},
         year = 2000,
        month = oct,
       volume = {542},
       number = {2},
        pages = {925-945},
          doi = {10.1086/317056},
       adsurl = {https://ui.adsabs.harvard.edu/abs/2000ApJ...542..925L}
}

@ARTICLE{Bjerkeli2016,
       author = {{Bjerkeli}, Per and {van der Wiel}, Matthijs H.~D. and {Harsono}, Daniel and {Ramsey}, Jon P. and {J{\o}rgensen}, Jes K.},
        title = "{Resolved images of a protostellar outflow driven by an extended disk wind}",
      journal = {\nat},
         year = 2016,
        month = dec,
       volume = {540},
       number = {7633},
        pages = {406-409},
          doi = {10.1038/nature20600},
archivePrefix = {arXiv},
       eprint = {1612.05148},
 primaryClass = {astro-ph.SR},
       adsurl = {https://ui.adsabs.harvard.edu/abs/2016Natur.540..406B}
}

@ARTICLE{Tabone2020,
       author = {{Tabone}, B. and {Cabrit}, S. and {Pineau des For{\^e}ts}, G. and {Ferreira}, J. and {Gusdorf}, A. and {Podio}, L. and {Bianchi}, E. and {Chapillon}, E. and {Codella}, C. and {Gueth}, F.},
        title = "{Constraining MHD disk winds with ALMA. Apparent rotation signatures and application to HH212}",
      journal = {\aap},
         year = 2020,
        month = aug,
       volume = {640},
          eid = {A82},
        pages = {A82},
          doi = {10.1051/0004-6361/201834377},
archivePrefix = {arXiv},
       eprint = {2004.08804},
 primaryClass = {astro-ph.SR},
       adsurl = {https://ui.adsabs.harvard.edu/abs/2020A&A...640A..82T}
}

@ARTICLE{Lee2017,
       author = {{Lee}, Chin-Fei and {Ho}, Paul. T.~P. and {Li}, Zhi-Yun and {Hirano}, Naomi and {Zhang}, Qizhou and {Shang}, Hsien},
        title = "{A rotating protostellar jet launched from the innermost disk of HH 212}",
      journal = {Nature Astronomy},
         year = 2017,
        month = jul,
       volume = {1},
          eid = {0152},
        pages = {0152},
          doi = {10.1038/s41550-017-0152},
archivePrefix = {arXiv},
       eprint = {1706.06343},
 primaryClass = {astro-ph.GA},
       adsurl = {https://ui.adsabs.harvard.edu/abs/2017NatAs...1E.152L}
}

@ARTICLE{Shu1987,
       author = {{Shu}, Frank H. and {Adams}, Fred C. and {Lizano}, Susana},
        title = "{Star formation in molecular clouds: observation and theory.}",
      journal = {\araa},
         year = 1987,
        month = jan,
       volume = {25},
        pages = {23-81},
          doi = {10.1146/annurev.aa.25.090187.000323},
       adsurl = {https://ui.adsabs.harvard.edu/abs/1987ARA&A..25...23S}
}

@ARTICLE{Ray2023,
       author = {{Ray}, T.~P. and {McCaughrean}, M.~J. and {Caratti o Garatti}, A. and {Kavanagh}, P.~J. and {Justtanont}, K. and {van Dishoeck}, E.~F. and {Reitsma}, M. and {Beuther}, H. and {Francis}, L. and {Gieser}, C. and {Klaassen}, P. and {Perotti}, G. and {Tychoniec}, L. and {van Gelder}, M. and {Colina}, L. and {Greve}, Th. R. and {G{\"u}del}, M. and {Henning}, Th. and {Lagage}, P.~O. and {{\"O}stlin}, G. and {Vandenbussche}, B. and {Waelkens}, C. and {Wright}, G.},
        title = "{Outflows from the youngest stars are mostly molecular}",
      journal = {\nat},
         year = 2023,
        month = oct,
       volume = {622},
       number = {7981},
        pages = {48-52},
          doi = {10.1038/s41586-023-06551-1},
       adsurl = {https://ui.adsabs.harvard.edu/abs/2023Natur.622...48R}
}

@INPROCEEDINGS{Pudritz2007,
       author = {{Pudritz}, R.~E. and {Ouyed}, R. and {Fendt}, Ch. and {Brandenburg}, A.},
        title = "{Disk Winds, Jets, and Outflows: Theoretical and Computational Foundations}",
    booktitle = {Protostars and Planets V},
         year = 2007,
       editor = {{Reipurth}, Bo and {Jewitt}, David and {Keil}, Klaus},
        month = jan,
        pages = {277},
          doi = {10.48550/arXiv.astro-ph/0603592},
archivePrefix = {arXiv},
       eprint = {astro-ph/0603592},
 primaryClass = {astro-ph},
       adsurl = {https://ui.adsabs.harvard.edu/abs/2007prpl.conf..277P}
}

@ARTICLE{deValon2020,
       author = {{de Valon}, A. and {Dougados}, C. and {Cabrit}, S. and {Louvet}, F. and {Zapata}, L.~A. and {Mardones}, D.},
        title = "{ALMA reveals a large structured disk and nested rotating outflows in DG Tauri B}",
      journal = {\aap},
         year = 2020,
        month = feb,
       volume = {634},
          eid = {L12},
        pages = {L12},
          doi = {10.1051/0004-6361/201936950},
archivePrefix = {arXiv},
       eprint = {2001.09776},
 primaryClass = {astro-ph.SR},
       adsurl = {https://ui.adsabs.harvard.edu/abs/2020A&A...634L..12D}
}

@ARTICLE{Kim2026,
       author = {{Kim}, Chul-Hwan and {Lee}, Jeong-Eun and {Johnstone}, Doug and {Herczeg}, Gregory J. and {Lee}, Chin-Fei and {Francis}, Logan and {Sheehan}, Patrick D.},
        title = "{Direct evidence for magnetohydrodynamic disk winds driving rotating outflows in protostar HOPS 358}",
      journal = {Nature Communications},
         year = 2026,
        month = apr,
       volume = {17},
       number = {1},
          eid = {2957},
        pages = {2957},
          doi = {10.1038/s41467-026-71142-3},
       adsurl = {https://ui.adsabs.harvard.edu/abs/2026NatCo..17.2957K}
}

@ARTICLE{Gaudel2020,
       author = {{Gaudel}, M. and {Maury}, A.~J. and {Belloche}, A. and {Maret}, S. and {Andr{\'e}}, Ph. and {Hennebelle}, P. and {Galametz}, M. and {Testi}, L. and {Cabrit}, S. and {Palmeirim}, P. and {Ladjelate}, B. and {Codella}, C. and {Podio}, L.},
        title = "{Angular momentum profiles of Class 0 protostellar envelopes}",
      journal = {\aap},
         year = 2020,
        month = may,
       volume = {637},
          eid = {A92},
        pages = {A92},
          doi = {10.1051/0004-6361/201936364},
archivePrefix = {arXiv},
       eprint = {2001.10004},
 primaryClass = {astro-ph.SR},
       adsurl = {https://ui.adsabs.harvard.edu/abs/2020A&A...637A..92G}
}

@ARTICLE{Dunham2015,
       author = {{Dunham}, Michael M. and {Allen}, Lori E. and {Evans}, II, Neal J. and {Broekhoven-Fiene}, Hannah and {Cieza}, Lucas A. and {Di Francesco}, James and {Gutermuth}, Robert A. and {Harvey}, Paul M. and {Hatchell}, Jennifer and {Heiderman}, Amanda and {Huard}, Tracy L. and {Johnstone}, Doug and {Kirk}, Jason M. and {Matthews}, Brenda C. and {Miller}, Jennifer F. and {Peterson}, Dawn E. and {Young}, Kaisa E.},
        title = "{Young Stellar Objects in the Gould Belt}",
      journal = {\apjs},
         year = 2015,
        month = sep,
       volume = {220},
       number = {1},
          eid = {11},
        pages = {11},
          doi = {10.1088/0067-0049/220/1/11},
archivePrefix = {arXiv},
       eprint = {1508.03199},
 primaryClass = {astro-ph.GA},
       adsurl = {https://ui.adsabs.harvard.edu/abs/2015ApJS..220...11D}
}

@ARTICLE{Bai2016,
       author = {{Bai}, Xue-Ning and {Ye}, Jiani and {Goodman}, Jeremy and {Yuan}, Feng},
        title = "{Magneto-thermal Disk Winds from Protoplanetary Disks}",
      journal = {\apj},
         year = 2016,
        month = feb,
       volume = {818},
       number = {2},
          eid = {152},
        pages = {152},
          doi = {10.3847/0004-637X/818/2/152},
archivePrefix = {arXiv},
       eprint = {1511.06769},
 primaryClass = {astro-ph.EP},
       adsurl = {https://ui.adsabs.harvard.edu/abs/2016ApJ...818..152B}
}

@ARTICLE{Ferreira1997AA,
       author = {{Ferreira}, J.},
        title = "{Magnetically-driven jets from Keplerian accretion discs.}",
      journal = {\aap},
         year = 1997,
        month = mar,
       volume = {319},
        pages = {340-359},
          doi = {10.48550/arXiv.astro-ph/9607057},
archivePrefix = {arXiv},
       eprint = {astro-ph/9607057},
 primaryClass = {astro-ph},
       adsurl = {https://ui.adsabs.harvard.edu/abs/1997A&A...319..340F}
}

@ARTICLE{Ai2024,
       author = {{Ai}, Tsung-Han and {Liu}, Chun-Fan and {Shang}, Hsien and {Johnstone}, Doug and {Krasnopolsky}, Ruben},
        title = "{A Unified Model for Bipolar Outflows from Young Stars: Kinematic and Mixing Structures in HH 30}",
      journal = {\apj},
         year = 2024,
        month = apr,
       volume = {964},
       number = {2},
          eid = {147},
        pages = {147},
          doi = {10.3847/1538-4357/ad2355},
archivePrefix = {arXiv},
       eprint = {2402.02529},
 primaryClass = {astro-ph.SR},
       adsurl = {https://ui.adsabs.harvard.edu/abs/2024ApJ...964..147A}
}

@ARTICLE{Andrews2005,
       author = {{Andrews}, Sean M. and {Williams}, Jonathan P.},
        title = "{Circumstellar Dust Disks in Taurus-Auriga: The Submillimeter Perspective}",
      journal = {\apj},
         year = 2005,
        month = oct,
       volume = {631},
       number = {2},
        pages = {1134-1160},
          doi = {10.1086/432712},
archivePrefix = {arXiv},
       eprint = {astro-ph/0506187},
 primaryClass = {astro-ph},
       adsurl = {https://ui.adsabs.harvard.edu/abs/2005ApJ...631.1134A}
}

@ARTICLE{Polnitzky2026,
       author = {{Polnitzky}, Fabian A. and {Ratzenb{\"o}ck}, Sebastian and {Gro{\ss}schedl}, Josefa and {Alves}, Jo{\~a}o},
        title = "{Precise determination of circumstellar disk lifetimes: Disk evolution in a single star-forming region}",
      journal = {\aap},
         year = 2026,
        month = mar,
       volume = {707},
          eid = {A216},
        pages = {A216},
          doi = {10.1051/0004-6361/202554921},
archivePrefix = {arXiv},
       eprint = {2512.06873},
 primaryClass = {astro-ph.SR},
       adsurl = {https://ui.adsabs.harvard.edu/abs/2026A&A...707A.216P}
}

@ARTICLE{Guillermo2026,
       author = {{Bl{\'a}zquez-Calero}, Guillermo and {Anglada}, Guillem and {Cabrit}, Sylvie and {Osorio}, Mayra and {Raga}, Alejandro C. and {Fuller}, Gary A. and {G{\'o}mez}, Jos{\'e} F. and {Estalella}, Robert and {Diaz-Rodriguez}, Ana K. and {Torrelles}, Jos{\'e} M. and {Rodr{\'\i}guez}, Luis F. and {Mac{\'\i}as}, Enrique and {de Gregorio-Monsalvo}, Itziar and {Megeath}, S. Thomas and {Zapata}, Luis and {Ho}, Paul T.~P.},
        title = "{Bowshocks driven by the pole-on molecular jet of outbursting protostar SVS 13}",
      journal = {Nature Astronomy},
         year = 2026,
        month = jan,
       volume = {10},
        pages = {105-123},
          doi = {10.1038/s41550-025-02716-2},
archivePrefix = {arXiv},
       eprint = {2512.14458},
 primaryClass = {astro-ph.GA},
       adsurl = {https://ui.adsabs.harvard.edu/abs/2026NatAs..10..105B}
}

@ARTICLE{Hodapp2014,
       author = {{Hodapp}, Klaus W. and {Chini}, Rolf},
        title = "{The Launch Region of the SVS 13 Outflow and Jet}",
      journal = {\apj},
         year = 2014,
        month = oct,
       volume = {794},
       number = {2},
          eid = {169},
        pages = {169},
          doi = {10.1088/0004-637X/794/2/169},
archivePrefix = {arXiv},
       eprint = {1408.5940},
 primaryClass = {astro-ph.SR},
       adsurl = {https://ui.adsabs.harvard.edu/abs/2014ApJ...794..169H}
}

@ARTICLE{Lesur2021AA,
       author = {{Lesur}, Geoffroy R.~J.},
        title = "{Systematic description of wind-driven protoplanetary discs}",
      journal = {\aap},
         year = 2021,
        month = jun,
       volume = {650},
          eid = {A35},
        pages = {A35},
          doi = {10.1051/0004-6361/202040109},
archivePrefix = {arXiv},
       eprint = {2101.10349},
 primaryClass = {astro-ph.SR},
       adsurl = {https://ui.adsabs.harvard.edu/abs/2021A&A...650A..35L}
}

@ARTICLE{Tobin16,
       author = {{Tobin}, John J. and {Looney}, Leslie W. and {Li}, Zhi-Yun and {Chandler}, Claire J. and {Dunham}, Michael M. and {Segura-Cox}, Dominique and {Sadavoy}, Sarah I. and {Melis}, Carl and {Harris}, Robert J. and {Kratter}, Kaitlin and et al.},
        title = "{The VLA Nascent Disk and Multiplicity Survey of Perseus Protostars (VANDAM). II. Multiplicity of Protostars in the Perseus Molecular Cloud}",
      journal = {\apj},
         year = 2016,
        month = feb,
       volume = {818},
       number = {1},
          eid = {73},
        pages = {73},
          doi = {10.3847/0004-637X/818/1/73},
       adsurl = {https://ui.adsabs.harvard.edu/abs/2016ApJ...818...73T}
}

@ARTICLE{Tobin22,
       author = {{Tobin}, John J. and {Offner}, Stella S.~R. and {Kratter}, Kaitlin M. and {Megeath}, S. Thomas and {Sheehan}, Patrick D. and {Looney}, Leslie W. and {Diaz-Rodriguez}, Ana Karla and {Osorio}, Mayra and {Anglada}, Guillem and {Sadavoy}, Sarah I. and et al.},
        title = "{The VLA/ALMA Nascent Disk And Multiplicity (VANDAM) Survey of Orion Protostars. V. A Characterization of Protostellar Multiplicity}",
      journal = {\apj},
         year = 2022,
        month = jan,
       volume = {925},
       number = {1},
          eid = {39},
        pages = {39},
          doi = {10.3847/1538-4357/ac36d2},
       adsurl = {https://ui.adsabs.harvard.edu/abs/2022ApJ...925...39T}
}

@ARTICLE{Antoniucci2008,
       author = {{Antoniucci}, S. and {Nisini}, B. and {Giannini}, T. and {Lorenzetti}, D.},
        title = "{Accretion and ejection properties of embedded protostars: the case of HH26, HH34, and HH46 IRS}",
      journal = {\aap},
         year = 2008,
        month = feb,
       volume = {479},
       number = {2},
        pages = {503-514},
          doi = {10.1051/0004-6361:20077468},
archivePrefix = {arXiv},
       eprint = {0710.5609},
 primaryClass = {astro-ph},
       adsurl = {https://ui.adsabs.harvard.edu/abs/2008A&A...479..503A}
}

@ARTICLE{Siess2000,
       author = {{Siess}, L. and {Dufour}, E. and {Forestini}, M.},
        title = "{An internet server for pre-main sequence tracks of low- and intermediate-mass stars}",
      journal = {\aap},
         year = 2000,
        month = jun,
       volume = {358},
        pages = {593-599},
          doi = {10.48550/arXiv.astro-ph/0003477},
archivePrefix = {arXiv},
       eprint = {astro-ph/0003477},
 primaryClass = {astro-ph},
       adsurl = {https://ui.adsabs.harvard.edu/abs/2000A&A...358..593S}
}

@ARTICLE{Bacciotti2002,
       author = {{Bacciotti}, Francesca and {Ray}, Thomas P. and {Mundt}, Reinhard and {Eisl{\"o}ffel}, Jochen and {Solf}, Josef},
        title = "{Hubble Space Telescope/STIS Spectroscopy of the Optical Outflow from DG Tauri: Indications for Rotation in the Initial Jet Channel}",
      journal = {\apj},
         year = 2002,
        month = sep,
       volume = {576},
       number = {1},
        pages = {222-231},
          doi = {10.1086/341725},
archivePrefix = {arXiv},
       eprint = {astro-ph/0206175},
 primaryClass = {astro-ph},
       adsurl = {https://ui.adsabs.harvard.edu/abs/2002ApJ...576..222B},
}

@ARTICLE{Jorgensen2022,
       author = {{J{\o}rgensen}, Jes K. and {Kuruwita}, Rajika L. and {Harsono}, Daniel and {Haugb{\o}lle}, Troels and {Kristensen}, Lars E. and {Bergin}, Edwin A.},
        title = "{Binarity of a protostar affects the evolution of the disk and planets}",
      journal = {\nat},
         year = 2022,
        month = may,
       volume = {606},
       number = {7913},
        pages = {272-275},
          doi = {10.1038/s41586-022-04659-4},
       adsurl = {https://ui.adsabs.harvard.edu/abs/2022Natur.606..272J}
}

@INPROCEEDINGS{Briggs1995,
       author = {{Briggs}, D.~S.},
        title = "{High Fidelity Interferometric Imaging: Robust Weighting and NNLS Deconvolution}",
    booktitle = {American Astronomical Society Meeting Abstracts},
         year = 1995,
       series = {American Astronomical Society Meeting Abstracts},
       volume = {187},
        month = dec,
          eid = {112.02},
        pages = {112.02},
       adsurl = {https://ui.adsabs.harvard.edu/abs/1995AAS...18711202B}
}

@ARTICLE{Tabone2017,
       author = {{Tabone}, B. and {Cabrit}, S. and {Bianchi}, E. and {Ferreira}, J. and {Pineau des For{\^e}ts}, G. and {Codella}, C. and {Gusdorf}, A. and {Gueth}, F. and {Podio}, L. and {Chapillon}, E.},
        title = "{ALMA discovery of a rotating SO/SO$_{2}$ flow in HH212. A possible MHD disk wind?}",
      journal = {\aap},
         year = 2017,
        month = nov,
       volume = {607},
          eid = {L6},
        pages = {L6},
          doi = {10.1051/0004-6361/201731691},
archivePrefix = {arXiv},
       eprint = {1710.01401},
 primaryClass = {astro-ph.SR},
       adsurl = {https://ui.adsabs.harvard.edu/abs/2017A&A...607L...6T}
}

@ARTICLE{Coffey2007,
       author = {{Coffey}, Deirdre and {Bacciotti}, Francesca and {Ray}, Thomas P. and {Eisl{\"o}ffel}, Jochen and {Woitas}, Jens},
        title = "{Further Indications of Jet Rotation in New Ultraviolet and Optical Hubble Space Telescope STIS Spectra}",
      journal = {\apj},
         year = 2007,
        month = jul,
       volume = {663},
       number = {1},
        pages = {350-364},
          doi = {10.1086/518100},
archivePrefix = {arXiv},
       eprint = {astro-ph/0703271},
 primaryClass = {astro-ph},
       adsurl = {https://ui.adsabs.harvard.edu/abs/2007ApJ...663..350C}
}
\bibliographystyle{aasjournalv7}



\newpage

\end{document}